\documentclass[a4paper,12pt]{article}
\usepackage[dvipdfmx]{graphicx}
\usepackage{amsfonts,amsthm,amsmath,amssymb,graphicx,float,mathtools}
\usepackage[T1]{fontenc}
\usepackage{cite}
\numberwithin{equation}{section}
\usepackage{subfigure}
\usepackage{color}
\usepackage{ulem}
\usepackage{bbm} 
\usepackage{tensor} 
\usepackage{verbatim} 
\usepackage{subfigure}
\usepackage[
	  pagebackref=false,
	  colorlinks=true,
      linkcolor=blue,
      urlcolor=blue,
      filecolor=black,
      citecolor=red,
      pdfstartview=FitV,
      pdftitle={},
        pdfauthor={},
        pdfsubject={},
        pdfkeywords={},
        pdfpagemode=None,
        bookmarksopen=true
      ]{hyperref}
\usepackage{ascmac}

\begin{document}

\newtheorem{definition}{Definition}[section]
\newcommand{\be}{\begin{equation}}
\newcommand{\ee}{\end{equation}}
\newcommand{\bea}{\begin{eqnarray}}
\newcommand{\eea}{\end{eqnarray}}
\newcommand{\LE}{\left[}
\newcommand{\R}{\right]}
\newcommand{\nn}{\nonumber}
\newcommand{\Tr}{\text{Tr}}
\newcommand{\N}{\mathcal{N}}
\newcommand{\G}{\Gamma}
\newcommand{\vf}{\varphi}
\newcommand{\LL}{\mathcal{L}}
\newcommand{\Op}{\mathcal{O}}
\newcommand{\HH}{\mathcal{H}}
\newcommand{\arctanh}{\text{arctanh}}
\newcommand{\up}{\uparrow}
\newcommand{\down}{\downarrow}
\newcommand{\ket}[1]{\left| #1 \right>}
\newcommand{\bra}[1]{\left< #1 \right|}
\newcommand{\ketbra}[1]{\left|#1\right>\left<#1\right|}
\newcommand{\rd}{\partial}
\newcommand{\de}{\partial}
\newcommand{\ba}{\begin{eqnarray}}
\newcommand{\ea}{\end{eqnarray}}
\newcommand{\db}{\bar{\partial}}
\newcommand{\we}{\wedge}
\newcommand{\ca}{\mathcal}
\newcommand{\lr}{\leftrightarrow}
\newcommand{\f}{\frac}
\newcommand{\s}{\sqrt}
\newcommand{\vp}{\varphi}
\newcommand{\hvp}{\hat{\varphi}}
\newcommand{\tvp}{\tilde{\varphi}}
\newcommand{\tp}{\tilde{\phi}}
\newcommand{\ti}{\tilde}
\newcommand{\ap}{\alpha}
\newcommand{\pr}{\propto}
\newcommand{\mb}{\mathbf}
\newcommand{\ddd}{\cdot\cdot\cdot}
\newcommand{\no}{\nonumber \\}
\newcommand{\la}{\langle}
\newcommand{\lb}{\rangle}
\newcommand{\ep}{\epsilon}
 \def\we{\wedge}
 \def\lr{\leftrightarrow}
 \def\f {\frac}
 \def\ti{\tilde}
 \def\ap{\alpha}
 \def\pr{\propto}
 \def\mb{\mathbf}
 \def\ddd{\cdot\cdot\cdot}
 \def\no{\nonumber \\}
 \def\la{\langle}
 \def\lb{\rangle}
 \def\ep{\epsilon}
\newcommand{\mcl}{\mathcal}
 \def\g{\gamma}
\def\Tr{\text{tr}}

\begin{titlepage}
\thispagestyle{empty}

\begin{flushright}

\end{flushright}
\bigskip

\begin{center}
  \noindent{\large \textbf{Inhomogeneous quenches as state preparation in two-dimensional conformal field theories}}\\
\vspace{2cm}

\renewcommand\thefootnote{\mbox{$\fnsymbol{footnote}$}}
Masahiro Nozaki\footnote{mnozaki@ucas.ac.cn}${}^{1,2}$,
 Kotaro Tamaoka\footnote{tamaoka.kotaro@nihon-u.ac.jp}${}^{3}$ and Mao Tian Tan\footnote{maotian.tan@apctp.org}${}^{4}$\\

\vspace{1cm}
${}^{1}${\small \sl Kavli Institute for Theoretical Sciences, University of Chinese Academy of Sciences,
Beijing 100190, China}\\
${}^{2}${\small \sl RIKEN Interdisciplinary Theoretical and Mathematical Sciences (iTHEMS), \\Wako, Saitama 351-0198, Japan}\\
${}^{3}${\small \sl Department of Physics, College of Humanities and Sciences, Nihon University, \\Sakura-josui, Tokyo 156-8550, Japan}\\
${}^{4}${\small \sl Asia Pacific Center for Theoretical Physics, Pohang, Gyeongbuk, 37673, Korea}\\

\vskip 4em
\end{center}
\begin{abstract}
The non-equilibrium process where the system does not evolve to the featureless state is one of the new central objects in the non-equilibrium phenomena. 
In this paper, starting from the short-range entangled state in the two-dimensional conformal field theories ($2$d CFTs), the boundary state with a regularization, we evolve the system with the inhomogeneous Hamiltonians called M\"obius/SSD ones.
Regardless of the details of CFTs considered in this paper, during the M\"obius evolution, the entanglement entropy exhibits the periodic motion called quantum revival.
During SSD time evolution, except for some subsystems, in the large time regime,  entanglement entropy and mutual information are approximated by those for the vacuum state.
We argue the time regime for the subsystem to cool down to vacuum one is $t_1 \gg \mathcal{O}(L\sqrt{l_A})$, where $t_1$, $L$, and $l_A$ are time, system, and subsystem sizes.
This finding suggests the inhomogeneous quench induced by the SSD Hamiltonian may be used as the preparation for the approximately-vacuum state.
We propose the gravity dual of the systems considered in this paper, furthermore, and generalize it.
In addition to them, we discuss the relation between the inhomogenous quenches and continuous multi-scale entanglement renormalization ansatz (cMERA).
\end{abstract}
\end{titlepage} 
\tableofcontents

\section{Introduction}

A major trend in twenty-first-century theoretical physics is the extensive utilization of quantum information-theoretic ideas and techniques in a wide range of seemingly disparate sub-fields of theoretical physics. For example, in the study of holography \cite{Maldacena:1997re}, an understanding of the quantum mechanical properties possessed by the system is expected to lead to a deeper grasp of the subject \cite{2006PhRvL..96r1602R,2006JHEP...08..045R}. One such quantum information-theoretic concept is that of quantum error correction which originated from the field of quantum computing but has since been used for bulk reconstruction in holography \cite{Almheiri:2014lwa, Pastawski:2015qua} as well as the study of measurement-induced phase transition \cite{Choi:2019nhg,2021PhRvB.103j4306L} by the condensed matter community. Another topic from the field of quantum computation that might be of relevance to condensed matter physics is the preparation of quantum states \cite{2019PhRvB..99j4308M,2018PhRvL.120u0604A,1995AmJPh..63..767G,PhysRevLett.101.170503,PhysRevE.84.061152,PhysRevA.86.052334,PhysRevLett.111.100502,PhysRevA.88.040101,1998Sci...280..421G,2017PNAS..114E3909S,2016NatSR...634187V,PhysRevLett.117.170501,PhysRevA.95.012317,PhysRevLett.116.230503,PhysRevE.95.012148,PhysRevLett.111.260501,2019ScPP....6...29H,2018PhRvA..97f2343B,PhysRevX.7.021027,2018PhRvX...8c1086B}. In particular, obtaining low entropy states will allow the simulation of exotic phases of matter ranging from antiferromagnetic spin liquids to high-temperature superconductors \cite{Zaletel_2021,2009arXiv0911.5506H}. One of the desired outcomes of state preparation is the preparation of these states faster than an adiabatic evolution of the system. In holography, preparing quantum states corresponds to the production of asymptotically AdS spacetimes.

Since two-dimensional conformal field theories ($2$d CFTs) possess the infinite-dimensional Virasoro symmetry, they may allow an analytic treatment of vacuum state preparation. 
In this paper, we consider the inhomogeneous quenches induced by the so-called M\"obius and sine-squared deformed (SSD) Hamiltonians in $2$d CFTs \cite{PhysRevB.97.184309,2019JPhA...52X5401M,Goto:2021sqx, PhysRevLett.118.260602,2018arXiv180500031W,2020PhRvX..10c1036F,Han_2020,2021PhRvR...3b3044W,2020arXiv201109491F,2021arXiv210910923W,PhysRevB.103.224303, PhysRevResearch.2.023085, Moosavi2021, PhysRevLett.122.020201,10.21468/SciPostPhys.3.3.019, Gaw_dzki_2018,10.21468/SciPostPhys.2.1.002} and explore the utility of these quenches in vacuum state preparation.
The densities of these inhomogeneous Hamiltonians are modulated by envelope functions that vary in space.
In addition, these inhomogeneous Hamiltonians can be thought of as the Hamiltonians of systems that exist on a curved spacetime described by a metric whose time component is determined by the envelope function.
Originally, the modulation of energy density was used to remove the effect of the boundaries of spin systems with finite size\cite{PhysRevB.83.060414,2009PThPh.122..953G,2011PhRvA..83e2118G}. 
Subsequently, these inhomogeneous deformations were generalized to $2$d CFTs \cite{PhysRevB.84.165132,PhysRevB.84.165132,2012JPhA...45k5003K,2015JPhA...48E5402I,2016IJMPA..3150170I,2016arXiv160309543O,2017arXiv170906238T,2018PTEP.2018f1B01T}.
These inhomogenous Hamiltonians were used in the implementation of $2$d Floquet CFTs \cite{2020PhRvX..10c1036F, Han_2020, PhysRevB.103.224303, PhysRevResearch.2.023085,fan2021floquet,wen2018floquet}.
In addition, they have also been used to explore a variety of non-equilibrium phenomena and quantum information-theoretic aspects of $2$d CFTs \cite{Goto:2023wai,deBoer:2023lrd,Caputa:2022zsr,10.21468/SciPostPhys.14.5.108,Kudler-Flam:2023ahk,Liu:2023tiq} and non CFTs \cite{Goto:2023yxb}.

In \cite{Goto:2021sqx}, the authors found that the time evolution induced by the SSD Hamiltonian can be used to approximately prepare the vacuum state.
In the setup considered in \cite{Goto:2021sqx}, the system begins in the thermal state and subsequently undergoes time evolution with the SSD Hamiltonian which has a spatial point where the envelope function vanishes.
If the subsystem does not include this point, the entanglement entropy associated with this subsystem evolves to that of the vacuum state.
Regardless of the spatial location and the size of subsystems, the mutual information evolves into the mutual information for the vacuum state.
In \cite{Wen:2022pyj}, this setup is generalized to a Floquet time evolution, and the authors consider the cooling effect of the inhomogeneous Hamiltonian. 
 
In previous studies, the SSD time evolution was found to deform the entanglement structure of the long-range entangled state, where the entanglement entropy is proportional to the subsystem size, so that the resulting reduced density matrix is approximately given by that of the vacuum state.
In this paper, we will explore whether this vacuum entanglement structure emerges during time evolution with the SSD Hamiltonian, starting from a short-range entangled state.
The short-range entangled state considered in this paper is the regularized boundary state \cite{Calabrese:2005in,Miyaji:2014mca}.
Since short-range entangled states are easily prepared in the laboratory, this study may pave the way for creating low entropy states using quantum quenches.

\subsubsection*{Summary}
In this paper, starting from a short-range entangled state which we take to be the regulated boundary state, we evolve the system with M\"obius/SSD Hamiltonians in $2$d CFTs.
The major difference from previous studies \cite{2019PhRvB..99j4308M,2018PhRvL.120u0604A} is that we aim to explore whether vacuum states occur in CFTs with differing ability to scramble information which may determine the speed of information processing \cite{Margolus:1997ih}.
Our findings are as follows:
{\bf Entanglement entropy:} During M\"obius time evolution, the time dependence of entanglement entropy exhibits a periodic behavior in time called quantum revival.
During the SSD time evolution, when the edges of subsystems are not located at the origin, regardless of the spatial location of the subsystems, at sufficiently late times, the entanglement entropies are approximated by that of the vacuum state.
If the subsystem size, $l_A$, is much smaller than the system size, $L$, the time regime for the entanglement entropy for the single intervals to cool down to the vacuum one is $t_1 \gg t_* \approx \mathcal{O}(L\sqrt{L_A/\epsilon})$, where $t_1$ and $\epsilon$ are time and the parameter determining the short-range entanglement of the initial state. 
We argue that $t_*$ characterizes the time for the subsystem to cool down to the vacuum one for the SSD time evolution.
In addition, when the edges of subsystems are located at the origin, for sufficiently large times, the entanglement entropy logarithmically grows with time in the holographic CFTs while it approaches the saturation value with a power law in the free fermion CFTs.
{\bf Mutual information:} During the SSD time evolution, in the large time regime, mutual information eventually saturates to that of the vacuum one for both the free fermion CFT and the holographic CFTs while it goes to zero in the quasiparticle picture. 
This indicates that non-local correlations emerge under time evolution by these inhomogeneous Hamiltonians.
The time dependence of entanglement entropy and mutual information suggests that the SSD time evolution prepares the state with the entanglement structure and non-local correlations possessed by the vacuum state.
{\bf Gravity dual:} We propose the gravity dual of the system considered and also generalize it.
In particular, the end of the world brane approaches to the asymptotic boundary and eventually collides with the cutoff surface.  We argue that this is a time scale over which holographic calculations become unreliable. Furthermore, we discuss an interpretation of the M\"obius/SSD time evolution as a type of tensor network called the continuous multi-scale entanglement renormalization ansatz (cMERA).  

\subsubsection*{Organization of this paper}
In Section \ref{Section:Preliminary}, we will describe the details of the inhomogeneous quench, how to compute entanglement entropy in the twist operator formalism, and the evolution of local operators induced by the M\"obius/SSD Hamiltonians.
In Section \ref{Section:Evolution-in-CFT}, we will present the time evolution of entanglement entropy in the $2$d Dirac fermion CFT and in the quasiparticle picture that captures the dynamics of entanglement in the Dirac fermion CFT, as well as the entanglement entropy for the single interval in the $2$d holographic CFTs which possess a gravitational dual.
In Section \ref{Section:Mutual-informaion-in-CFTs}, we will report the time evolution of mutual information in $2$d CFTs considered in this paper.
In Section \ref{Section:GD-and-cMERA}, we will present the gravity dual of the systems considered and give an interpretation of the quenches as a cMERA tensor network.
In Section \ref{Section:Discussions}, we discuss the relation between the inhomogeneous quenches and renormalization group, and comment on future directions.

\section{Inhomogeneous quenches from the boundary state \label{Section:Preliminary}}
Here, we will describe the inhomogeneous quenches considered in this paper.
Suppose that we prepare the system in the boundary state with a proper regularization, $e^{-\epsilon H}\ket{\Psi_0}$, \cite{Calabrese:2005in} and then evolve it with inhomogeneous Hamiltonian $H_{\text{Inho}}$:
\be
\ket{\Psi(t)}=\mathcal{N}e^{-itH_{\text{Inho}}}e^{-\epsilon H}\ket{\Psi_0},
\ee
where $\mathcal{N}^2$ is the normalization constant that guarantees that $\left\langle \Psi(t) | \Psi(t) \right \rangle=1$. The operator $H$ is the homogeneous Hamiltonian defined as $H=\int^L_0dx h(x)$, where $h(x)$ is Hamiltonian density and $L$ is the system size. 
One of the inhomogeneous Hamiltonians considered in this paper is the M\"obius Hamiltonian defined as \cite{wen2018floquet,Wen_2018,fan2021floquet,Han_2020,wen2021periodically},
\be
H_{\text{M\"obius}} =\int^{L}_0dx \left[1-\tanh{(2\theta)}\cos{\left(\f{2\pi x}{L}\right)}\right]h(x),
\ee
where  $\theta$ is a real parameter. 
The system considered in this paper is on the spatial circle with the circumference $L$.
For $\infty>\theta > 0$, $H_{\text{Inho}}=H_{\text{M\"obius}}$ is called M\"obius Hamiltonian, and in the SSD limit where $\theta \rightarrow \infty$, it reduces to so-called sine-squared deformed Hamiltonian (SSD Hamiltonian) defined as 
\be
H_{\text{SSD}} =\int^{L}_0dx~ 2 \sin^2{\left(\f{\pi x}{L}\right)}h(x).
\ee
On the other hand, when $\theta=0$, $H_{\text{M\"obius}}$ reduces to the homogeneous Hamiltonian.
In this paper, by using the time dependence of entanglement entropy and mutual information during the evolution induced by the M\"obius/SSD Hamiltonian in the two-dimensional conformal field theories ($2$d CFTs), we will explore the entanglement structure of the state. 
The entanglement entropy and mutual information are defined as the von Neumann entropy for the reduced density matrix and a linear combination of the entanglement entropies respectively. Denote the density matrix of the state as $\rho(t)=\ket{\Psi(t)}\bra{\Psi(t)}$. Divide the Hilbert space into $A$ and its complement $\overline{A}$, and define the reduced density matrix associated to $A$ as $\rho_A(t)=\Tr_{\overline{A}} \rho(t)$. 
The entanglement entropy for $A$ is defined as the von Neumann entropy of $\rho_A(t)$, 
\be
S_A(t)=-\Tr_A \rho_A(t) \log{\rho_A(t)}.
\ee
Let us now turn to the detailed definition of the mutual information. Divide the Hilbert space into $A\cup B$ and $\overline{A \cup B}$, define the entanglement entropies associated with $A,B,$ and $A\cup B$, and define the mutual information as the linear combination of these entanglement entropies,
\be
I_{A,B}=S_A+S_B-S_{A\cup B},
\ee
where $S_A$, $S_B$, and $S_{A\cup B}$ are the entanglement entropies associated with $A$, $B$ and $A\cup B$.
\subsubsection*{The parameter region considered in this paper}
The parameter regime where we explore the inhomogeneous non-equilibrium process in this paper is
\be
L\gg l_{\mathcal{V}}, t_1 \gg \epsilon \gg 1, 
\ee
where $l_{\mathcal{V}}$ denotes the size of the subsystem $\mathcal{V}$ and $t_1$ is the time associated with the M\"obius/SSD Hamiltonian.
\subsection{Entanglement entropy in the twist operator formalism}
To employ the Euclidean path-integral formalism suitable to the analytical computation of the entanglement entropy in $2$d CFTs, we define the Euclidean density operator as
\be
\rho_{E}=\mathcal{N}^2e^{-\tau_1 H_{\text{Inho}}}e^{-\epsilon H} \ket{\Psi_0}\bra{\Psi_0}e^{-\epsilon H} e^{\tau_1 H_{\text{Inho}}},
\ee
where $\epsilon$ is a regularization parameter, and $\tau_1$ is a real Euclidean time.
The normalization parameter $\mathcal{N}$ guarantees that $\Tr\rho_E=1$, and it satisfies $\mathcal{N}^{-2}=\bra{\Psi_0}e^{-2\epsilon H} \ket{\Psi_0}$. 

Divide the Hilbert space into $\mathcal{V}$ and $\overline{\mathcal{V}}$, the complement space to $\mathcal{V}$, and then define a reduced Euclidean density matrix associated with $\mathcal{V}$ as $\rho_{E,\mathcal{V}}=\Tr_{\overline{\mathcal{V}}}\rho_E$.  
Subsequently, define Euclidean entanglement entropy of $\mathcal{V}$ as the von Neumann entropy of $\rho_{E,\mathcal{V}}$:
\be
S_{E,\mathcal{V}}=\lim_{n\rightarrow 1}S^{(n)}_{E,\mathcal{V}}=\lim_{n\rightarrow 1}\f{1}{1-n}\log{\left[\Tr_{\mathcal{V}}\rho^n_{E,\mathcal{V}}\right]}=-\Tr_{\mathcal{V}}\left[\rho_{E,\mathcal{V}}\log{\rho_{E,\mathcal{V}}}\right],
\ee
where we call $S^{(n)}_{E,\mathcal{V}}$ the $n$-th R\'enyi entanglement entropy. 
In the path-integral formalism, $n$-th R\'enyi entanglement entropy is given by $\Tr_{\mathcal{V}}\rho^n_{E,\mathcal{V}}$ which is the partition function on a $n$-sheeted geometry where each of sheets is a finite cylinder.

Let us assume that $\mathcal{V}$ is a single interval.
In the twist operator formalism, $\Tr_{\mathcal{V}}\rho^n_{E,\mathcal{V}}$ is given by the two-point function of the twist and anti-twist operators.
As a consequence, the $n$-th R\'enyi entanglement entropy is given by 
\be
S^{(n)}_{E,\mathcal{V}}=\f{1}{1-n}\log{\left[\f{\bra{\Psi_0}e^{-\epsilon H}e^{\tau_1 H_{\text{Inho}}}\mathcal{T}_{n}(v_1)e^{-\tau_1 H_{\text{Inho}}}e^{\tau_1 H_{\text{Inho}}}\overline{\mathcal{T}}_n(v_2)e^{-\tau_1 H_{\text{Inho}}}e^{-\epsilon H}\ket{\Psi_0}}{\bra{\Psi_0}e^{-2\epsilon H}\ket{\Psi_0}}\right]},
\ee
where $v_1$ and $v_2$ denote the edges of $\mathcal{V}$, respectively. 
Here, $\mathcal{T}_{n}$ and $\overline{\mathcal{T}}_{n}$ are the primary operators called the twist and anti-twist operators.
Their conformal dimensions are $(h_n, \overline{h}_n)=(\f{c}{24}\left(n-\f{1}{n}\right),\f{c}{24}\left(n-\f{1}{n}\right))$.
Here, we assume that $L>v_1>v_2>0$.
As in \cite{Goto:2021sqx}, the evolution of the primary operator $\mathcal{O}(w,\overline{w})$ induced by M\"obius/SSD Hamiltonian is given by
\be
e^{\tau_1 H_{\text{Inho}}}\mathcal{O}(w,\overline{w})e^{-\tau_1 H_{\text{Inho}}}=\left(\f{dw^{\text{New},\alpha=0,1}}{dw}\right)^{h_n}\left(\f{d\overline{w}^{\text{New},\alpha=0,1}}{d\overline{w}}\right)^{h_n}\mathcal{O}(w^{\text{New},\alpha=0,1},\overline{w}^{\text{New},\alpha=0,1}),
\ee
where $(w^{\text{New},\alpha=0,1},\overline{w}^{\text{New},\alpha=0,1})$ is the location of the operator during the Euclidean time evolution induced by the M\"obius ($\alpha=1$) and SSD ($\alpha=0$) Hamiltonians, respectively. The details of $(w^{\text{New},\alpha},\overline{w}^{\text{New},\alpha})$ are reported in Appendix \ref{App:thelocofop}.
The Euclidean R\'enyi entanglement entropy is given by the ``free energy'' of two point function on the finite cylinder where the length along Euclidean time is $2\epsilon$, and the circumference of the spatial circle is $L$,
\be \label{EREE}
\begin{split}
S^{(n)}_{E,\mathcal{V}}&
=\f{h_n}{1-n}\log{\left[\prod_{i=1,,2}\left(\f{dw^{\text{New},\alpha}_{v_i}}{dw_{v_i}}\right)\left(\f{d\overline{w}^{\text{New},\alpha}_{v_i}}{d\overline{w}_{v_i}}\right)\right]}\\
&+\f{1}{(1-n)}\log{\left[\f{\bra{\Psi_0}e^{-2\epsilon H}\mathcal{T}_{n}\left(w^{\text{New},\alpha}_{v_1}+\epsilon, \overline{w}^{\text{New},\alpha}_{v_1}+\epsilon\right)\overline{\mathcal{T}}_n\left(w^{\text{New},\alpha}_{v_2}+\epsilon, \overline{w}^{\text{New},\alpha}_{v_2}+\epsilon\right)\ket{\Psi_0}}{\bra{\Psi_0}e^{-2\epsilon H}\ket{\Psi_0}}\right]}.\\
\end{split}
\ee
In the von Neumann limit where $n \rightarrow 1$, the Euclidean R\'enyi entanglement entropy reduces to the Euclidean entanglement entropy,
\be \label{eq:EEE}
\begin{split}
S_{E,\mathcal{V}}&=-\f{c}{12}\log{\left[\prod_{i=1,,2}\left(\f{dw^{\text{New},\alpha}_{v_i}}{dw_{v_i}}\right)\left(\f{d\overline{w}^{\text{New},\alpha}_{v_i}}{d\overline{w}_{v_i}}\right)\right]}\\
&+\lim_{n\rightarrow 1}\f{1}{(1-n)}\log{\left[\f{\bra{\Psi_0}e^{-2\epsilon H}\mathcal{T}_{n}\left(w^{\text{New},\alpha}_{v_1}+\epsilon, \overline{w}^{\text{New},\alpha}_{v_1}+\epsilon\right)\overline{\mathcal{T}}_n\left(w^{\text{New},\alpha}_{v_2}+\epsilon, \overline{w}^{\text{New},\alpha}_{v_2}+\epsilon\right)\ket{\Psi_0}}{\bra{\Psi_0}e^{-2\epsilon H}\ket{\Psi_0}}\right]}.\\
\end{split}
\ee

\subsection{Entanglement entropy for the single interval in $2$d massless free fermion}

The entanglement entropy of the free Dirac fermion after an inhomogeneous quench of a boundary state can be computed using bosonization similar to that in \cite{Takayanagi2010,Takayanagi2022} but with the time evolution Hamiltonian replaced with the M\"{o}bius Hamiltonian. The n-th moment of the reduced density matrix for a single interval is
\begin{align}
    &\langle \Psi_0|e^{-2\epsilon H} \mathcal{T}_n(w_i,\bar{w}_i) \overline{\mathcal{T}}_n(w_j,\bar{w}_j)|\Psi_0\rangle_{\text{Cylinder}}  \nonumber \\
    =&\left(\frac{2\pi}{L}\right)^{4h_n} \prod_{a=-\frac{n-1}{2}}^{\frac{n-1}{2}}
    \frac{\langle \Psi_0 |e^{-2\epsilon H}\mathcal{T}^{(a)}(y_i,\bar{y}_i)\mathcal{T}^{(-a)}(y_j,\bar{y}_j)|\Psi_0\rangle}{\langle \Psi_0 |e^{-2\epsilon H}|\Psi_0\rangle}
\end{align}
where $|\Psi_0\rangle$ is the boundary state that lives on the ends of the cylinder and there is a conformal factor that comes from rescaling the correlation function so that the spatial coordinate has a periodicity of $L$ instead of $2\pi$. Let $y = \tau-i\sigma$ where $0\leq \tau \leq 2\epsilon$ and $0\leq \sigma \leq 2\pi$ be the holomorphic coordinate on this rescaled cylinder. Applying the bosonization dictionary, the twist operators can be written as twisted vertex operators that depend on the boundary condition,
\begin{equation}
    \mathcal{T}^{(a)}(y,\bar{y}) = \begin{cases}
    \mathcal{T}^{(a)}_1(y,\bar{y})=V_{(\frac{a}{n},-\frac{a}{n})}(y,\bar{y})=e^{i\frac{a}{n}(X_L(y)-X_R(\bar{y}))} & \text{Neumann B.C.} \\
    \mathcal{T}^{(a)}_2(y,\bar{y})=V_{(\frac{a}{n},\frac{a}{n})}(y,\bar{y})=e^{i\frac{a}{n}(X_L(y)+X_R(\bar{y}))} & \text{Dirichlet B.C.} 
    \end{cases}.
\end{equation}
The correlation function of vertex operators in the boundary state can be computed by writing the vertex operators and the boundary state in terms of the bosonic modes and by repeatedly applying the bosonic commutation and the Baker-Campbell-Haursdorff relations \cite{Takayanagi2010}. The denominator is the partition function for the finite cylinder with boundary states $|\Psi_0\rangle$ on both ends as is given by
\begin{equation}
    \langle \Psi_0 |e^{-2\epsilon H}|\Psi_0\rangle = \begin{cases}
    \frac{\theta_3(0| \frac{i4\epsilon}{L})+\theta_2(0| \frac{i4\epsilon}{L})}{\eta\left(\frac{i4\epsilon}{L}\right)},& \text{Neumann B.C.} \\
    \frac{\theta_3(0| \frac{i4\epsilon}{L})}{\eta\left(\frac{i4\epsilon}{L}\right)},& \text{Dirichlet B.C.}
    \end{cases}
\end{equation}
Therefore, the correctly normalized Euclidean R\'enyi entanglement entropy to leading order in $\frac{L}{\epsilon}$ is 
\begin{align}\label{RenyiEntropySingleInterval}
    &S_{\mathcal{V}}^{(n)} \nonumber \\
    =& -\frac{c}{24}\frac{n+1}{n}\log\left|\prod_{i=1,2}\left(\frac{dw_{v_i}^{\text{New},\alpha}}{dw_{v_i}}\right)
    \left(\frac{d\bar{w}_{v_i}^{\text{New},\alpha}}{d\bar{w}_{v_i}}\right)
    \right|\nonumber \\
    +&
    \frac{n+1}{12n}\log\left(
    \frac{L}{2\pi}
    \right)^2 \Bigg|\bigg[\theta_1\left(i\frac{w_{v_2}^{\text{New},\alpha}-w_{v_1}^{\text{New},\alpha}}{L}\bigg| \frac{4i\epsilon}{L}\right) \theta_1\left(i\frac{\bar{w}_{v_2}^{\text{New},\alpha}-\bar{w}_{v_1}^{\text{New},\alpha}}{L}\bigg|\frac{4i\epsilon}{L}\right)\nonumber \\ 
    \times&\theta_1\left(i\frac{w_{v_1}^{\text{New},\alpha}+\bar{w}_{v_1}^{\text{New},\alpha}}{L}\bigg|\frac{4i\epsilon}{L}\right)\theta_1\left(i\frac{w_{v_2}^{\text{New},\alpha}+\bar{w}_{v_2}^{\text{New},\alpha}}{L}\bigg|\frac{4i\epsilon}{L}\right)\bigg]\bigg/\bigg[\eta\left(i\frac{4\epsilon}{L}\right)^6 \nonumber \\
    \times&\theta_1\left(i\frac{w_{v_1}^{\text{New},\alpha}+\bar{w}_{v_2}^{\text{New},\alpha}}{L}\bigg| i\frac{4\epsilon}{L}\right)\theta_1\left(i\frac{w_{v_2}^{\text{New},\alpha}+\bar{w}_{v_1}^{\text{New},\alpha}}{L}\bigg| i\frac{4\epsilon}{L}\right)\bigg]
    \Bigg| \nonumber\\
    +&
    \begin{cases}
    \frac{1}{1-n}\sum\limits_{a=-\frac{n-1}{2}}^{\frac{n-1}{2}} \log\Bigg|
 \bigg[\theta_2\left(i\frac{a(w_{v_1}^{\text{New},\alpha}-w_{v_2}^{\text{New},\alpha}+\bar{w}_{v_1}^{\text{New},\alpha}-\bar{w}_{v_2}^{\text{New},\alpha})}{Ln}\bigg| \frac{i4\epsilon}{L}\right) &\\
 +\theta_3\left(i\frac{a(w_{v_1}^{\text{New},\alpha}-w_{v_2}^{\text{New},\alpha}+\bar{w}_{v_1}^{\text{New},\alpha}-\bar{w}_{v_2}^{\text{New},\alpha})}{Ln}\bigg| \frac{4i\epsilon}{L}\right)\bigg]\bigg/\bigg[\theta_2\left(0\bigg|\frac{4i\epsilon}{L}\right)+\theta_3\left(0\bigg| \frac{i4\epsilon}{L}\right)\bigg]
    \Bigg| & \text{Neumann} \\ 
    \frac{1}{1-n}\sum\limits_{a=-\frac{n-1}{2}}^{\frac{n-1}{2}} \log\left|
 \theta_3\left(i\frac{a(w_{v_1}^{\text{New},\alpha}-w_{v_2}^{\text{New},\alpha}+\bar{w}_{v_1}^{\text{New},\alpha}-\bar{w}_{v_2}^{\text{New},\alpha})}{Ln}\bigg| \frac{4i\epsilon}{L}\right)\bigg/
    \theta_3\left(0\bigg| \frac{4i\epsilon}{L}\right)
    \right| & \text{Dirichlet} 
    \end{cases}
\end{align}
In this expression, as well as the subsequent expressions for Euclidean R\'{e}nyi entropy and mutual information, the coordinates $w_{v_i}^{\text{New},\alpha}$ and $\bar{w}_{v_i}^{\text{New},\alpha}$ are analytically continued to real time $\tau_1 \rightarrow it_1$. In the uniform $\theta=0$ case with spatial periodicity $L=2\pi$, we recover the corresponding expressions in \cite{Takayanagi2010,Takayanagi2022}. The entanglement entropy for a boundary state at the initial time $t=0$ for a subsystem of size $|\mathcal{V}|\gg \epsilon$ $\rightarrow v_1-v_2\gg \epsilon$ in the limit where $L\gg \epsilon$ is given by 
\begin{equation}
    S^{(n)}_\mathcal{V}(t=0) = \frac{n+1}{6n} \log \frac{4\epsilon}{\pi}
\end{equation}
which does not depend on the total system subsystem sizes since the boundary state possesses a low amount of entanglement \cite{Miyaji:2014mca}. To compute the entanglement entropy of two intervals, the four-point function of the vertex operators is required. The calculation is a generalization of that in \cite{Takayanagi2010} and can be found in Appendix \ref{VertexFourPointFunctionCalculation}.

If we have two intervals $A=[v_2,v_1]$ and $B=[v_4,v_3]$, then repeating the same calculation as for the single interval case gives
\begin{align}\label{SAB} 
S_{A\cup B}^{(n)}
=S_{AB,\text{univ.}}^{(n)}+S_{AB,\text{non-univ.}}^{(n)}
\end{align}
where the universal part that does not depend on the boundary condition is  
\begin{align}  \label{SABuniv} 
&S_{A \cup B,\text{univ.}}^{(n)}\nonumber \\
    =& -\frac{c}{24}\frac{n+1}{n}\log \prod_{i=1}^4\left|\frac{dw_{v_i}^{\text{New},\alpha}}{dw_{v_i}}\frac{d\bar{w}_{v_i}^{\text{New},\alpha}}{d\bar{w}_{v_i}}\right|+\frac{n+1}{12n}\log\Bigg| \left(\frac{L}{2\pi}\right)^4  \nonumber \\
    \times&\frac{\theta_1\left(i\frac{w_{v_4}^{\text{New},\alpha}-w_{v_3}^{\text{New},\alpha}}{L}\big|\frac{4i\epsilon}{L}\right)\theta_1\left(i\frac{w_{v_4}^{\text{New},\alpha}-w_{v_1}^{\text{New},\alpha}}{L}\big|\frac{4i\epsilon}{L}\right)\theta_1\left(i\frac{w_{v_3}^{\text{New},\alpha}-w_{v_2}^{\text{New},\alpha}}{L}\big|\frac{4i\epsilon}{L}\right)\theta_1\left(i\frac{w_{v_2}^{\text{New},\alpha}-w_{v_1}^{\text{New},\alpha}}{L}\big|\frac{4i\epsilon}{L}\right)
    }{\eta\left(\frac{4i\epsilon}{L}\right)^{12}\theta_1\left(i\frac{w_{v_4}^{\text{New},\alpha}-w_{v_2}^{\text{New},\alpha}}{L}\big|\frac{4i\epsilon}{L}\right) \theta_1\left(i\frac{w_{v_3}^{\text{New},\alpha}-w_{v_1}^{\text{New},\alpha}}{L}\big|\frac{4i\epsilon}{L}\right)
    } \nonumber \\
    \times&\frac{\theta_1\left(i\frac{\bar{w}_{v_4}^{\text{New},\alpha}-\bar{w}_{v_3}^{\text{New},\alpha}}{L}\big|\frac{4i\epsilon}{L}\right)\theta_1\left(i\frac{\bar{w}_{v_4}^{\text{New},\alpha}-\bar{w}_{v_1}^{\text{New},\alpha}}{L}\big|\frac{4i\epsilon}{L}\right)\theta_1\left(i\frac{\bar{w}_{v_3}^{\text{New},\alpha}-\bar{w}_{v_2}^{\text{New},\alpha}}{L}\big|\frac{4i\epsilon}{L}\right)\theta_1\left(i\frac{\bar{w}_{v_2}^{\text{new},\alpha}-\bar{w}_{v_1}^{\text{New},\alpha}}{L}\big|\frac{4i\epsilon}{L}\right)
    }{\theta_1\left(i\frac{\bar{w}_{v_4}^{\text{New},\alpha}-\bar{w}_{v_2}^{\text{New},\alpha}}{L}\big|\frac{4i\epsilon}{L}\right) \theta_1\left(i\frac{\bar{w}_{v_3}^{\text{New},\alpha}-\bar{w}_{v_1}^{\text{New},\alpha}}{L}\big|\frac{4i\epsilon}{L}\right)
    }\nonumber \\ 
    \times& 
    \frac{\theta_1\left(i\frac{w_{v_1}^{\text{New},\alpha}+\bar{w}_{v_1}^{\text{New},\alpha}}{L}\big|\frac{4i\epsilon}{L}\right)\theta_1\left(i\frac{w_{v_1}^{\text{New},\alpha}+\bar{w}_{v_3}^{\text{New},\alpha}}{L}\big|\frac{4i\epsilon}{L}\right)}{\theta_1\left(i\frac{w_{v_1}^{\text{New},\alpha}+\bar{w}_{v_2}^{\text{New},\alpha}}{L}\big|\frac{4i\epsilon}{L}\right)\theta_1\left(i\frac{w_{v_1}^{\text{New},\alpha}+\bar{w}_{v_4}^{\text{New},\alpha}}{L}\big|\frac{4i\epsilon}{L}\right)}
    \frac{\theta_1\left(i\frac{w_{v_2}^{\text{New},\alpha}+\bar{w}_{v_2}^{\text{New},\alpha}}{L}\big|\frac{4i\epsilon}{L}\right)\theta_1\left(i\frac{w_{v_2}^{\text{New},\alpha}+\bar{w}_{v_4}^{\text{New},\alpha}}{L}\big|\frac{4i\epsilon}{L}\right)}{\theta_1\left(i\frac{w_{v_2}^{\text{New},\alpha}+\bar{w}_{v_1}^{\text{New},\alpha}}{L}\big|\frac{4i\epsilon}{L}\right)\theta_1\left(i\frac{w_{v_2}^{\text{New},\alpha}+\bar{w}_{v_3}^{\text{New},\alpha}}{L}\big|\frac{4i\epsilon}{L}\right)} \nonumber\\
\times&\frac{\theta_1\left(i\frac{w_{v_3}^{\text{New},\alpha}+\bar{w}_{v_1}^{\text{New},\alpha}}{L}\big|\frac{4i\epsilon}{L}\right)\theta_1\left(i\frac{w_{v_3}^{\text{New},\alpha}+\bar{w}_{v_3}^{\text{New},\alpha}}{L}\big|\frac{4i\epsilon}{L}\right)}{\theta_1\left(i\frac{w_{v_3}^{\text{New},\alpha}+\bar{w}_{v_2}^{\text{New},\alpha}}{L}\big|\frac{4i\epsilon}{L}\right)\theta_1\left(i\frac{w_{v_3}^{\text{New},\alpha}+\bar{w}_{v_4}^{\text{New},\alpha}}{L}\big|\frac{4i\epsilon}{L}\right)}
    \frac{\theta_1\left(i\frac{w_{v_4}^{\text{New},\alpha}+\bar{w}_{v_2}^{\text{New},\alpha}}{L}\big|\frac{4i\epsilon}{L}\right)\theta_1\left(i\frac{w_{v_4}^{\text{New},\alpha}+\bar{w}_{v_4}^{\text{New},\alpha}}{L}\big|\frac{4i\epsilon}{L}\right)}{\theta_1\left(i\frac{w_{v_4}^{\text{New},\alpha}+\bar{w}_{v_1}^{\text{New},\alpha}}{L}\big|\frac{4i\epsilon}{L}\right)\theta_1\left(i\frac{w_{v_4}^{\text{New},\alpha}+\bar{w}_{v_3}^{\text{New},\alpha}}{L}\big|\frac{4i\epsilon}{L}\right)} 
\Bigg|
\end{align}
and the non-universal part that depends on the boundary condition is
\begin{align}
&S_{AB,\text{non-univ.}}^{(n)}\nonumber\\
=& \begin{cases}
 \frac{1}{1-n}\sum\limits_{a=-\frac{n-1}{2}}^{\frac{n-1}{2}}\log\bigg|\big[\theta_2\left(i\frac{a}{n}\frac{w_{v_1}^{\text{New},\alpha}-w_{v_2}^{\text{New},\alpha}+w_{v_3}^{\text{New},\alpha}-w_{v_4}^{\text{New},\alpha}
+\bar{w}_{v_1}^{\text{New},\alpha}-\bar{w}_{v_2}^{\text{New},\alpha}+\bar{w}_{v_3}^{\text{New},\alpha}-\bar{w}_{v_4}^{\text{New},\alpha}
 }{L}\big|\frac{4i\epsilon}{L}\right) & \\
+\theta_3\left(i\frac{a}{n}\frac{w_{v_1}^{\text{New},\alpha}-w_{v_2}^{\text{New},\alpha}+w_{v_3}^{\text{New},\alpha}-w_{v_4}^{\text{New},\alpha}
+\bar{w}_{v_1}^{\text{New},\alpha}-\bar{w}_{v_2}^{\text{New},\alpha}+\bar{w}_{v_3}^{\text{New},\alpha}-\bar{w}_{v_4}^{\text{New},\alpha}
 }{L}\big|\frac{4i\epsilon}{L}\right)\big]& \\
 /\big[\theta_2\left(\frac{4i\epsilon}{L}\right)+\theta_3\left(\frac{4i\epsilon}{L}\right)\big]\bigg|,&\text{N} \\
 \frac{1}{1-n}\sum\limits_{a=-\frac{n-1}{2}}^{\frac{n-1}{2}}\log\left|\frac{\theta_3\left(i\frac{a}{n}\frac{w_{v_1}^{\text{New},\alpha}-w_{v_2}^{\text{New},\alpha}+w_{v_3}^{\text{New},\alpha}-w_{v_4}^{\text{New},\alpha}
+\bar{w}_{v_1}^{\text{New},\alpha}-\bar{w}_{v_2}^{\text{New},\alpha}+\bar{w}_{v_3}^{\text{New},\alpha}-\bar{w}_{v_4}^{\text{New},\alpha}
 }{L}\big|\frac{4i\epsilon}{L}\right)}{\theta_3\left(\frac{4i\epsilon}{L}\right)}\right|,&\text{D} 
\end{cases}
\end{align}
where N and D stand for the Neumann and Dirichlet boundary conditions, respectively.

\subsection{Entanglement entropy for the single interval in $2$d holographic CFT \label{Section:NUPin2dhCFT}}
Now, we turn to the details of the computation on entanglement entropy for the single interval in $2$d holographic CFT.
\subsubsection{Method of image}
In this paper, we employ the method of images to compute the Euclidean R\'enyi entanglement entropy for the holographic CFTs.
Consider (\ref{EREE}) for the single interval in this method.
In this method, the two point function for the boundary state is given by the four point function of the original operators and their images. In the von Neumann limit, $n\rightarrow 1$, the Euclidean entanglement entropy is given by 
\be \label{eq:EEE}
\begin{split}
S_{E,\mathcal{V}}&=\lim_{n\rightarrow 1}\f{h_n}{1-n}\log{\left[\left(\f{dw^{\text{New},\alpha}_{v_1}}{dw_{v_1}}\right)\left(\f{d\overline{w}^{\text{New},\alpha}_{v_1}}{d\overline{w}_{v_1}}\right)\left(\f{dw^{\text{New},\alpha}_{v_2}}{dw_{v_2}}\right)\left(\f{d\overline{w}^{\text{New},\alpha}_{v_2}}{d\overline{w}_{v_2}}\right)\right]}\\
&+\lim_{n\rightarrow 1}\f{1}{2(1-n)}\log\Bigg{[}\big{\langle} \overline{\mathcal{T}}_{n}\left(4\epsilon-\tau_{v_1,\tau_1,\alpha}+i\f{L\varphi_{v_1,\tau_1,\alpha}}{2\pi}, 4\epsilon-\tau_{v_1,\tau_1,\alpha}+i\f{L\overline{\varphi}_{v_1,\tau_1,\alpha}}{2\pi}\right)\\
&\times\mathcal{T}_n\left(4\epsilon-\tau_{v_2,\tau_1,\alpha}+i\f{L\varphi_{v_2,\tau_1,\alpha}}{2\pi}, 4\epsilon-\tau_{v_2,\tau_1,\alpha}+i\f{L\overline{\varphi}_{v_2,\tau_1,\alpha}}{2\pi}\right)\\ &\times\mathcal{T}_{n}\left(\tau_{v_1,\tau_1}+i\f{L\varphi_{v_1,\tau_1}}{2\pi}, \tau_{v_1,\tau_1,\alpha}+i\f{L\overline{\varphi}_{v_1,\tau_1,\alpha}}{2\pi}\right)\overline{\mathcal{T}}_n\left(\tau_{v_2,\tau_1,\alpha}+i\f{L\varphi_{v_2,\tau_1,\alpha}}{2\pi}, \tau_{v_2,\tau_1,\alpha}+i\f{L\overline{\varphi}_{v_2,\tau_1,\alpha}}{2\pi} \right)\big{\rangle}_{\text{Torus}}\Bigg{]},\\
&=-\f{c}{12}\log{\left[\left(\f{dw^{\text{New},\alpha}_{v_1}}{dw_{v_1}}\right)\left(\f{d\overline{w}^{\text{New},\alpha}_{v_1}}{d\overline{w}_{v_1}}\right)\left(\f{dw^{\text{New},\alpha}_{v_2}}{dw_{v_2}}\right)\left(\f{d\overline{w}^{\text{New},\alpha}_{v_2}}{d\overline{w}_{v_2}}\right)\right]}\\
    &+\lim_{n\rightarrow 1}\f{1}{2(1-n)}\log\Bigg{[}\big{\langle} \overline{\mathcal{T}}_{n}\left(3\epsilon-\overline{w}^{\text{New},\alpha}_{v_1},3\epsilon-w^{\text{New},\alpha}_{v_1}\right)\mathcal{T}_n\left(3\epsilon-\overline{w}^{\text{New},\alpha}_{v_2}, 3\epsilon-w^{\text{New},\alpha}_{v_2}\right)\\ &~~~~\times\mathcal{T}_{n}\left(\epsilon+w^{\text{New},\alpha}_{v_1}, \epsilon+\overline{w}^{\text{New},\alpha}_{v_1}\right)\overline{\mathcal{T}}_n\left(\epsilon+w^{\text{New},\alpha}_{v_2}, \epsilon+\overline{w}^{\text{New},\alpha}_{v_2}\right)\big{\rangle}_{\text{Torus}}\Bigg{]},
\end{split}
\ee
where the length of the thermal cycle is $4\epsilon$, and the mirrors are located at $\tau=0$ and $\tau=2\epsilon$. 
Here, $\tau_{v_i,\tau_1,\alpha}$ denotes the real part of $w^{\text{New},\alpha}_{v_i}$ and $\overline{w}^{\text{New},\alpha}_{v_i}$, while $\f{L\varphi_{v_i,\tau_1,\alpha}}{2\pi}$ and $\f{L\overline{\varphi}_{v_i,\tau_1,\alpha}}{2\pi}$ denote the imaginary parts, respectively. 
The first term in the last equation of (\ref{eq:EEE}) is independent of the details of $2$d CFTs, while the second term depends on those.
We call the first and second terms the universal and non-universal pieces, respectively.
In $2$d holographic CFTs, the non-universal piece is determined by the geodesic length in on the background dual of the system considered.

\subsubsection{Non-universal piece in $2$d holographic CFT}
Now, we focus on the non-universal piece of the entanglement entropy in $2$d holographic CFT. 
By employing the method of images, the system to is equivalent to that on a spatial circle with circumference $L$ in the thermal state with the inverse temperature $4\epsilon ~(\ll L)$. 
In $2$d holographic CFTs, the gravity dual of the system in the high temperature region, $\epsilon \ll L$, is given by the BTZ black hole, while that in the low temperature region, $\epsilon \gg L$, is given by the thermal AdS \cite{Witten:1998zw}. 
Therefore, the non-universal piece in the high temperature region is given by the minimal length of geodesics in the BTZ black hole geometry \cite{Ryu:2006bv, Headrick:2013zda}:
\be
S_{\mathcal{V};\text{Non-uni.}}=\text{Min}\left[S_{\mathcal{V};\text{con}},S_{\mathcal{V};\text{dis}}\right],
\ee
where $S_{\text{con}}$ is the length of geodesics connecting the points on the different Euclidean time slices, while $S_{\text{dis}}$ is the length of geodesics connecting the points on the same Euclidean time slice.
We present the details of $S_{\text{con}}$ and $S_{\text{dis}}$, and they are given by
\be \label{eq:EE_for_bds}
\begin{split}
S_{\mathcal{V};\text{con}} &\approx \f{c}{3}\log{\left(\f{4\epsilon}{\pi}\right)}+\f{c}{6}\times\left[\sum_{i=1,2}\log{\left(\cos{\left(\f{\pi}{4\epsilon}\left(w^{\text{New},\alpha}_{v_i}+\overline{w}^{\text{New},\alpha}_{v_i}\right)\right)}\right)}\right]\\
S_{\mathcal{V};\text{dis}} &\approx \f{c}{3}\log{\left(\f{4\epsilon}{\pi}\right)}+\text{Min}\left[S_{\mathcal{V};\text{dis},1}, S_{\mathcal{V};\text{dis},2}\right]\\
S_{\mathcal{V};\text{dis},1}&=\f{c}{12}\log{\left|\sin{\left(\f{\pi}{4\epsilon}(w^{\text{New},\alpha}_{v_1}-w^{\text{New},\alpha}_{v_2}\pm iL)\right)}\right|^2}+\f{c}{12}\log{\left|\sin{\left(\f{\pi}{4\epsilon}(\overline{w}^{\text{New},\alpha}_{v_1}-\overline{w}^{\text{New},\alpha}_{v_2}\mp iL)\right)}\right|^2}\\
S_{\mathcal{V};\text{dis},2}&=\f{c}{12}\log{\left|\sin{\left(\f{\pi}{4\epsilon}(w^{\text{New},\alpha}_{v_1}-w^{\text{New},\alpha}_{v_2})\right)}\right|^2}\bigg{]}+\f{c}{12}\log{\left|\sin{\left(\f{\pi}{4\epsilon}(\overline{w}^{\text{New},\alpha}_{v_1}-\overline{w}^{\text{New},\alpha}_{v_2})\right)}\right|^2},\\
\end{split}
\ee

\subsection{The trajectory of the local operator during the time evolution induced by SSD/M\"obius evolution \label{Section:evolution-of-local-operators}}
\begin{figure}[htbp]
    \begin{tabular}{cc}
      \begin{minipage}[t]{0.5\hsize}
        \centering
        \includegraphics[keepaspectratio, scale=0.18]{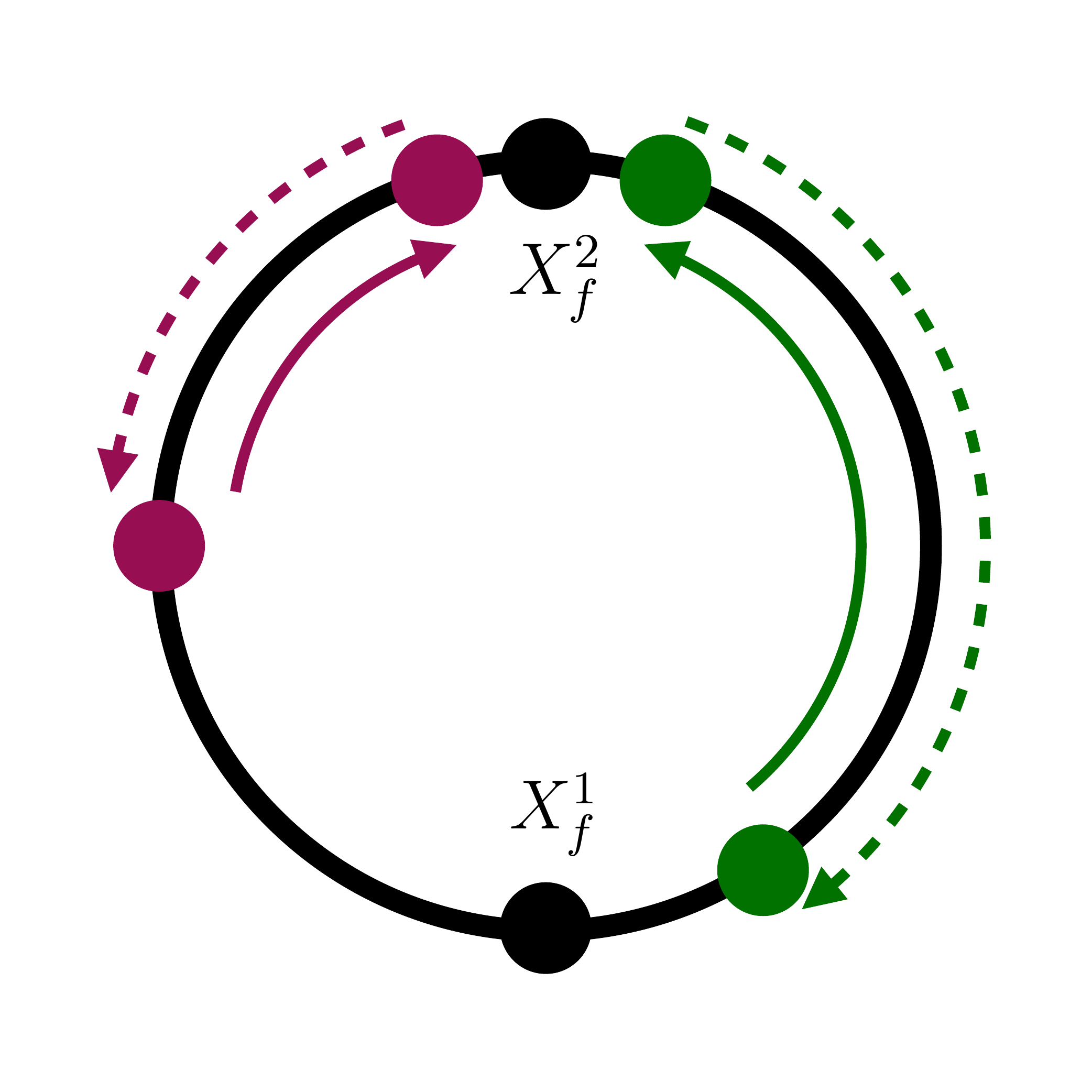}
        
    [a] M\"obius evolution  
    
      \end{minipage}
      &\begin{minipage}[t]{0.5\hsize}
        \centering
        \includegraphics[keepaspectratio, scale=0.18]{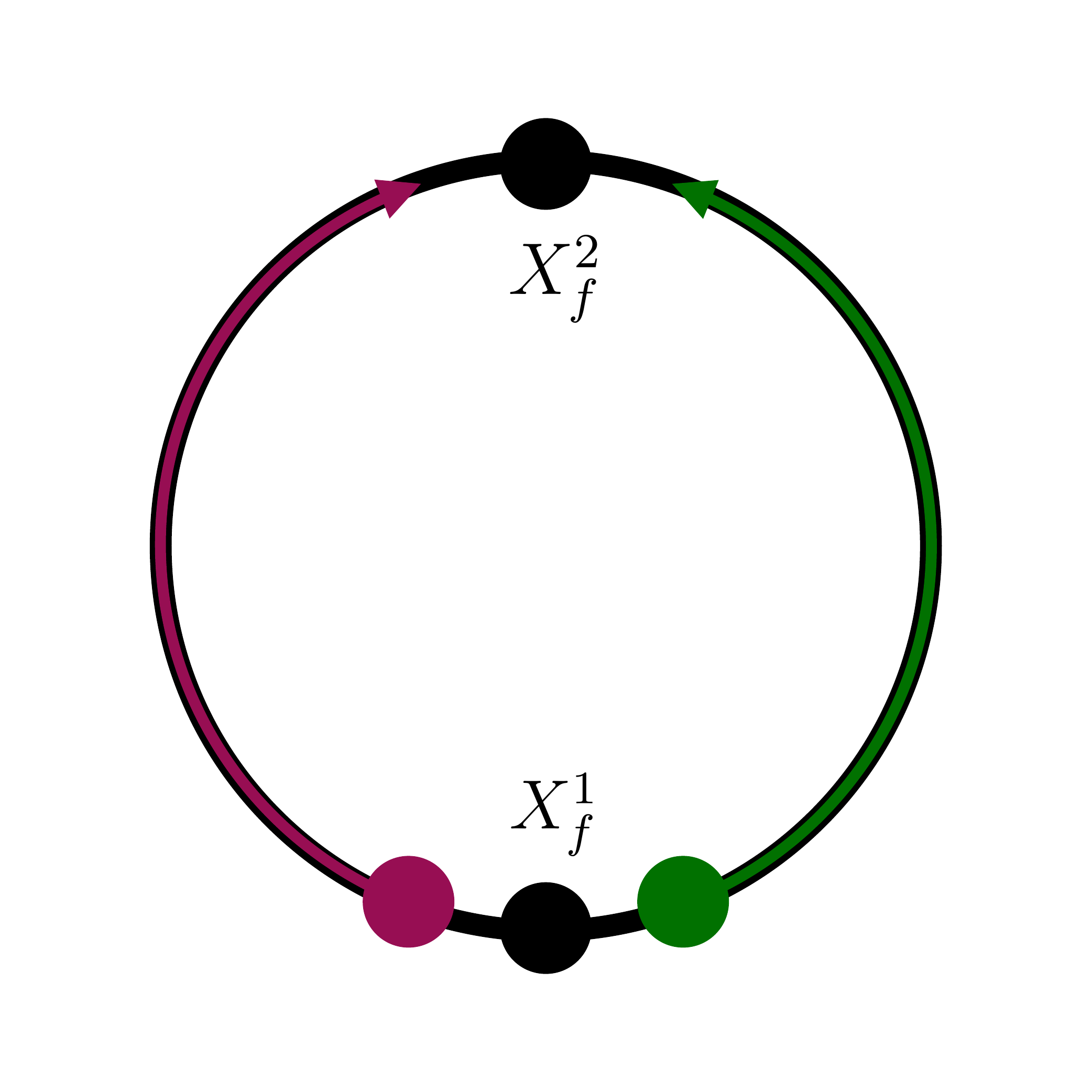}
        
          [b] SSD evolution 
          
      \end{minipage} \\
      
    \end{tabular}
      \caption{The evolution of operators during the evolution induced by SSD/M\"obius Hamiltonian. 
      The black dots illustrate the fixed point of $H_{\text{SSD}}$.  
      The green (purple) dot illustrates the operator that is initially inserted at $L/2>x>0$ ($L>x>L/2$) and $t_1=0$, and its trajectory during the time evolution is illustrated by the green (purple) curve. 
      In [a], we show the evolution of operators during the M\"obius time evolution. In [b], we show the evolution of operators during the SSD time evolution.}
        \label{Fig:OperatorEvolution}
  \end{figure}
After performing the analytic continuation, $\tau_1 =it_1$, we consider the trajectory of the twist and anti-twist operators during the time evolution induced by the SSD/M\"obius Hamiltonian.
Define the spatial position of these operators in the Heisenberg picture as 
\be
X^{\text{New},\alpha}_{v_i}= \f{w^{\text{New},\alpha}_{v_i}-\overline{w}^{\text{New},\alpha}_{v_i}}{2i},
\ee
where $v_i$ is the insertion point of these operators in the Schr\"odinger picture.
If these operators are inserted at $v_i=0, L/2$, then during the SSD time evolution, $X^{\text{New},\alpha}_{v_i}$ does not vary in time.
Therefore, call them fixed points, and let $X^1_f$ and $X^2_f$ denote $0$ and $\f{L}{2}$, respectively.
During the M\"obius evolution, these operators in the Heisenberg picture periodically move in time between  $X^1_f$ and $X^2_f$ (see [a] of Fig. \ref{Fig:OperatorEvolution}), while during the SSD evolution, those move to $X^2_f$ and accumulate around that spatial point (see [b] of Fig. \ref{Fig:OperatorEvolution}). 
Consequently, if the subsystem is a single interval including $X^1_f$, then during the SSD evolution, the effective subsystem size in the Heisenberg picture defined as $V_{\text{eff}} =X^{\text{New},\alpha}_{v_1}- X^{\text{New},\alpha}_{v_2}$, monotonically grows with time.
If the subsystem is a single interval excluding $X^1_f$, then during the SSD evolution, $V_{\text{eff}} =X^{\text{New},\alpha}_{v_1}- X^{\text{New},\alpha}_{v_2}$ eventually shrinks with time.

\section{The time dependence of entanglement entropy in $2$d CFTs \label{Section:Evolution-in-CFT}}
In this section, we will report the time dependence of entanglement entropy in $2$d CFTs.
\subsection{The time evolution of entanglement entropy in $2$d free fermion \label{Section:Evolution-in-freefermion}}
\begin{figure}
    \centering
\includegraphics[width=0.45\textwidth]{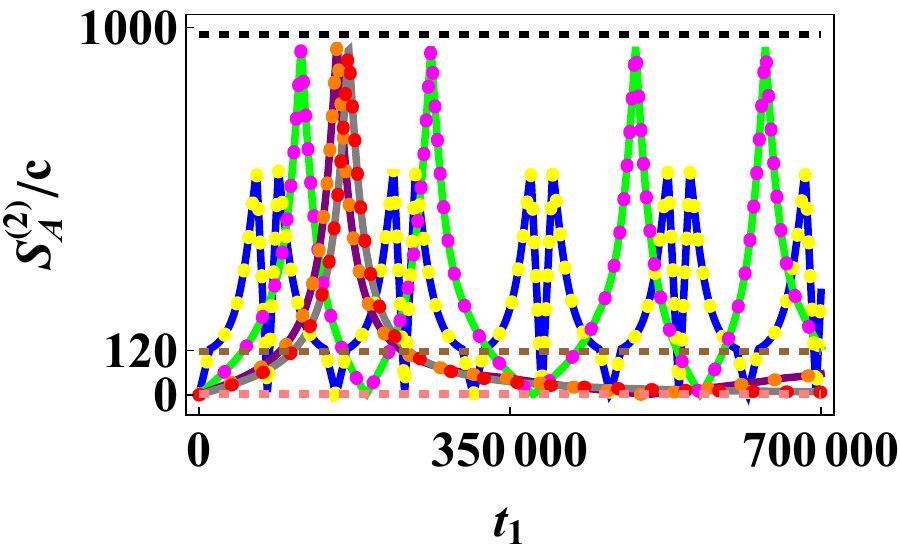}\includegraphics[width=0.45\textwidth]{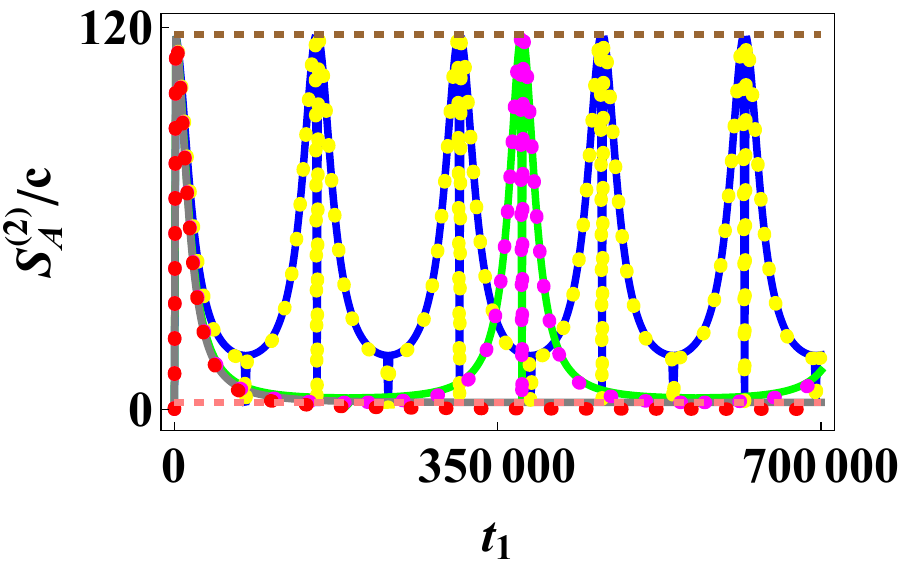}
\text{
[a] $P_c=X^1_f$\hspace{5.5cm} [b] $P_c=X^2_f$ } \par\medskip
\includegraphics[width=0.45\textwidth]{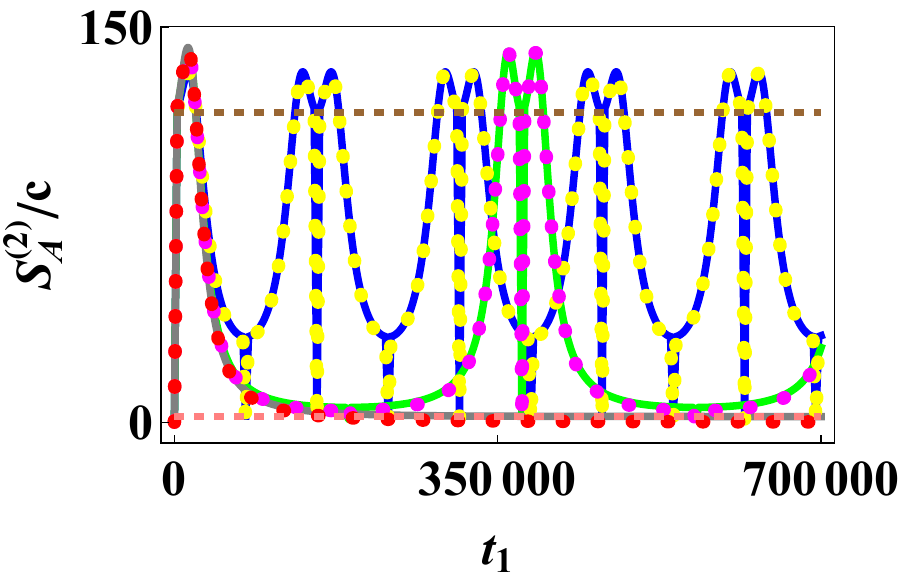}
\includegraphics[width=0.45\textwidth]{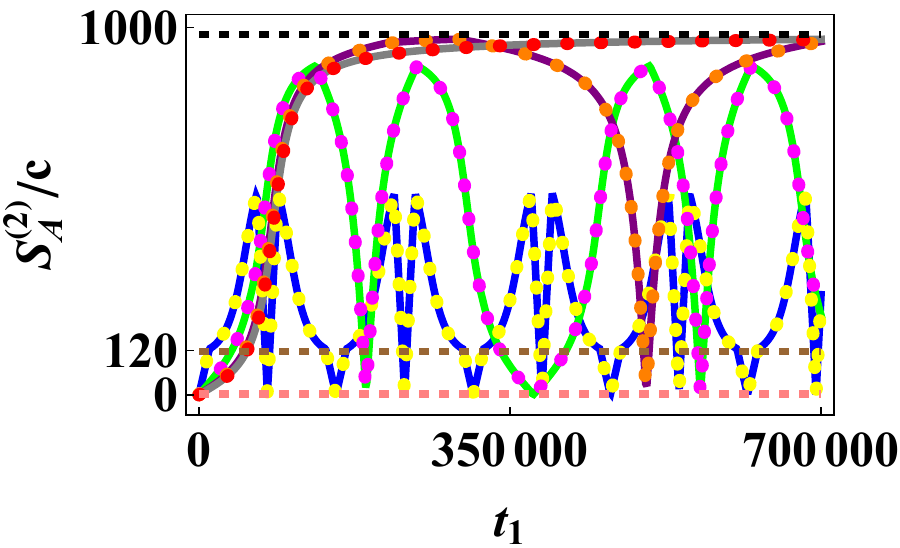}
\text{
[c] $P_c=\frac{L}{4}$\hspace{5.5cm} [d] $A=[X_f^1,X_f^1+l_A]$ } \par\medskip
\includegraphics[width=\textwidth]{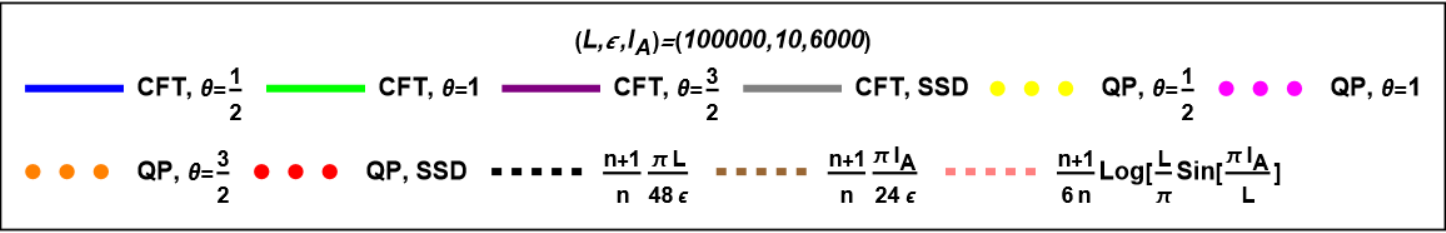}
\caption{Plots of the second R\'{e}nyi entropy for the 2d free fermion CFT as well as the quasiparticle picture for a total system size of $L=100000$, $\epsilon=10$ and a subsystem size of $l_A=6000$ with a center $P_c$ located at various positions.}
\label{BoundaryStateEntropyCFTvsQP}
\end{figure}

As examples of non-chaotic dynamics, we compare the R\'{e}nyi entropy of a single interval in the free fermionic CFT as well as in the quasiparticle picture which will be discussed later in Section \ref{QuasiparticlePictureSection}. The plots for three different subsystems centered at three different points along the spatial circle are shown in Fig. \ref{BoundaryStateEntropyCFTvsQP}. Let us just summarize the salient points of these plots. First, the CFT result and the quasiparticle entanglement agree to the leading order in $\frac{1}{\epsilon}$. The subleading difference is most apparent when the R\'{e}nyi entropy after an SSD quench decays to the vacuum value for the CFT but it decays to zero for the quasiparticles since all quasiparticles will accumulate at the fixed point $X_f^1$ at late times and hence the entanglement entropy will asymptotically decay to zero. The behavior of these free theories during M\"obius evolution is very similar to that for the holographic CFTs except for the fact that the entanglement entropy of quasiparticles drops down to zero at integer multiples of $\frac{L}{2}\cosh{2\theta}$ and the entanglement entropy of the free fermion CFT similarly drop downs to a small value. The entanglement entropy of quasiparticles must decay back to zero at integer multiples of $L\cosh{2\theta}$ because the quasiparticles return to their initial positions at those times. At half-integer multiples of $\frac{L}{2}\cosh{2\theta}$, the left and right moving partner of each Bell pair meet up and so give no contribution to the entanglement entropy. Secondly, the plots for both boundary conditions for the free fermion CFT appear to be identical. The boundary condition dependent terms in the free fermion CFT R\'{e}nyi entropy are either zero or tiny compared to the total entropy in most cases except for the case where the interval ends exactly on the fixed point $X_f^1$ in which case the boundary condition-dependent component is actually larger than the boundary condition-independent part for the larger values of $\theta$. In any case, the upshot is that the plots appear identical for both boundary conditions in the free fermion CFT and hence do not appear to depend on the boundary condition imposed on the Dirac fermions.

Just as in the holographic case, to obtain a non-zero value of the entanglement entropy for the quasiparticles and free fermion CFTs at late times, place the interval so that one of the endpoints sits exactly on the SSD fixed point $X_f^1$ so that exactly one member of each Bell pair in the quasiparticle picture is contained inside the subsystem at late enough times.

\subsection{Quasiparticle Picture}\label{QuasiparticlePictureSection}
The quasiparticle picture for the uniform global quench \cite{2016JSMTE..06.4003C,doi:10.1073/pnas.1703516114,2005JSMTE..04..010C} can be extended to the inhomogeneous case by assuming that the quasiparticle are not moving with uniform speed but instead with a speed that is determined by the inhomogeneous envelope function that appears in the inhomogeneous Hamiltonian. This might be due to the fact that the CFTs considered in this paper are defined on curved backgrounds where the time component of metric, which determines the speed of moving objects, is given by the inhomogeneous envelope function. During the M\"obius evolution, a quasiparticle that begins at position $x_0$ at time $t_0$ is located at $x$ at time $t_1$ which is given by
\begin{equation}
    \frac{\pi(t_1-t_0)}{L\cosh{2\theta}}=\pm\left[\tan^{-1}\left(e^{2\theta}\tan\frac{\pi x}{L}\right)-\tan^{-1}\left(e^{2\theta}\tan\frac{\pi x_0}{L}\right)\right]
\end{equation}
where the $+(-)$ sign correspond to right(left) moving quasiparticles. For simplicity, set the initial time $t_0=0$.

\begin{figure}
    \centering
\includegraphics[width=\textwidth]{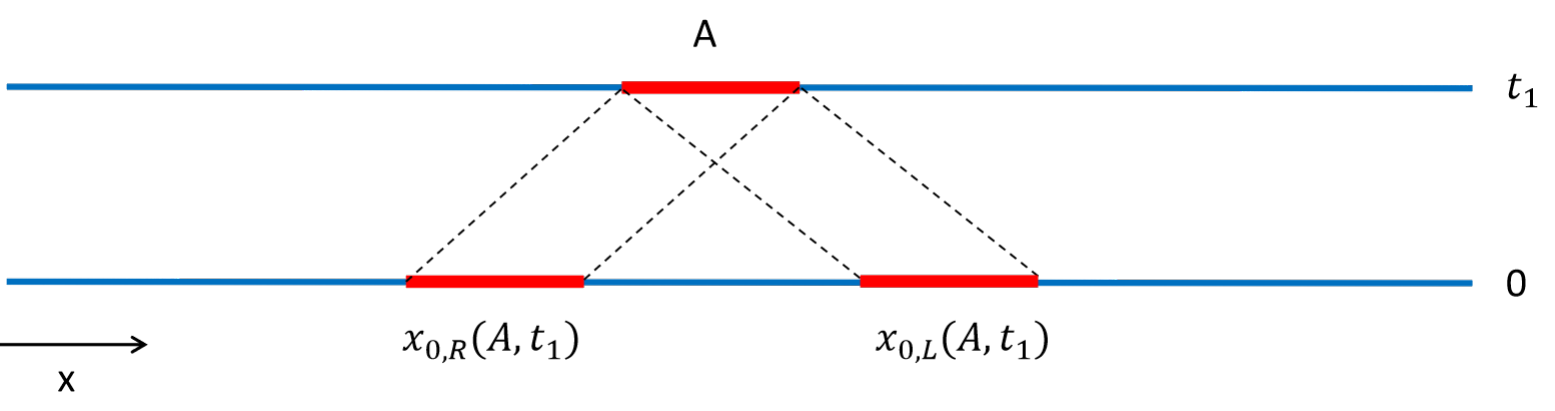}
    \caption{For a given interval $A$, $x_{0,R}(A,t_1)$ and $x_{0,L}(A,t_1)$ are the intervals that will move respectively to the right and left and will coincide with $A$ at time $t_1$. Shown here is the simplest case where the Hamiltonian is uniform so the intervals are translated with unit speed.}
\label{BoundaryStateQuasiparticle_InverseLightCone}
\end{figure}

For a subsystem $A$, which could generally be a union of intervals, the entanglement entropy as predicted by the quasiparticle picture is given by the number of Bell pairs shared between $A$ and its complement. Let $x_{0,i}(x,t_1)$ be the initial position of a quasiparticle situated at position $x$ at time $t_1$, where $i=R,L$ denotes the chirality of the quasiparticles. For additional details, see \cite{Goto:2021sqx}. Assuming that the quasiparticles are conserved, the right-moving and left-moving quasiparticles that are inside $A$ at a given time $t_1$ are therefore initially situated inside $x_{0,R}(A,t_1)$ and $x_{0,L}(A,t_1)$ at $t=0$, respectively. A simple example for the uniform Hamiltonian is depicted in Fig. \ref{BoundaryStateQuasiparticle_InverseLightCone}. Similarly, the left and right moving quasiparticles that end up in the complement of $A$, $\overline{A}$, initially began in the subsystems $x_{0,L}(\overline{A},t_1)$ and $x_{0,R}(\overline{A},t_1)$, respectively.

While the distribution of quasiparticles after a time evolution with an inhomogeneous Hamiltonian is complicated \cite{Goto:2021sqx}, the quasiparticles are initially distributed uniformly. Since the entanglement entropy of $A$ is given by the Bell pairs that it shares with its complement, the entanglement entropy can simply be written as 
\begin{equation}
    S_A^{(n)}(t_1) = \rho_0^{(n)}\left[ \text{length of } x_{0,L}(\overline{A},t_1)\cap x_{0,R}(A,t_1)+
    \text{length of } x_{0,R}(\overline{A},t_1)\cap x_{0,L}(A,t_1)
    \right]
\end{equation}
where $\rho_0^{(n)}=\frac{n+1}{n}\frac{\pi c}{48\epsilon}$ can be fixed by equating the saturation value of $S_A^{(n)}$ for the uniform Hamiltonian for a single finite interval on the real line with the known CFT result \cite{2016JSMTE..06.4003C} to leading order in $\frac{1}{\epsilon}$ and using the fact that for the uniform Hamiltonian, $x_{0,R}(A,t_1)$ and $x_{0,L}(A,t_1)$ are entirely contained inside $x_{0,L}(\overline{A},t_1)$ and $x_{0,R}(\overline{A},t_1)$ respectively at sufficiently late times.

The quasiparticle picture entanglement entropy appears to vanish when the time $t_1$ is a multiple of $\frac{L}{2}\cosh{2\theta}$. This is obvious when $t_1$ is a multiple of $L\cosh{2\theta}$ since all the quasiparticles would have returned to their original position so the entanglement entropy reverts back to its original value of 0. When $t_1$ is a half-integer multiple of $L\cosh{2\theta}$, the quasiparticle entanglement entropy vanishes not because the quasiparticles have returned to their initial positions but because the left and right moving partners of each Bell pair have met up after traversing the spatial circle on opposite directions. 

We can also compute the mutual information as predicted by the quasiparticle picture. The mutual information between two subsystems $A$ and $B$ is given by twice the number of Bell pairs that have one partner in each system since these Bell pairs contribute to both $S_A^{(n)}$ and $S_B^{(n)}$ but not to $S_{AB}^{(n)}$. Therefore, 
\begin{equation}
    I_{AB}^{(n)}(t_1) = 2\rho_0^{(n)}\left[ \text{length of } x_{0,L}(A,t_1)\cap x_{0,R}(B,t_1)+
    \text{length of } x_{0,R}(A,t_1)\cap x_{0,L}(B,t_1)
    \right]
\end{equation}

\subsubsection{Late time behaviour of free theories when the interval ends on $X_f^1$}
The R\'{e}nyi entropy for a subsystem that ends on the fixed point $X_f^1$, $[0,X_1]$, in the quasiparticle picture is
\begin{equation}\label{QPEndFixedPointLateTime}
    S_A^{(n)} \stackrel{t_1\gg L}{\approx} \frac{n+1}{48n} \frac{\pi c L}{\epsilon} \left(1-\frac{L}{\pi^2 t_1}\right)
\end{equation}
In the late time regime of interest $t_1\gg L\gg \epsilon$, the boundary condition-dependent terms in the free fermion CFT entanglement entropy are exponentially suppressed and the late time entanglement entropy of a single interval that ends on the fixed point when either boundary state is quenched by the SSD Hamiltonian is 
\begin{equation}
    S_{[0,X_1]}^{(n)}(t_1) \approx \frac{n+1}{6n} \log\left( \frac{L}{\pi}\sin{\frac{\pi X_1}{L}}\right)+\frac{n+1}{48n}\frac{\pi L}{\epsilon}\left(1-\frac{2L}{\pi^2 t_1}\right), \hspace{1cm}t_1\gg L\gg\epsilon,\frac{t_1}{L}\gg \frac{L}{\epsilon}
\end{equation}
This is very similar to the late time behaviour of the entanglement entropy of a single interval that ends on the fixed point $X_f^1$ for the quasiparticles undergoing going a SSD quench. The only difference is the additional factor of $2$ in the power law $\frac{1}{t_1}$ decay to the saturation value which is subleading in $\frac{L}{\epsilon}$. The saturation value also has the additional vacuum R\'{e}nyi entropy which is zero in the quasiparticle case as discussed earlier.


The discrepancy between the quasiparticle entanglement entropy and the CFT R\'{e}nyi entropy can be attributed to the order of the limits taken. The term that gives rise to the late time saturation is
\begin{align}\label{TermThatGivesLateTimeSaturation}
    S_{[0,X_1]}^{(n)}(t_1) =& \frac{n+1}{12n} \log \theta_1\left(\frac{w_{X_1}^{\text{New},\alpha}+\overline{w}_{X_1}^{\text{New},\alpha}}{4\epsilon }\bigg| i\frac{L}{4\epsilon}\right) +\ldots \nonumber \\
    \stackrel{t_1\gg L\gg\epsilon}{\approx}& \frac{n+1}{12n}\frac{\pi L}{4\epsilon}\left(1-\frac{L}{\pi^2 t_1}\right) + \frac{n+1}{12n}\log \left(1+e^{-\frac{L^2}{2\epsilon \pi t_1}}\right)+\ldots
\end{align}
If we first send $\frac{L}{\epsilon}\rightarrow \infty$ in \eqref{TermThatGivesLateTimeSaturation}
\begin{equation}
    S_{[0,X_1]}^{(n)}(t_1) \approx \frac{n+1}{48n}\frac{\pi L}{\epsilon}\left(1-\frac{L}{\pi^2 t_1}\right)+\ldots
\end{equation}
This is the same answer as for the quasiparticles \eqref{QPEndFixedPointLateTime}. On the other hand, if we first send $\frac{t_1}{L}\rightarrow \infty$ in \eqref{TermThatGivesLateTimeSaturation}, we get
\begin{equation}
    S_{[0,X_1]}^{(n)}(t_1) \approx \frac{n+1}{48n}\frac{\pi L}{\epsilon}\left(1-\frac{2L}{\pi^2 t_1}\right)+\ldots
\end{equation}
which corresponds to the asymptotic expression we found earlier. Therefore, we see that the discrepancy comes from the fact that the quasiparticles correspond to taking the $\epsilon\rightarrow0$ limit of the CFT first and so would not truly capture the $t_1\rightarrow \infty$ behaviour of the CFT.

\subsection{The time evolution of entanglement entropy in $2$d holographic CFT\label{Section:Evolution-in-holographic}}
We focus on the analysis on the time dependence of entanglement entropy and mutual information in $2$d holographic CFTs.
\subsubsection{The analysis of non-universal piece}
Now, we perform the analytic continuation to real time, $\tau_1=it_1$, and report the time dependence of the non-universal piece of entanglement entropy in $2$d holographic CFTs.
As explained in Section \ref{Section:NUPin2dhCFT}, the time dependence of the non-universal piece is determined by that of the geodesic length on the BTZ black hole geometry.
The $t_1$-dependence of $S_{\mathcal{V};\text{con}}$ is determined by the length of the geodesic connecting the points at the different Euclidean time slices in the method of image.
Without employing the method of the image, the $t_1$-dependence of this geodesic length should be equal to that of the geodesic length ending at the end of the world brane (EoW).  
In the high temperature regime, the location of EoW in the radial direction is near the boundary. 
In the parameter regime, where $v_1-v_2 \gg \epsilon$, the geodesic length is given by the one ending at the EoW brane. 
Consequently, in the small time region, $t_P>t_1>0$, the $t_1$-dependence of the non-universal piece of $S_A$ is determined by the motion of the EoW.
Here, we define $t_P$ as the time when $S_{\mathcal{V};\text{dis}}$ exchanges the dominance with $S_{\mathcal{V};\text{con}}$.
We will present the detailed motion of the EoW in Section \ref{Section:GD-and-cMERA}. 
We divide the system into $A$ and $\overline{A}$, the complement to $A$, let $X_{i=1,2}$ denote the edges of $A$, and assume $L>X_1>X_2\ge0$.
We will report on the time dependence of entanglement entropy during the M\"obius/SSD time evolution in the four cases:
(1) $X^{1}_f \in A$; (2) $L/2>X_1>X_2>0$; (3) $X^{2}_f \in A$; (4) $X_2=X^{1}_f$.
In the cases considered in this paper, $S_{\mathcal{V};\text{con}}$ monotonically grows in time according to the motion of EoW \cite{Goto:2023wai}.
For the large $t_1$-regime, $t_1>t_P$, the $t_1$-dependence of the non-universal piece is determined by the trajectory of twist and anti-twist operators during the time evolution induced by the inhomogeneous Hamiltonians.
As a simple example, let us consider the $t_1$-dependence of the non-universal piece for the reduced density matrix associated with the subsystem including $X^1_f$ during the evolution induced by $H_{\text{SSD}}$. 
In $t_P>t_1>0$, the $t_1$-dependence of the non-universal piece is determined by that of $S_{\mathcal{V};\text{con}}$.
In $t_1>t_P$, the $t_1$-dependence of the non-universal piece is determined by the smaller one of  $S_{\mathcal{V};\text{dis},1}$ and $S_{\mathcal{V};\text{dis},2}$.
We define an effective subsystem size in the Heisenberg picture as
\be
\begin{split}
    V_{\text{eff.}}=\begin{cases}
 L-\left(X^{\text{New},\alpha}_{v_1}-X^{\text{New},\alpha}_{v_2}\right) ~&~ \text{for}~ X^1_f \notin \mathcal{V}\\
    X^{\text{New},\alpha}_{v_1}-X^{\text{New},\alpha}_{v_2} ~&~ \text{for}~ X^1_f \in \mathcal{V}\\
\end{cases}.
\end{split}
\ee
In the time-regime where $L/2>V_{\text{eff.}}$, the non-universal piece is given by $S_{\mathcal{V};\text{dis},1}$, while in $V_{\text{eff.}}>L/2$, it is determined by $S_{\mathcal{V};\text{dis},2}$.

\subsubsection{During the M\"obius/SSD time evolution}
Now, let us present the time dependence of $S_A$ during the M\"obius/SSD time evolution.
For all the cases considered, $S_{A;\text{con}}$ monotonically grows with time during the M\"obius/SSD evolution (see Appendix \ref{App:earlytimeEE}).
The exchanging time $t_1=t_P$ should be determined by 
\be
\begin{split}
&w^{\text{New},\alpha}_{X_1}+\overline{w}^{\text{New},\alpha}_{X_2}=iL~ \text{For}~ S_{\text{con}}=S_{\text{dis},1} (<S_{\text{dis},2}),\\
&\overline{w}^{\text{New},\alpha}_{X_1}+w^{\text{New},\alpha}_{X_2}=0~ \text{For}~ S_{\text{con}}=S_{\text{dis},2} (<S_{\text{dis},1})\\
\end{split}.
\ee
After $t_1=t_P$, the non-universal piece of $S_A$ is determined by $S_{\text{dis},i}$.

\begin{figure}[htbp]
    \begin{tabular}{cc}
      \begin{minipage}[t]{0.5\hsize}
        \centering
        \includegraphics[keepaspectratio, scale=0.5]{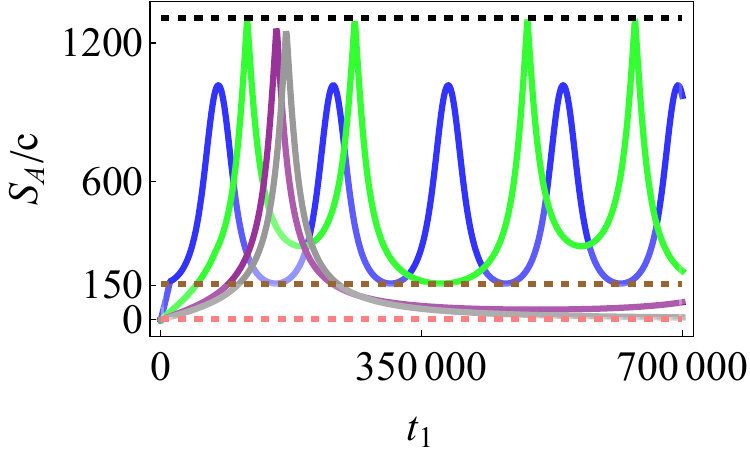}
        
    [a] $P_c=X^1_f$    
    
      \end{minipage}
      &\begin{minipage}[t]{0.5\hsize}
        \centering
        \includegraphics[keepaspectratio, scale=0.5]{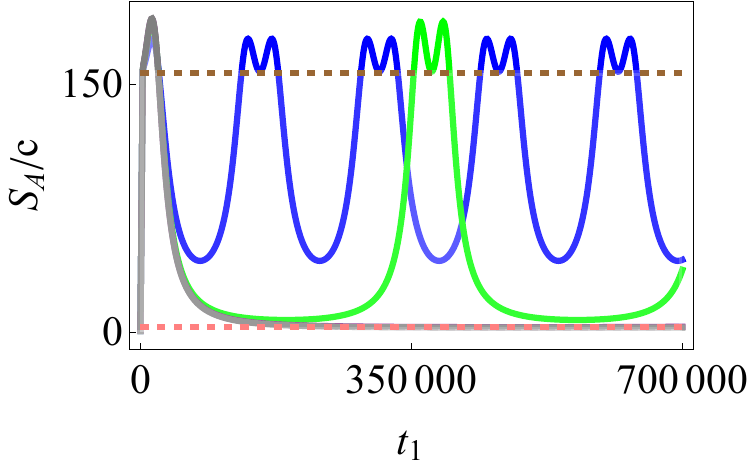}
        
          [b] $P_c=\f{L}{4}$    
          
      \end{minipage} \\
      
      \begin{minipage}[t]{0.5\hsize}
        \centering
        \includegraphics[keepaspectratio, scale=0.5]{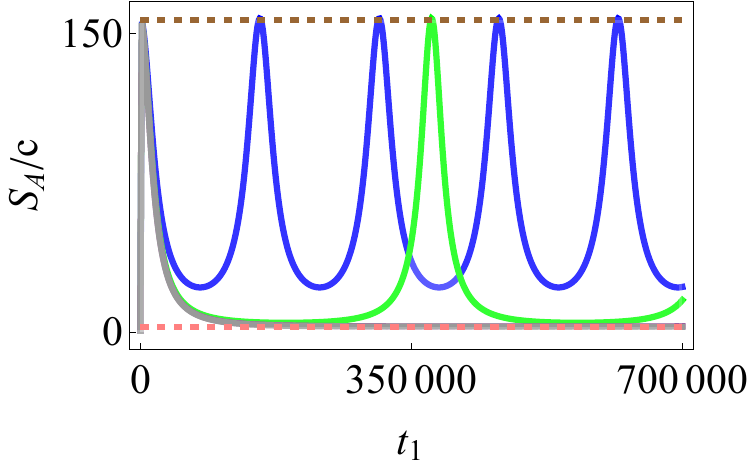}

          [c] $P_c=X^2_f$    
          
      \end{minipage} 

      &\begin{minipage}[t]{0.5\hsize}
        \centering
        \includegraphics[keepaspectratio, scale=0.5]{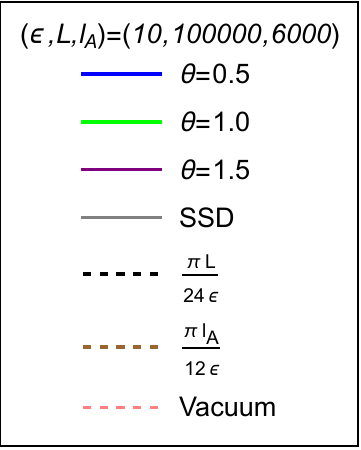}
      \end{minipage} 
    \end{tabular}
      \caption{The $t_1$-dependence of $S_{A}$ during the evolution induced by the M\"obius and SSD Hamiltonians. The panels, [a], [b], and [c], correspond to cases (1), (2), and (3).
      Here, $P_c$ denotes the center of $A$. For simplicity, in [a], [b], and [c], $P_c$ is taken to be $P_c=X^1_f, L/4$, and $X^2_f$, respectively.
      The black dashed line illustrates the entanglement entropy of the thermal state with $4\epsilon$, the inverse temperature, for half of the total space. The gray and pink dashed lines illustrate $S_A$ of this thermal state and the vacuum state.} 
        \label{Fig:entanglement-entropy-SSD-M0bious}
  \end{figure}
We depict $S_A$ in cases (1), (2), and (3) as a function of $t_1$ in Fig. \ref{Fig:entanglement-entropy-SSD-M0bious}.
During the M\"obius evolution in the cases (1), (2), and (3), the time-dependence of $S_A$ exhibits the two time-regimes: a heating one for $0<t<t_P$, and an oscillating one of $t_P<t$.
In the heating regime, the growth of $S_A$ depends on the inhomogeneous parameter $\theta$, the subsystem size $l_A$, and the system size $L$. 
Regardless of time evolution considered, in the limit where $\pi\left(w^{\text{New},\alpha}_{X_i}+\overline{w}^{\text{New},\alpha}_{X_i}\right)/4i\epsilon \gg 1$, the early time behavior of $S_A$ should be determined by the non-universal piece of $S_A$,
\be \label{eq:early-time-behavior}
\begin{split}
    S_A\approx S_{A,\text{Non-uni}}\approx \f{cL}{24\epsilon}\times \sum_{i=1,2}\left[\varphi_{X_i,\tau_1,\alpha}+\overline{\varphi}_{X_i,\tau_1,\alpha}\right],
\end{split}
\ee
where the details of $\varphi$ and $\overline{\varphi}$ are reported in Appendix \ref{App:thelocofop}.
During the M\"obius time evolution, $\varphi_{X_1,\tau_1,1}+\overline{\varphi}_{X_1,\tau_1,1}$ monotonically grows with time.
In the time interval, $1 \gg \varphi_{X_i,\tau_1,0}+\overline{\varphi}_{X_i,\tau_1,0}\gg \f{4\pi \epsilon}{L}$, 
the early-time growth of $S_A$ is approximated by
\be\label{EarlyTimeGrowthSAMobius}
S_A\approx \f{c\pi t_1}{12\epsilon}\sum_{i=1,2}\left[1-\tanh{2\theta}\cos{\left(\f{2\pi x_i}{L}\right)}\right].
\ee
Thus, in this time interval, $S_A$ linearly grows with $t_1$ as in the homogeneous quench \cite{2005JSMTE..04..010C,Hartman:2013qma}.
However, the coefficient of the linear growth is given by multiplying that of the homogeneous quench by an additional factor of the envelope function, $1-\tanh{2\theta}\cos{\left(\f{2\pi X_i}{L}\right)}$.
In the oscillating regime, $S_A$ periodically behaves in time with the period $L\cosh{2\theta}$.
During the evolution induced by the SSD Hamiltonian ($\alpha=0$), the time dependence of $S_A$ exhibits the two time-regimes: a heating one for $0<t<t_P$, and a cooling one for $t_P<t$.
As in the case of the time evolution in cases (1), (2), and (3), $S_A$ in the heating regime monotonically grows with $t_1$.
In the limit where $\pi\left(w^{\text{New},0}_{X_i}+\overline{w}^{\text{New},0}_{X_i}\right)/4i\epsilon \gg 1$, the early time behavior of $S_A$ is given by (\ref{eq:early-time-behavior}) for $\alpha=0$.
During the SSD time evolution, $\varphi_{X_i,\tau_1,0}+\overline{\varphi}_{X_i,\tau_1,0}$ monotonically increases with $t_1$. 
In the time interval, $1 \gg \varphi_{X_i,\tau_1,0}+\overline{\varphi}_{X_i,\tau_1,0}\gg \f{4\pi \epsilon}{L}$, 
the early-time growth of $S_A$ is approximated by
\be
S_A\approx \f{c\pi t_1}{6\epsilon}\sum_{i=1,2}\sin^2{\left(\f{\pi X_i}{L}\right)}.
\ee
Thus, in this time interval, $S_A$ linearly grows with $t_1$.
The coefficient of the linear growth is given by multiplying that for homogeneous quench by the envelope function, $2\sin^2{\left(\f{\pi X_i}{L}\right)}$.
In the cooling regime, the universal and non-universal pieces of $S_A$ are asymptotically given by
\be
\begin{split}
&S_{A,\text{Uni.}} \approx \f{c}{3}\log{\left[\f{2\pi t_1 \sin{\left(\f{\pi X_1}{L}\right)}}{L}\right]}+\f{c}{3}\log{\left[\f{2\pi t_1 \sin{\left(\f{\pi X_2}{L}\right)}}{L}\right]},\\
&S_{A,\text{Non-uni.}} \approx S_{\text{dis},2}\approx \f{c}{3}\log{\left[\f{L}{\pi}\cdot\f{L^2}{4\pi^2t_1^2}\cdot\f{\sin{\left[\f{\pi (X_1-X_2)}{L}\right]}}{\sin{\left[\f{\pi X_1}{L}\right]}\sin{\left[\f{\pi X_2}{L}\right]}}\right]}.
\end{split}
\ee
Thus, the universal and non-universal pieces logarithmically grow and decrease with $t_1$, respectively. 
Since the logarithmic growth cancels with the logarithmic decrease, $S_A$ for large $t_1$ becomes independent of $t_1$.
As a consequence, $S_A$ for the large $t_1$ is approximated by the vacuum entanglement entropy,
\be
S_A =S_{A,\text{Uni.}}+S_{A,\text{Non-uni.}}\approx \f{c}{3}\log{\left[\f{L}{\pi}\sin{\left[\f{\pi (X_1-X_2)}{L}\right]}\right]}.
\ee
Unlike the case of the thermal state in \cite{Goto:2021sqx}, even if the subsystem includes $x=X^1_f$, the entanglement entropy saturates to that for the vacuum one, not the thermal one.
\subsubsection{Thermal configuration \label{subsubsection:thermal_configuration}}
\begin{figure}[htbp]
    \begin{tabular}{cc}
      \begin{minipage}[t]{0.5\hsize}
        \centering
        \includegraphics[keepaspectratio, scale=0.6]{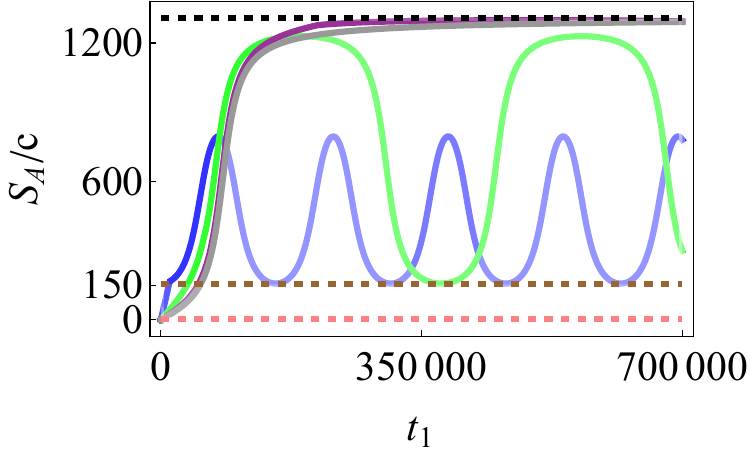}

      \end{minipage}

      &\begin{minipage}[t]{0.5\hsize}
        \centering
        \includegraphics[keepaspectratio, scale=0.5]{Time_evolution_of_EE_single/News/Legend.pdf}
      \end{minipage} 
    \end{tabular}
      \caption{The $t_1$-dependence of $S_{A}$ during the evolution induced by the M\"obius and SSD Hamiltonians. In this figure, we take $X_2$ to be $X^1_f$.
      The black dashed line illustrates the entanglement entropy for half of the total space of the thermal state with $4\epsilon$, the inverse temperature. The gray and pink dashed lines illustrate $S_A$ of this thermal state and the vacuum state.}
        \label{Fig:entanglement-entropy-SSD-M0bious-thermal}
  \end{figure}

As in the cases (1)-(3), if the edge of $A$ is not at $x=X^1_f$, then $S_A$ is asymptotically approximated by the vacuum entanglement entropy.
Now, we consider the case (4), the case where the edge of $A$ is at $x=X^1_f$.
During the M\"obius evolution, the $t_1$-dependence of $S_A$ is similar to the one in the cases (1)-(3).
During the SSD evolution, that of $S_A$ in case (4) is different from that in (1)-(3).
In the SSD limit where $\theta \rightarrow \infty$, $S_A$ monotonically grows with $t_1$, and then it is, for large $t_1$, approximated by
\be
S_A \approx \f{c}{6}\log{\left( t_1 \right)}+\f{c \pi L}{24 \epsilon}+\f{c}{6}\log{\left[ \f{8\epsilon}{L}\cdot\left|\sin{\left(\f{\pi X_1}{L}\right)}\right|^2\right]},
\ee
where the first term is the logarithmic function of $t_1$, while the second term is the entanglement entropy of thermal entropy with $4\epsilon$ for the half space.
Thus, in case (4), $S_A$ is not asymptotically approximated by the vacuum entanglement entropy.
In Fig. \ref{Fig:entanglement-entropy-SSD-M0bious-thermal}, we depict $S_A$ as a function of $t_1$\footnote{One may worry that the entanglement entropy of a finite-size subsystem grows forever with time. 
However, there is a time scale over which holographic calculations become unreliable.
This scale is determined by the time when the EoW brane collides with the cutoff surface. In particular, it is given by a polynomial in $L$, not an exponent in $L$. For more details, see section \ref{section:gravitation}.}.

\subsection{Cooling time\label{Sec:cooling-time}}
We close this section by defining the cooling time that describes how fast the subsystem evolves to the vacuum state during the SSD time evolution.
In $2$d holographic CFT, except for case (4), the entanglement entropy for the single intervals, $A$, is approximated by the vacuum one in the late time regime that is defined by
\be\label{CoolingTime}
t\gg t_*= \f{L}{4\pi} \sqrt{\f{L}{\epsilon}\cdot \f{\sin{\left(\f{\pi (X_1-X_2)}{L}\right)}}{\sin{\left(\f{\pi X_1}{L}\right)}\sin{\left(\f{\pi X_2}{L}\right)}}} \underset{L\gg X_1-X_2}{\approx}\f{L}{4\pi} \sqrt{\f{\pi l_A}{\epsilon\sin{\left(\f{\pi X_1}{L}\right)}\sin{\left(\f{\pi X_2}{L}\right)}}},
\ee
where $l_A$ denotes the size of the subsystem, $A$.
Thus, we call $t_*$ the cooling time.
In this late time regime, $t\gg t_*$, the quasiparticle picture does not work well because the entanglement entropy following this picture is smaller than $\mathcal{O}(1)$. 
This suggests that even during the SSD time evolution of $2$d free fermion, $t_*$ characterizes the time for the subsystem to cool down to the vacuum state.

Incidentally, this is also the time scale it takes for the quasiparticle R\'{e}nyi entropy to decay to an $\mathcal{O}(1)$ value. For intervals that are located away from the origin $X_f^1$ so that $0<X_2<X_1<L$, at late times, when $t_1\gg L$, the R\'{e}nyi entropy of the quasiparticles are approximately given by
\begin{equation}
    S_A^{(n)}(t)\approx\frac{n+1}{n}\frac{cL}{24\epsilon} \left(\frac{L}{2\pi t_1}\right)^2 \frac{\sin{\frac{\pi(X_1-X_2)}{L}}}{\sin{\frac{\pi X_1}{L}}\sin{\frac{\pi X_2}{L}}}
\end{equation}
For a subsystem that is much smaller than the total system, the quasiparticle R\'{e}nyi entropy becomes $\mathcal{O}(1)$ when $t\sim t_*$ as defined in \eqref{CoolingTime}.

\if[0]
\subsection{Summary on the entanglement entropy for the single interval \label{sum_single_interval}}
Except when the endpoints of the subsystem are at $x=0$, the late-time behavior of entanglement entropy for the SSD quenched boundary state can eventually be approximated by the entanglement entropy for the vacuum state. 
Therefore, the late-time structure of entanglement in the SSD quenched boundary state is almost the same as that of the entanglement in a vacuum state.
\fi

\section{The time evolution of mutual information in $2$d CFTs \label{Section:Mutual-informaion-in-CFTs}}
Now, we consider the time dependence of mutual information during the SSD/M\"obius evolution to see if the non-local correlation measured by the mutual information is also asymptotically approximated by that of the vacuum state.
Divide the system into the subsystems, $A$ and $B$, and $\overline{A\cup B}$, the compliment to the union of $A$ and $B$, and then define the mutual information as the linear combination of entanglement entropies,
\be \label{eq:mutualinformation}
I_{A,B}=S_A+S_B-S_{A\cup B},
\ee
where $S_{\mathcal{V}=A,B,A\cup B}$ denotes the entanglement entropies for the reduced density matrix associated with $\mathcal{V}=A,B,A\cup B$, respectively.
Let $X_1$ and $X_2$ denote the edges of $A$, and let $Y_1$ and $Y_2$ denote the edges of $B$.
Here, we assume that $L>Y_1>Y_2>X_1>X_2>0$ or $L>X_1>Y_1>Y_2>X_2>0$, and $A$ does not overlap with $B$.
In this case, only the non-universal pieces of $S_{\mathcal{V}}$ contribute to the mutual information because the universal pieces cancel out with each other.
\subsection{Time dependence of $I_{A,B}$ in $2$d free fermion}
\begin{figure}
    \centering
    \includegraphics[width=0.45\textwidth]{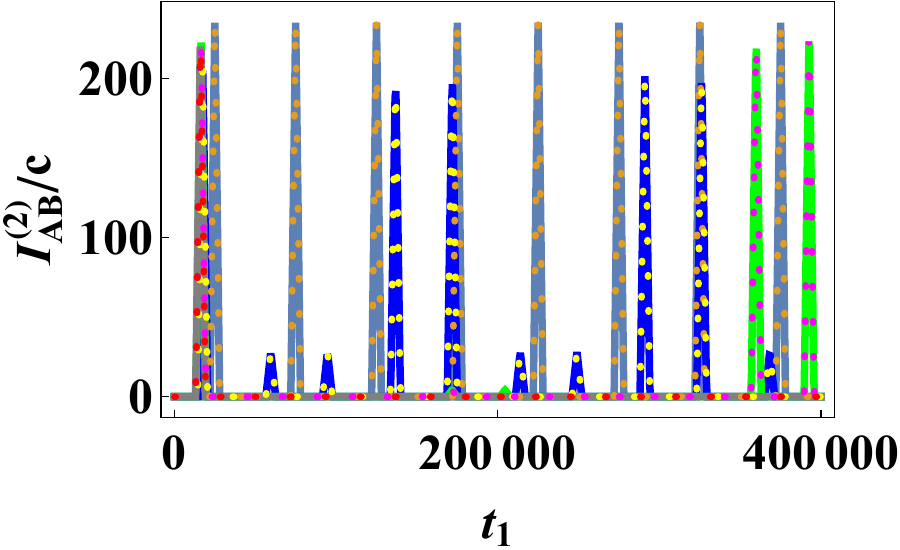}
    \includegraphics[width=0.45\textwidth]{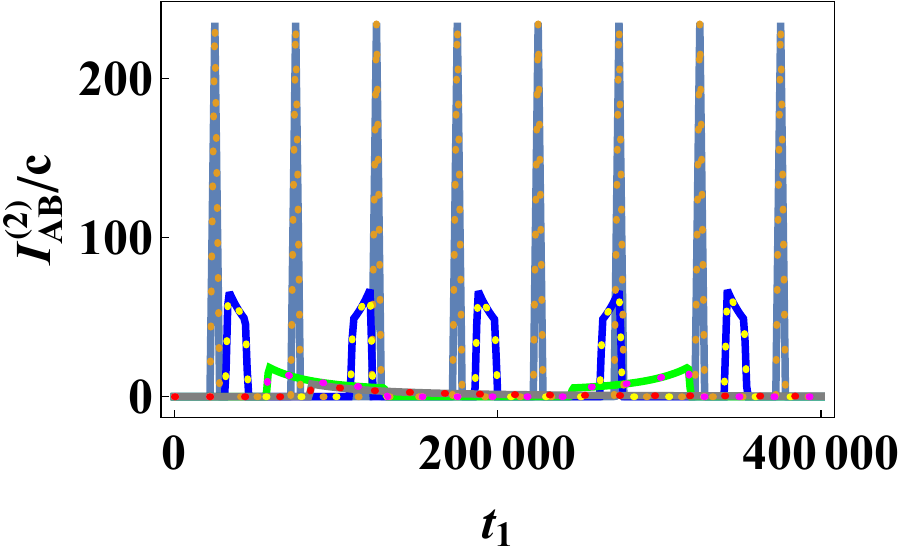}
    \text{    [a] $A=[\frac{L}{4}-X,\frac{L}{4}+X], B=[\frac{3L}{4}-X,\frac{3L}{4}+X]$\hspace{1cm} [b] $A=[L-X,X], B=[\frac{L}{2}-X,\frac{L}{2}+X]$ } \par\medskip
    \includegraphics[width=0.45\textwidth]{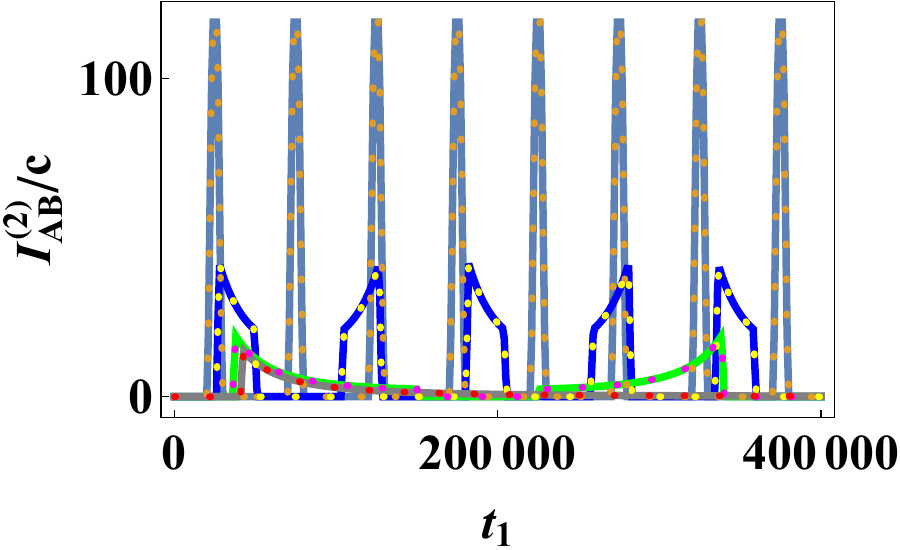}\hspace{2.6cm}
    \raisebox{1.5cm}{\includegraphics[width=0.3\textwidth]{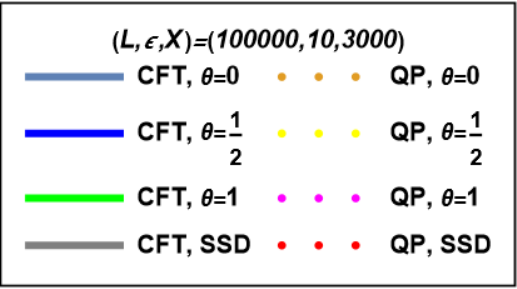}}
    \text{ [c] $A=[0,2X], B=[\frac{L}{2}-X,\frac{L}{2}+X]$ \hspace{7cm}} 
    \caption{Plots of the second R\'{e}nyi mutual information for the free Diract fermion as well as the quasiparticles for different placements of the subsystems of equal size $2X$. The total system size is fixed at $L=100000$ while the regulator is set at $\epsilon=10$. The continuous curves correspond to the CFT result while the dashed plots are the quasiparticle plots. }
    \label{MobiusQuenchMI}
\end{figure}
Consider the mutual information of the boundary state under these inhomogeneous quenches. The quasiparticle picture describes the second R\'{e}nyi mutual information of the free Dirac CFT well as seen in Fig. \ref{MobiusQuenchMI}.
Just as for the second R\'{e}nyi entropy, the second R\'{e}nyi mutual information does not depend on the boundary condition. The mutual information also shows the periodicity of $L\cosh{2\theta}$ which goes to infinity in the SSD limit. For the setups shown in figure \ref{MobiusQuenchMI}, the late time second R\'{e}nyi mutual information of the free fermion boundary states after the SSD quench is given by the second R\'{e}nyi mutual information for the vacuum state which is similar to the SSD quench of the spatially uniform thermal state \cite{Goto:2021sqx}.

When both subsystems are symmetrically placed away from the fixed point $X_1^f$ as seen in the first plot [a], the uniform quench produces non-zero mutual information when the Bell pairs that are initially nearly equidistant from both subsystems enter these subsystems. As these quasiparticles and their partners make their way around the spatial circle, they pass through both subsystems giving rise to two peaks in the first period. The quasiparticles that begin at around $L/2$ and $X_1^f$ reach both subsystems at about the same time which is no longer the case for M\"{o}bius quenches where the Bell pairs that begin at $L/2$ reach the subsystems earlier than those that begin near $X_1^f$, leading to a small second peak. The members of the quasiparticle pairs that begin near $X_1^f$ then go on to enter their second subsystem, leading to a third small peak. Finally, when the members of the quasiparticles that begin near $L/2$ enter their second subsystem, we see a fourth peak in the first period. In the SSD limit, we only observe a single peak since the quasiparticles are not able to go past the fixed point $X_f^1$. For both M\"{o}bius and SSD quenches, as the deformation parameter $\theta$ is increased, the speed about $L/2$ is greater so the first peak occurs earlier. The mutual information for the other two setups has peaks that are less symmetric but are nevertheless well-described by the quasiparticle picture.

\subsection{Time dependence of $I_{A,B}$ in $2$d holographic CFT}
Now, we focus on the non-universal piece of $S_{A \cup B}$ in $2$d holographic CFT.
We begin by computing the non-universal piece of $S_{A \cup B}$ in the Euclidean space.
In AdS/CFT correspondence, $S_{A \cup B}$ is determined by the minimal geodesic length \cite{Ryu:2006ef,Ryu:2006bv}:
\be
S_{A\cup B, \text{Non-uni.}}\approx \f{2c}{3}\log{\left(\f{4\epsilon}{\pi}\right)}+\f{c}{12}\text{Min}\left[ \mathcal{L}_{A\cup B, \text{con}}, \mathcal{L}_{A\cup B, \text{dis}}\right],
\ee
where $\mathcal{L}_{A\cup B, \text{con}}$ is the length of geodesic ending at the EoW brane, while $\mathcal{L}_{A\cup B, \text{dis}}$ is that of geodesic connecting two points at the same Euclidean time slice.

These non-universal pieces, $\mathcal{L}_{A\cup B, \text{con}}$ and $\mathcal{L}_{A\cup B, \text{dis}}$, are given by
\be
\begin{split}
    &\mathcal{L}_{A\cup B, \text{con}} =2\times\left[\sum_{i=1,2}\log{\left\{\cos{\left(\f{\pi\left(w^{\text{New},\alpha}_{X_i}+\overline{w}^{\text{New},\alpha}_{X_i}\right)}{4\epsilon}\right)}\cos{\left(\f{\pi\left(w^{\text{New},\alpha}_{Y_i}+\overline{w}^{\text{New},\alpha}_{Y_i}\right)}{4\epsilon}\right)}\right\}}\right]\\
    &\mathcal{L}_{A\cup B, \text{dis}}=\text{Min}\left[\mathcal{L}_{A;\text{dis},1}+\mathcal{L}_{B;\text{dis},1},\mathcal{L}_{A;\text{dis},2}+\mathcal{L}_{B;\text{dis},2}\right],\\
    \end{split}
\ee
where $\mathcal{L}_{A;\text{dis},i=1,2}$ and $\mathcal{L}_{A;\text{con},i=1,2}$ are defined in Appendix \ref{App:non-universalpiece-SAB}.
After performing the analytic continuation, $\tau_1=i t_1$, the minimal geodesic length determines the time dependence of $I_{A,B}$.
In the early time-regime where the non-universal pieces of $S_{\mathcal{V}=A,B,A\cup B}$ are determined by $S_{\mathcal{V},\text{con}}$, $S_{A\cup B}$ cancels out with $S_A+S_B$.
Consequently, in this early time-regime, $I_{A,B}$ is zero. 
For the late time-regime where the non-universal pieces of $S_{\mathcal{V}}$ are determined by $S_{\mathcal{V},\text{dis}}$, the time-evolution of $I_{A,B}$ is determined by
\be
I_{A,B}\approx \sum_{\mathcal{V}=A,B}\text{Min}\left[S_{\mathcal{V};\text{dis},1}, S_{\mathcal{V};\text{dis},2}\right]-\f{c}{12}\text{Min}\left[ \mathcal{L}_{A\cup B, \text{con}}, \mathcal{L}_{A\cup B, \text{dis}}\right].
\ee
 For the large $t_1$-regime in the SSD limit,  $I_{A,B}$ asymptotically reduces to the mutual information for the vacuum state,
\be \label{IAB-late-time}
\begin{split}
    I_{A,B} \approx \begin{cases}
\text{Max}\left[0, \f{c}{3}\log{\left\{\f{\sin{\left[\f{\pi(X_1-X_2)}{L}\right]}\sin{\left[\f{\pi(Y_1-Y_2)}{L}\right]}}{\sin{\left[\f{\pi(Y_1-X_2)}{L}\right]}\sin{\left[\f{\pi(Y_2-X_1)}{L}\right]}}\right\}}\right]~\text{for}~ X^1_f \notin A\\
\text{Max}\left[0, \f{c}{3}\log{\left\{\f{\sin{\left[\f{\pi(X_1-X_2)}{L}\right]}\sin{\left[\f{\pi(Y_1-Y_2)}{L}\right]}}{\sin{\left[\f{\pi(X_1-Y_1)}{L}\right]}\sin{\left[\f{\pi(Y_2-X_2)}{L}\right]}}\right\}}\right]~\text{for}~ X^1_f \in A\\
    \end{cases}.
\end{split}
\ee
When the endpoint of A or B is at $x=X^1_f$, $I_{A,B}$ is given by (\ref{IAB-late-time}) for $X^1_f \notin A$.
In conclusion, when starting from a boundary state and time-evolving it with the SSD Hamiltonian, the entanglement entropy does not strictly approach that of the vacuum state.
However, during the SSD time evolution, the mutual information may exactly approach that of the vacuum state.
Furthermore, unless $x=X^1_f$ at the edges of A and B, the reduced density matrices for $\mathcal{V}=A,B,A\cup B$ are asymptotically approximated by the vacuum reduce density matrices
\be
\rho_{\mathcal{V}=A,B,A\cup B} (t\gg 1)\approx \rho^{\text{Vacuum}}_{\mathcal{V}=A,B,A\cup B}, 
\ee
where $\rho^{\text{Vacuum}}_{\mathcal{V}}$ is the vacuum reduced density matrices associated to $\mathcal{V}$.

\section{Gravitational description and cMERA interpretation\label{Section:GD-and-cMERA}}
In this section, we will report on the gravity dual of the system considered in this paper.
In addition to it, we will discuss an interpretation for the SSD time evolution operator as a continuous multi-scale entanglement normalization ansatz (cMERA) \cite{Haegeman:2011uy,Nozaki:2012zj,Mollabashi:2013lya}.
\subsection{Gravitational description \label{section:gravitation}}

The early-time dependence of $S_A$ is determined by that of geodesic ending at the end of the world (EoW) brane.
Here, we discuss the gravity dual of the deformed boundary state, especially the trajectory of the EoW brane. In the AdS/BCFT correspondence~\cite{Takayanagi:2011zk,Fujita:2011fp}, the boundary effects we have discussed are explained by the insertion of the EoW brane into the bulk spacetime. The EoW brane is characterized by the brane tension $T$, which determines the boundary entropy in the calculation of the holographic entanglement entropy. 

We consider three-dimensional Einstein gravity on asymptotically AdS space $M$ with the dynamical brane located on $Q$,
\begin{align}
S=\dfrac{1}{16\pi G_N}\int_M \sqrt{-g}(R+2)+\dfrac{1}{8\pi G_N}\int_{\partial M} \sqrt{-h}K+\dfrac{1}{8\pi G_N}\int_{Q} \sqrt{-h}(K-T),
\end{align}
where we fixed the AdS radius to be unity, and $G_N$ is Newton's constant. On the one hand, we impose the Dirichlet boundary conditions on the metric at the asymptotic boundary $\partial M$. On the other hand, we impose the Neumann boundary condition on $Q$ which is necessary to consider the dynamical brane. Note that we have already fixed the matter profile on $Q$, characterized by the constant brane tension $T$, so that the boundary conformal symmetry is preserved. 

As a solution of Einstein's equation, we obtain the BTZ black hole metric
\begin{align}
ds^2=(r^2-r_+^2)d\tau^2+\dfrac{dr^2}{r^2-r_+^2}+r^2dx^2
\end{align}
with the EoW brane, whose (Euclidean) trajectory is given by
\begin{align}
r(\tau)=\dfrac{r_+}{\sqrt{1-T^2}}\sqrt{T^2\tan^2(r_+\tau)+1},
\end{align}
where $-1<T<1$. Throughout this subsection, we assume the length of circumference is $2\pi$. In the CFT part, we have discussed only the case $T=0$, while here we will discuss the more general value of $T$. For positive tension $T>0$, the EoW brane is located on the other asymptotic boundary of a maximally extended solution. On the other hand, for negative tension $T<0$, the EoW brane is outside of the horizon. In both cases, the brane eliminates the spacetime behind it, as the name suggests. In what follows, we mainly focus on the point of view of the boundary observer. Therefore, we shall discuss non-positive tension brane. 

\begin{figure}[tbp]
    \begin{tabular}{cc}
      \begin{minipage}[t]{0.33\hsize}
        \centering
        \includegraphics[keepaspectratio, scale=0.1]{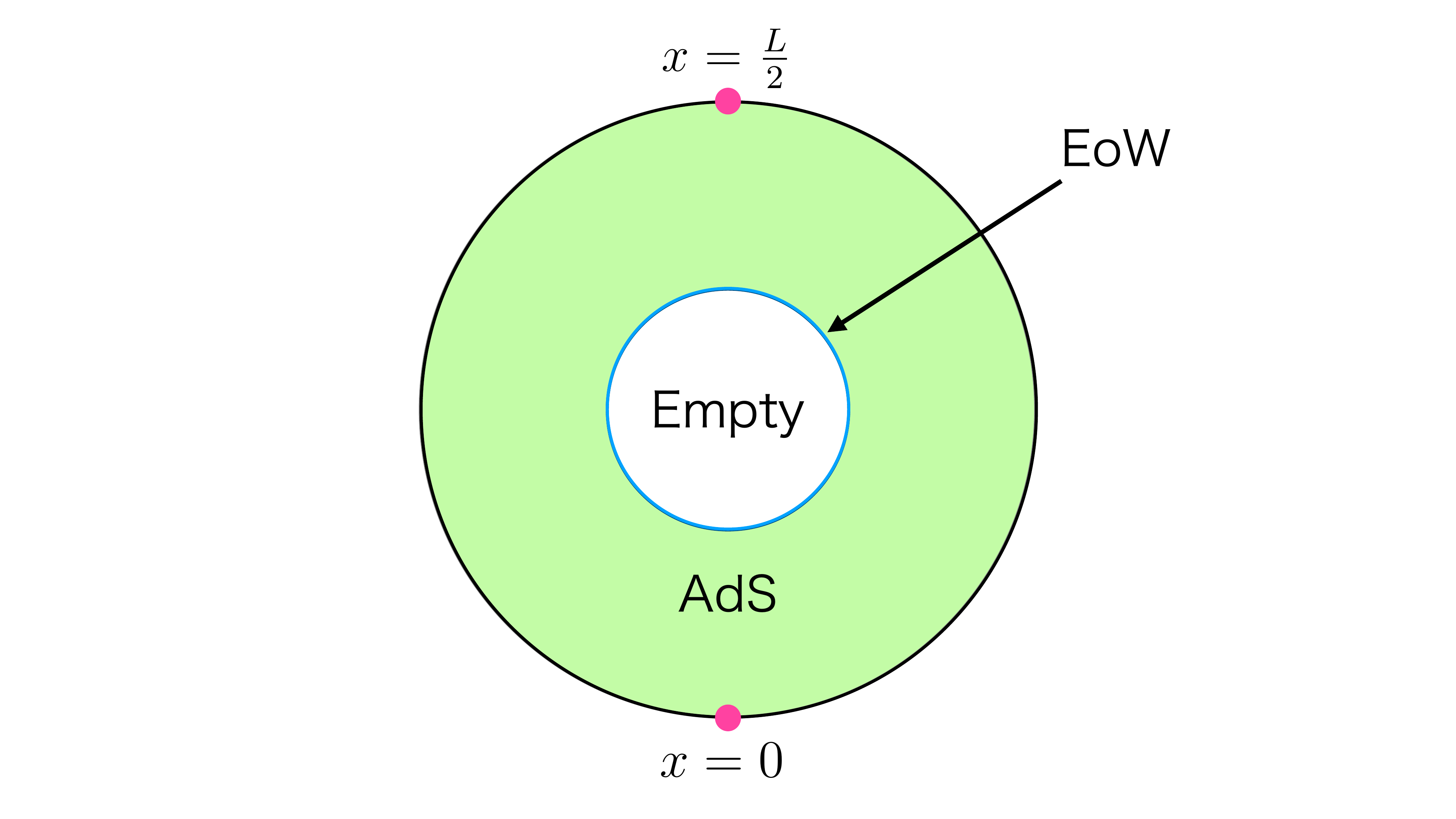}
        
    [a] Initial region
    
      \end{minipage}
      \begin{minipage}[t]{0.33\hsize}
        \centering
        \includegraphics[keepaspectratio, scale=0.1]{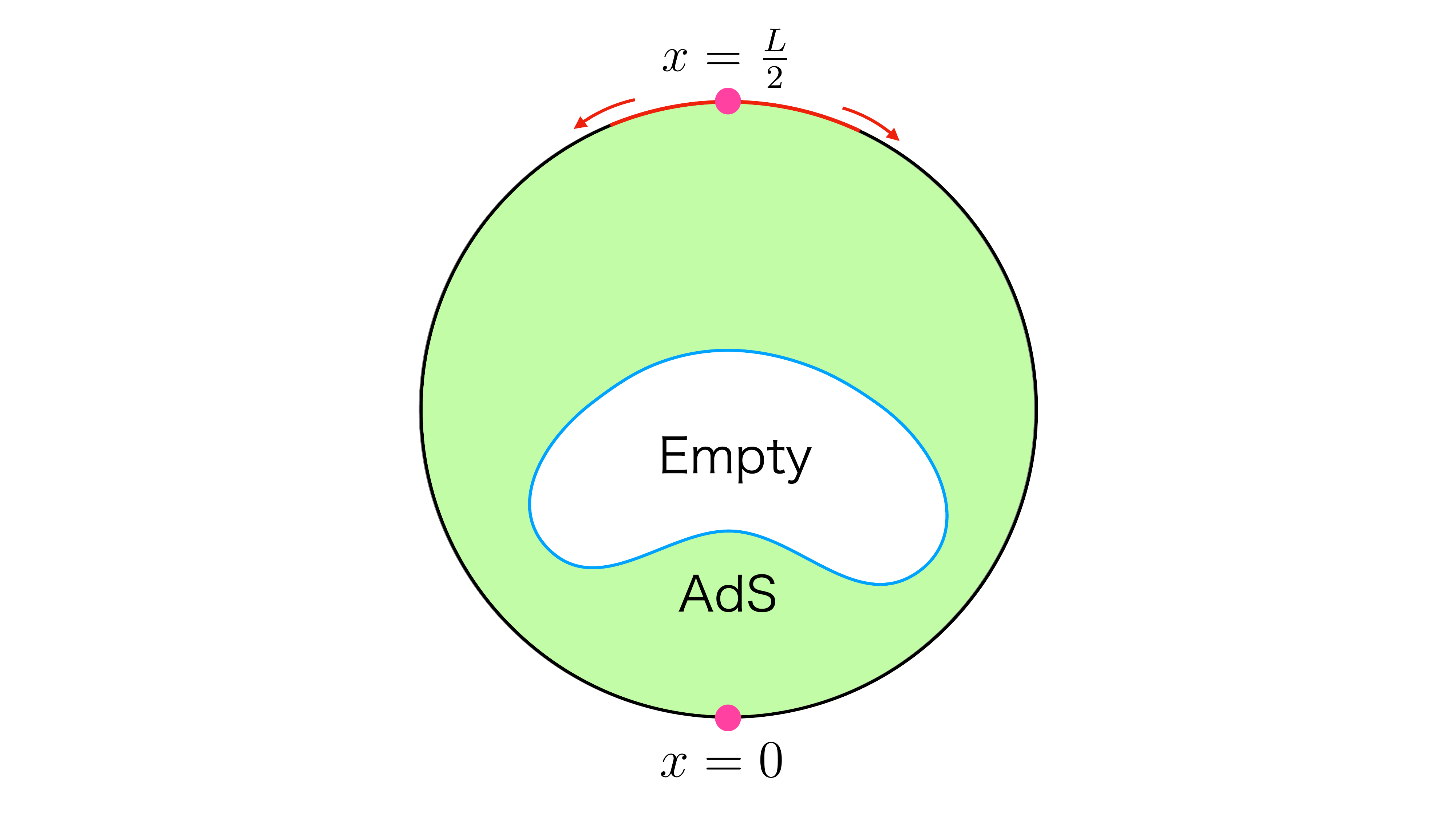}
        
          [b] Intermediate region
          
      \end{minipage} 
      \begin{minipage}[t]{0,33\hsize}
        \centering
        \includegraphics[keepaspectratio, scale=0.1]{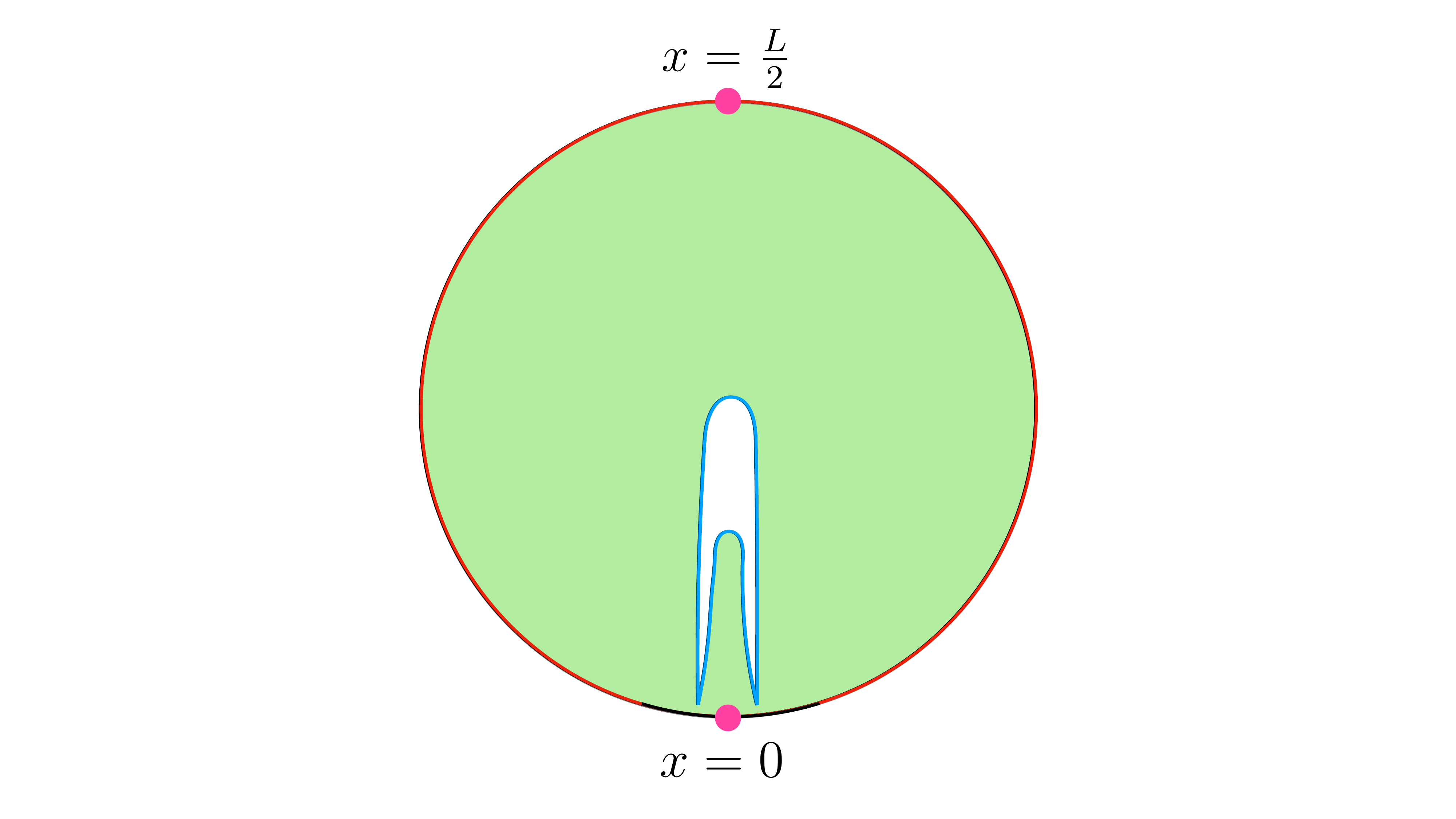}
        
        [c] Late time region 
        
      \end{minipage} 
    \end{tabular}
      \caption{A schematic view of motion of EoW brane under the SSD evolution ($\alpha=0$) with non-positive brane tension $T\leq0$. In this picture, time flows from the right panel to the left panel. The reduced density matrix in the red region is approximated by that for the vacuum state. The size of the red region grows with time following the deformation of EoW brane. }
        \label{The_motion_of_EoW}
  \end{figure}

\begin{figure}[t]
    \centering
    \subfigure[]{\resizebox{70mm}{!}{\includegraphics{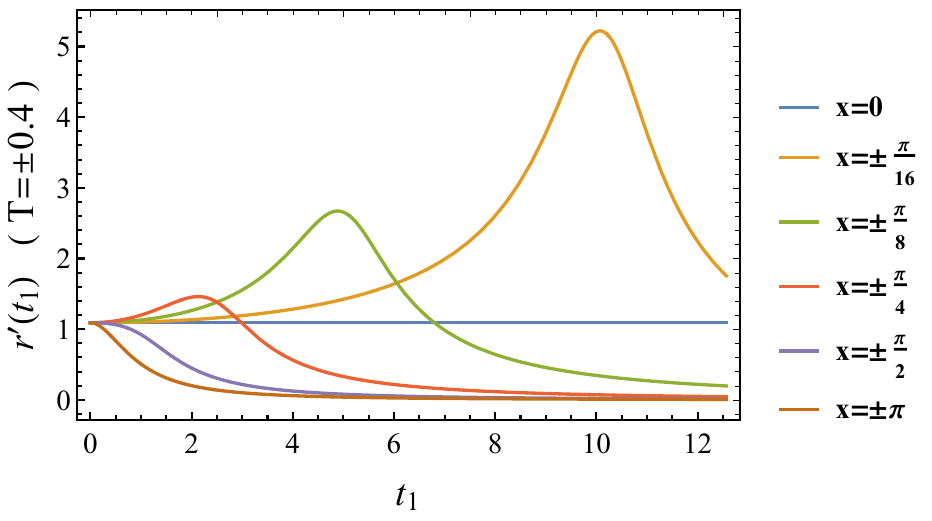}}}
    \subfigure[]{\resizebox{78mm}{!}{\includegraphics{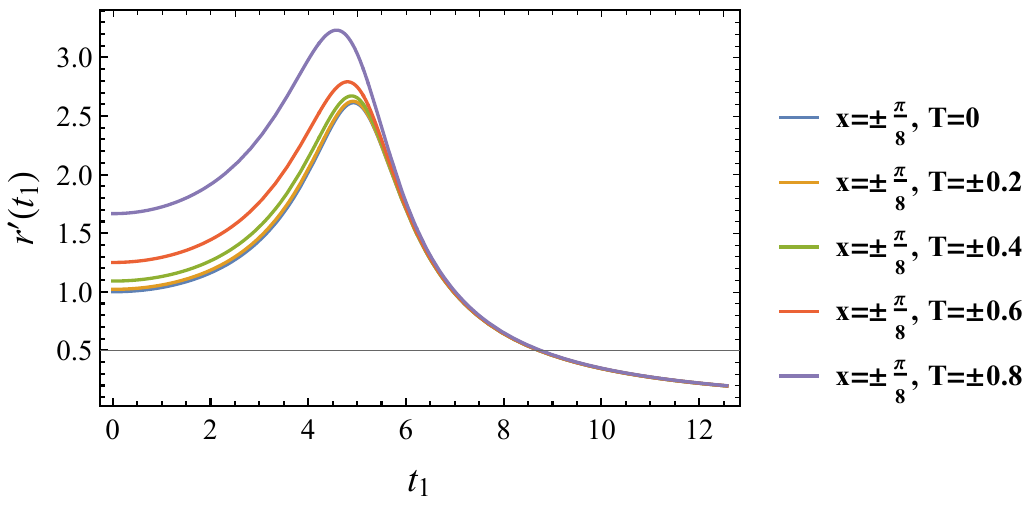}}}
    \caption{(a) Time-dependence of brane trajectory in SSD limit with $T=\pm 0.4$. (b) Time-dependence of brane trajectory in SSD limit with $x=\pm \pi/8$. Here, we set $r_+=1$.}
    \label{fig:trajectory1}
\end{figure}

\begin{figure}[t]
    \centering
    \subfigure[]{\resizebox{70mm}{!}{\includegraphics{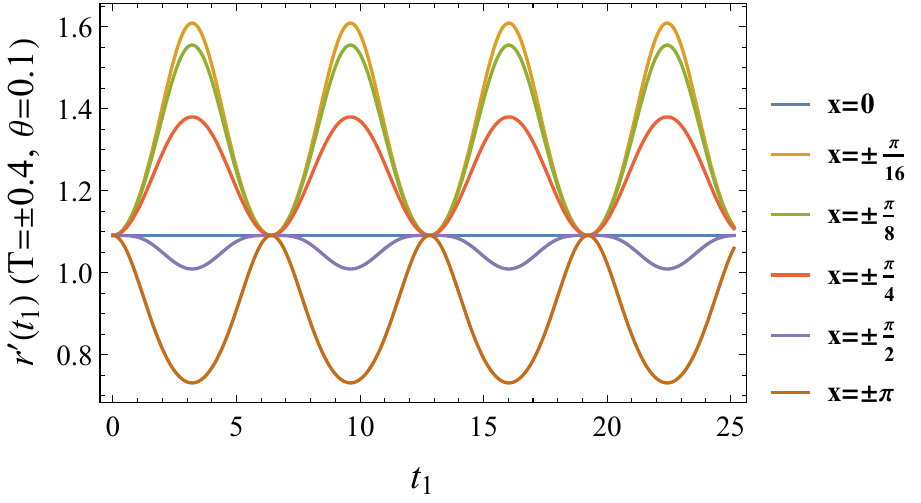}}}
    \subfigure[]{\resizebox{78mm}{!}{\includegraphics{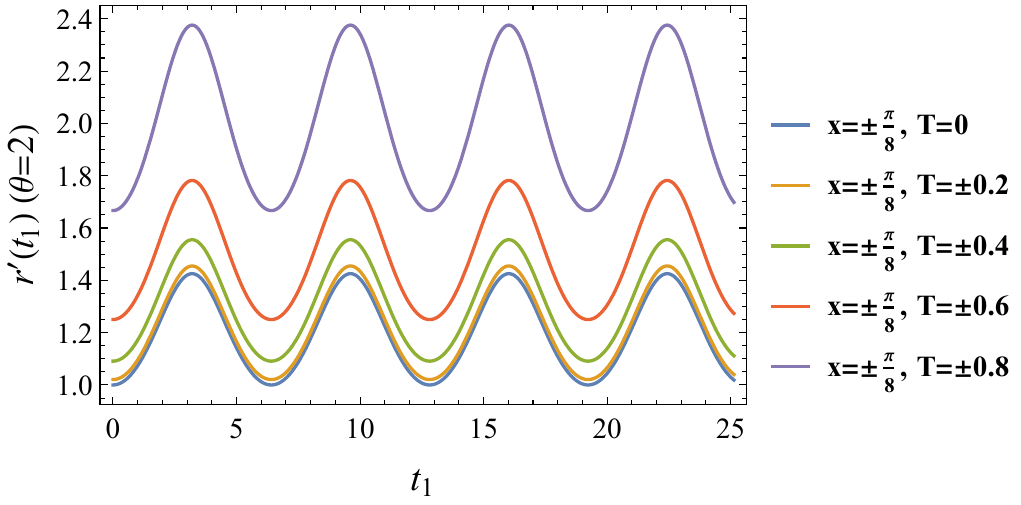}}}
    \caption{(a) time-dependence of brane trajectory with $\theta=0.1$ and $T=\pm 0.4$. (b) time-dependence of brane trajectory with $\theta=0.1$ and $x=\pm \pi/8$. Here we set $r_+=1$.}
    \label{fig:trajectory2}
\end{figure}

Let us describe the trajectory of EoW brane on the deformed black hole geometry. To this end, we follow the prescription discussed in~\cite{Goto:2021sqx}. Namely, we rescale the radial coordinate $r$ by the conformal factor and use the $w^{\text{New}}$ coordinates discussed in Appendix \ref{App:thelocofop}. After this replacement, we perform the analytic continuation to the Lorentzian time $t_1$. Consequently, the $t_1$-dependence of radial location is determined by
\begin{align}
r(\tau)\rightarrow r^\prime(t_1,x)=r(\tau_{\text{New}}(t_1,x))\sqrt{\left(\dfrac{dw_x^{\text{New},\alpha}}{dw_x}\right)\left(\dfrac{d\bar{w}_x^{\text{New},\alpha}}{d\bar{w}_x}\right)}, \label{eq:replacement}
\end{align}
where $\tau_{\text{New}}(t_1,x)=(w_x^{\text{New},\alpha}+\bar{w}_x^{\text{New},\alpha})/2$. Note that since the new coordinates depend on both $t_1$ and $x$, the brane trajectory also acquires the spatial inhomogeneity. See Fig. \ref{The_motion_of_EoW} for a schematic picture of the motion of the EoW brane from the outside observer. In Figs.\ \ref{fig:trajectory1} and \ref{fig:trajectory2}, we plot the spatial and time dependence of brane trajectories determined by \eqref{eq:replacement} in SSD limit ($\alpha=0$, see Fig. \ref{fig:trajectory1}) and M\"obius Hamiltonian ($\alpha=1$, see Fig.\ \ref{fig:trajectory2}) with small $\theta$. It is worth noting that for sufficiently large $\theta$, the time dependence reduces to the one in the SSD limit. 

The existence of EoW brane becomes clear when we discuss holographic entanglement entropy with a large subsystem compared with the inverse temperature $\beta=2\pi/r_+(=4\epsilon)$. In this case, the holographic entanglement entropy is calculated from the phase where the minimal surfaces end on the EoW brane. 

\begin{figure}[t]
 \begin{center}
  \resizebox{120mm}{!}{\includegraphics{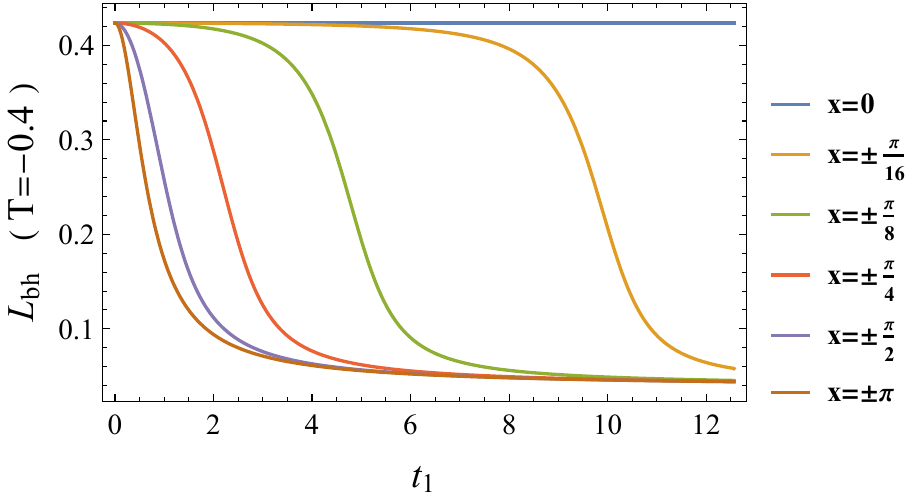}}
 \end{center}
 \caption{Geodesic distance between EoW with negative tension (T=-0.4) and black hole horizon for the original geometry. Here we show the case of SSD limit.}
 \label{fig:diffr}
\end{figure}

As a reference, we also plot the geodesic distance between the location of the EoW brane with negative tension and horizon as $L_{bh}$, although our true geometry ends at the EoW brane. See Fig. \ref{fig:diffr}. These figures suggest that the size of the domain eliminated by the EoW brane comes to depend on the location. In particular, such regions are more likely to be eliminated in the early time. 


\subsubsection*{Comments on cut off surface}

As we have seen so far, the EoW brane approaches the asymptotic boundary. At sufficiently late time, $t\gg L$, the location of ``peak of horizon’’ is linearly grows in time~\cite{Goto:2021sqx},
\begin{align}
r^\prime_{\text{horizon}}= \dfrac{r_+ \sqrt{L^2+\pi^2t^2}}{L} \simeq \dfrac{\pi r_+ t}{L}.
\end{align}
This is also the case for the brane because the distance between the brane and horizon becomes a constant value at the late time. 

On the other hand, we should introduce a cutoff surface on a certain radial scale $r^\prime=r_\infty$ that determines the UV cutoff for the dual CFT. It means that the EoW will eventually collide with the cutoff surface at some time. We interpret this as the time scale on which the holographic calculation in Section \ref{subsubsection:thermal_configuration} becomes unreliable\footnote{If the cutoff surface is defined globally, as it is in much of the literature, this collision problem should occur in all holographic calculations. Nevertheless, for all holographic calculations except Section \ref{subsubsection:thermal_configuration}, any immediate problems do not occur as we have seen. This observation suggests that we have to define the cutoff surface locally around the entangling surface. Note that this is a natural prescription from the CFT side~\cite{Ohmori:2014eia} and its gravity dual~\cite{Kusuki:2022ozk}.}. More concretely, we cannot trust our calculation after $ t\sim L/r_+ \epsilon_{UV}$. Here we introduced the cutoff length scale in the dual CFT $\epsilon_{UV}$ such that $r_\infty=1/\epsilon_{UV}$. 
\subsection{cMERA interpretation}
Here, we discuss an interpretation of the M\"obius/SSD time evolution operator as a continuous multi-scale entanglement normalization ansatz (cMERA).
\subsubsection{A brief review of MERA and cMERA}
In this section, we begin by reviewing the MERA shortly, and subsequently review the cMERA.
The MERA is the scheme of the renormalization group in terms of a tensor network constructed of the two kinds of tensors \cite{2007PhRvL..99v0405V,2009arXiv0912.1651V,2011arXiv1109.5334E}.
This is suitable for the (numerical) computation in the discrete system at the critical point. 
The tensor network in the MERA has a discrete layered structure where we have the discrete energy scale direction perpendicular to the space-time direction.
We label this discrete energy scale direction by $u$, and we assume that $u=0,-1,\cdots, -\infty$. 
On each layer at $u$, the tensor network is constructed of two types of tensors, called isometry and (dis-)entangler.
Let $\mathcal{L}$ and $K$ denote the isometry and entangler.
The isometry is a linear map where two nearest spins at $u$ reduce to a single spin at $u-1$.
Since this resembles the coarse-graining or the scale transformation, the isometry is considered as these operations in MERA.
If the dimension of Hilbert space at $u=0$ is $2^L$, where $L$ is the system size, then the one at $u$ is estimated with $2^{2^{u}L}$.
The entangler is the unitary operator acting the two nearest spins at $u$.
This tensor endows the state with short-range entanglement.
We start from an un-entangled state $\ket{\Omega}$ that is invariant under the scale transformation generator $\mathcal{L}$,
\be
\mathcal{L}\ket{\Omega}=0,
\ee
and then non-unitarily evolve the system from $u=-\infty$ to $u=0$ with the circuit constituting of the isometries and entanglers.
We call $\ket{\Omega}$ the reference state, while we call the state at $u=0$ the target state.
We tune the parameters of isometries and entanglers so that the target state is the state of interest.

Let us begin the review of cMERA.
The cMERA is designed to be suitable for the (numerical) computation in the quantum field theories at the critical point. 
Define the state at $u=-\infty$ as 
\be
\mathcal{L}\ket{\Omega}=0,
\ee
where $\mathcal{L}$ is the scale transformation generator as in the MERA.
However, in the cMERA, the isometry is replaced with a unitary operator.
Since the isometry and entangler preserve the dimension of Hilbert space at each step $u$, the tensor network in the cMERA is a unitary evolution operator.
We assume that the energy scale is labeled by the continuous parameter $u$.
Consequently, under the cMERA, the system unitarily evolves with the unitary evolution operator to the target state,
\be
\ket{\Psi(0)}=U(0,-\infty)\ket{\Omega}=\mathcal{P}e^{-i\int^{0}_{-\infty}du(K(u)+\mathcal{L})}\ket{\Omega},
\ee
where $\mathcal{P}$ is defined as the path-ordering operator arranging the operators in order of increasing $u$ from right to left.
The symbols, $K(u)$ and $\mathcal{L}$, are defined as the integrals of entangler and isometry densities along the spatial directions, respectively.

Furthermore, define the state and entangler in the interaction picture, $\ket{\Psi_{\text{I}}(u)}$ and $\hat{K}(u)$, as
\be
\begin{split}
&\ket{\Psi_{\text{I}}(u)}=e^{i \mathcal{L} u}\ket{\Psi(u)},~\hat{K}(u)=e^{i\mathcal{L} u} K(u) e^{-i\mathcal{L} u}.
\end{split}
\ee
Consequently, the target state in the interaction picture is given by
\be
\begin{split}
  \ket{\Psi_{\text{I}}(0)}=  \ket{\Psi(0)}=\mathcal{P}e^{-i\int^{0}_{-\infty}du\hat{K}(u)}\ket{\Omega}.
\end{split}
\ee
Thus, $\ket{\Psi_{\text{I}}(0)}$ does not depend on the definition of $\mathcal{L}$.
\subsubsection{A cMERA interpretation on the M\"obius/SSD time evolution}
Now, we turn to the quenches induced by M\"obius/SSD Hamiltonians, and discuss the interpretation for these quenches as the cMERA.
As in \cite{2015JHEP...05..152M}, we employ the boundary state with the proper regularization as the reference state,
\be
\ket{\Omega}=e^{-\epsilon H} \ket{\Psi_0},
\ee
where the Hamiltonian $H$ is defined as $H=(2\pi\left(L_0+\overline{L}_0\right))/L-(c\pi)/6L$, where $L_n$ and $\overline{L}_n$ are chiral and anti-chiral Virasoro generators.
In addition to them, the boundary state is defined as 
\be
(L_n-\overline{L}_n)\ket{\Psi_0}=0.
\ee
In this paper, unlike the common procedure in cMERA, we define the operator that keeps the reference state invariant as $\mathcal{L}$.
We utilize the spin operator $\left(2\pi\left(L_0-\overline{L}_0\right)\right)/L$ as $\mathcal{L}$.

Subsequently, we consider $t_1$ during the time evolution as the energy scale $u$ in the tensor network as in \cite{Chandra:2022pgl,Chandra:2021kdv}.
We start from the boundary state $\ket{\Omega}$ at $t_1=-U_{\text{IR}}$ and evolve the system up to $t_1=0$.
This is equivalent to the time evolution from $t_1=0$ to $t=U_{\text{IR}}$ as considered in Sections \ref{Section:Evolution-in-freefermion} and \ref{Section:Evolution-in-holographic}.
The depth of the time evolution and tensor network in cMERA is determined by $U_{\text{IR}}$.

\subsubsection*{M\"obius/SSD time evolution}
We begin by considering the quantum quench induced by the M\"obius Hamiltonian.
Divide the M\"obius Hamiltonian into the entangler $K(\theta)$ and the isometry $\mathcal{L}$,
\be
\begin{split}
    &H_{\text{M\"obius}} =K(\theta)+\mathcal{L},\\
    &K(\theta)=\f{2\pi}{L}\left[L_0+\overline{L}_0-\f{\tanh{2\theta}}{2}\left(L_{1}+L_{-1}+\overline{L}_{1}+\overline{L}_{-1}\right)-(L_0-\overline{L}_0)\right],~\mathcal{L}=\f{2\pi}{L}(L_0-\overline{L}_0),
\end{split}
\ee
where we express $H_{\text{M\"obius}}$, $K(\theta)$, and $\mathcal{L}$ in terms of Virasoro generators.
In the interaction picture, $\ket{\Psi_{\text{I}}(0)}$ is given by
\be \label{Quenched-state}
\ket{\Psi_{\text{I}}(0)}=\ket{\Psi(0)}=\mathcal{T}e^{-i \int^0_{-U_{\text{IR}}}dt_1 \hat{K}(\theta,t_1)}\ket{\Omega},
\ee
where $\mathcal {T}$ is defined as the time-ordering operator arranging the operators in order of increasing $t_1$ from right to left, and $\hat{K}(\theta,t_1)$ is defined by $\hat{K}(\theta,t_1)=e^{i\mathcal{L} t_1} K(\theta) e^{-i\mathcal{L} t_1}$.
Thus, $\ket{\Psi_{\text{I}}(0)}$ does not depend on the definition of $\mathcal{L}$.
The entangler in the interaction picture is explicitly given by
\be  \label{eq:entangler_mobius_interaction}
\hat{K}(\theta,t_1)=H_0-\mathcal{L}+i \sin{\left(\f{2\pi t_1}{L}\right)}\tilde{H}(\theta)+\cos{\left(\f{2\pi t_1}{L}\right)}\left(H_{\text{M\"obius}}-H_0\right)
\ee
where $H_0$ and $\tilde{H}(\theta)$ are defined by
\be
H_0= \f{2\pi}{L}\left(L_0+\overline{L}_0\right),~ \tilde{H}(\theta)=\f{\tanh{2\theta}}{2}\left(\f{2\pi}{L}\right)\left(L_1-L_{-1}-\overline{L}_{1}+\overline{L}_{-1}\right).
\ee
This entangler in the interaction picture posses the periodicity with $t_1$,
\be
\hat{K}(\theta,t_1+L)=\hat{K}(\theta,t_1).
\ee
This period is independent of the parameter $\theta$.
In the SSD limit where $\theta \rightarrow \infty$, $\hat{K}(\theta)$ reduces to $\tilde{K}$, 
\be
\begin{split}
\tilde{K}=\f{2\pi}{L}\left[L_0+\overline{L}_0-\f{1}{2}\left(L_{1}+L_{-1}+\overline{L}_{1}+\overline{L}_{-1}\right)-(L_0-\overline{L}_0)\right].
\end{split}
\ee
Consequently, the entangler in the interaction picture is given by replacing $H_{\text{M\"obius}}$ in (\ref{eq:entangler_mobius_interaction}) with $H_{\text{SSD}}$.

\if[0]
\subsubsection*{SSD time evolution}
Now, we turn to the quench induced by the SSD Hamiltonian.
Divide the SSD Hamiltonian into $\tilde{K}$ and $\mathcal{L}$, where $\tilde{K}$ is defined as $\tilde{K}=\lim_{\theta\rightarrow \infty} K(\theta)$,
\be
\begin{split}
\tilde{K}=\f{2\pi}{L}\left[L_0+\overline{L}_0-\f{1}{2}\left(L_{1}+L_{-1}+\overline{L}_{1}+\overline{L}_{-1}\right)-(L_0-\overline{L}_0)\right],
\end{split}
\ee
In the interaction picture, $\ket{\Psi_{\text{I}}(0)}$ is given by
\be \label{Quenched-state}
\ket{\Psi_{\text{I}}(0)}=\ket{\Psi(0)}=\mathcal{T}e^{-i \int^0_{-U_{\text{IR}}}dt \hat{\tilde{K}}(t)}\ket{\Omega},
\ee
where 
$\hat{\tilde{K}}(t_1)$ is defined as $\hat{\tilde{K}}(t_1)=e^{i\mathcal{L} t_1} \tilde{K} e^{-i\mathcal{L} t_1}$.
Thus, as for M\"obius evolution, $\ket{\Psi_{\text{I}}(0)}$ does not depend on the definition of $\mathcal{L}$.
The entangler in the interaction picture is explicitly given by
\be\label{EntanglerInTermsOfVirasoro}
\hat{\tilde{K}}(t_1)=H_0 -\mathcal{L}+\cos{\left(\f{2\pi t_1}{L}\right)}\left(H_{\text{SSD}}-H_0\right)+i \sin{\left(\f{2\pi t_1}{L}\right)} \hat{H},
\ee
where $\hat{H}$ is defined as $\hat{H}=\lim_{\theta\rightarrow \infty}\tilde{H}(\theta)$, and it is given by
\be
 \hat{H}=\f{1}{2}\left(\f{2\pi}{L}\right)\left(L_1-L_{-1}-\overline{L}_{1}+\overline{L}_{-1}\right).
\ee
\fi

\subsubsection*{Expression in terms of spin variables}
We will rewrite the entanglers in the interaction picture in terms of spin variables that suit experimental research. 

As explained in \cite{Gainutdinov_2013}, these Virasoro generators can be realized in certain spin chains with $N$ sites governed by a Hamiltonian that can be expressed as a sum of Temperley-Lieb generators $e_i$,
\begin{equation}\label{HamiltonianTemperleyLieb}
    H = -\sum_{i=1}^{N-1}e_i
\end{equation}
The generators $e_i$ satisfy the Temperley-Lieb algebra whose exact form is not necessary for our purposes. For a spin-$\frac{1}{2}$ chain, one possible representation of this algebra is given by
\begin{equation}
    e_i = \frac{q+q^{-1}}{4} - \frac{1}{2}\left(X_i X_{i+1}+Y_i Y_{i+1}+\frac{q+q^{-1}}{2}Z_i Z_{i+1}\right) - \frac{q-q^{-1}}{4}(Z_i-Z_{i+1})
\end{equation}
where $X_i$, $Y_i$ and $Z_i$ are Pauli matrices acting on site $i$ and $q$ is an arbitrary complex parameter although the physically interesting cases are obtained when $q$ is a root of unity. With this representation of the Temperley-Lieb algebra, the Hamiltonian \eqref{HamiltonianTemperleyLieb} is, up to an inconsequential constant, the XXZ Hamiltonian that contains some additional boundary terms
\begin{equation}
    H = \frac{1}{2}\sum_{i=1}^{N-1}\left(X_i X_{i+1}+Y_i Y_{i+1}+\frac{q+q^{-1}}{2}Z_i Z_{i+1}\right)+\frac{q-q^{-1}}{4} (Z_1-Z_{N})
\end{equation}
The standard Heisenberg spin chain can be obtained by simply setting $q=1$. It is conjectured, with substantial evidence, that the lattice operators 
\begin{equation}
    L_n^{(N)} = \frac{N}{\pi}\left[-\frac{1}{v_F}\sum_{k=1}^{N-1}(e_k-e_{\infty})\cos{\left(\frac{n k\pi}{N}\right)}+\frac{1}{v_F^2}\sum_{k=1}^{N-2}[e_k,e_{k+1}]\sin{\left(\frac{nk\pi}{N}\right)}\right]+\frac{c}{24}\delta_{n,0}
\end{equation}
approaches the Virasoro generators $L_n$ in the $N\rightarrow \infty$ continuum limit. Here, $v_F = \frac{\pi \sin{\gamma}}{\gamma}$ is the Fermi velocity and is determined by the $q$ parameter via $2\cos{\gamma}=q+q^{-1}$. For treatments of the Ising model and the XX spin chain, see \cite{KOO1994459} and \cite{GAINUTDINOV2013245} respectively.

\section{Discussion and Future Directions \label{Section:Discussions}}

We will discuss the relation between the gravity dual in Heisenberg picture and renomalization group, and comment on future directions.

\subsection*{Renormalization group and SSD evolution }

In section \ref{section:gravitation}, we discussed the gravity dual in Schr$\ddot{o}$dinger picture.
In this picture, the EoW brane moves and is deformed during the M\"obius/SSD time evolution.
We can instead consider the gravity dual in the Heisenberg picture.
In this picture, the location of EoW is pinned at the horizon of the BTZ black hole, while the location of the surface (UV surface) where CFT lives moves and is deformed in time.
Let $r_{\text{UV}}$ denote the radial location of this UV surface at $t_1=0$.
The trajectory of $r_{\text{UV}}$ during the M\"obius/SSD time evolution is determined by
\be
r^{\text{BTZ}}_{\text{UV}}(x,t_1)=\f{r_{\text{UV}}}{\sqrt{\left(\f{dw^{\text{New}}_{x}}{dw_{x}}\f{d\overline{w}^{\text{New}}_{x}}{d\overline{w}_{x}}\right)}}.
\ee
Thus, the location of the UV surface depends on $(x,t_1)$. 

In the AdS/CFT correspondence, the radial direction in the gravity dual is considered as the energy scale in the $2$d CFT.
In the coordinate considered in this paper, the larger $r$ becomes, the larger the energy scale becomes. 
The spatial and temporal dependence of the $r_{\text{UV}}$ suggests that the energy scale in $2$d CFT depends on $(x,t_1)$.
Except for the case discussed in Section \ref{subsubsection:thermal_configuration}, during SSD time evolution, $S_A$ is eventually approximated by the vacuum entanglement entropy.
This suggests that for the large $t_1$, the location of the UV surface is further from the EoW than at $t_1=0$. 
In other words, in the large time regime, the energy scale is larger than the initial one.
The location of the UV surface near $x=X^2_f$ gets further from the EoW faster.
We can see from the spatial and time dependence of the UV surface that the reverse of SSD time evolution considered in this paper may be used as the renormalization group where the energy scale depends on the spatial location.

\subsection*{Future direction}
We close this section with comments on the entanglement entropy for the large time regime.
In this paper and \cite{Goto:2021sqx}, we propose the quasiparticle picture, an effective picture, describing the time dependence of entanglement entropy at $\mathcal{O}(\frac{1}{\epsilon})$. 
However, this does not describe the late-time entanglement entropy, the vacuum entanglement entropy, because the vacuum one is at $\mathcal{O}(1)$. 
The findings in the inhomogeneous quench show that inhomogeneous evolution may endow the states with two types of entanglement structure: One of them is a dynamical entanglement structure that can be described by the propagation of quasiparticles, while the other is a static entanglement structure that still remains after quasiparticles pass away \cite{2020arXiv200310106Y}.
Study on the static entanglement structure may lead to a deeper understanding of entanglement dynamics.
We leave this as a future problem.
\section*{Acknowledgements}
We thank useful discussions with Shinsei Ryu. Especially, his suggestion of a relationship between SSDs and cMERA is illuminating for us.
M.N.~is supported by funds from the University of Chinese Academy of Sciences (UCAS), funds from the Kavli
Institute for Theoretical Sciences (KITS).
K.T.~is supported by JSPS KAKENHI Grant No.~21K13920 and MEXT KAKENHI Grant No.~22H05265. M.T. is supported by an
appointment to the YST Program at the APCTP through the Science and Technology Promotion Fund and Lottery Fund of the Korean Government, as well as the Korean Local Governments -
Gyeongsangbuk-do Province and Pohang City.
\newpage
\appendix

\section{The location of evolved operators \label{App:thelocofop}}
We present the details of the location of operators during the M\"obius/SSD evolution.
\subsection{Before the analytic continuation \label{App:thelocofop}}
In terms of $\tau_1$ and $x$, $w^{\text{New},\alpha}_x$ and $\overline{w}^{\text{New},\alpha}_x$ are given by
\be
\begin{split}
    &w^{\text{New},\alpha}_x+\epsilon=\tau_{x,\tau_1,\alpha}+i\f{L \varphi_{x,\tau_1,\alpha}}{2\pi}, \overline{w}^{\text{New},\alpha}_x+\epsilon=\tau_{x,\tau_1,\alpha}+i\f{L \overline{\varphi}_{x,\tau_1,\alpha}}{2\pi},\\
    &\tau_{x,\tau_1,0}=\epsilon-\log{\left[2\left(\pi \tau_1\right)^2\left(1-\cos{\left(\f{2\pi x}{L}\right)}\right)+L^2-2\pi \tau_1 L\left(1-\cos{\left(\f{2\pi x}{L}\right)}\right)\right]}+\f{L}{2\pi}\log{r_{x,\tau_1,0}},\\
    &r_{x,\tau_1,0}=\sqrt{\left(2(\pi\tau_1)^2\left(1-\cos{\left(\f{2\pi x}{L}\right)}\right)+L^2\cos{\left(\f{2\pi x}{L}\right)}\right)^2+\left(L^2 \sin{\left(\f{2\pi x}{L}\right)}\right)^2},\\
    &\cos{\varphi_{x,\tau_1,0}}=\cos{\overline{\varphi}_{x,\tau_1,0}}=\f{2(\pi\tau_1)^2\left(1-\cos{\left(\f{2\pi x}{L}\right)}\right)+L^2\cos{\left(\f{2\pi x}{L}\right)}}{r_{x,\tau_1,0}},\\
    &\sin{\varphi_{x,\tau_1,0}}=-\sin{\overline{\varphi}_{x,\tau_1,0}}=\f{L^2\sin{\left(\f{2\pi x}{L}\right)}}{r_{x,\tau_1,0}}.\\
    &\tau_{x,\tau_1,1}=\epsilon+\f{L}{2\pi}\log{r_{x,\tau_1,1}}-\log\bigg{[}(1-\lambda_1)^2\sinh^2{\left(2\theta\right)}+\left((\lambda_1-1)\cosh{2\theta}-(\lambda_1+1)\right)^2\\
    &~~~~~~~~+2(1-\lambda_1)\sinh{(2\theta)}\left((\lambda_1-1)\cosh{2\theta}-(\lambda_1+1)\right)\cos{\left(\f{2\pi x}{L}\right)}\bigg{]}\\
    &r_{x,\tau_1,1}=\sqrt{\left[(1-\lambda_1)^2\sinh{4\theta}+\left(-(1-\lambda_1)^2\cosh{4\theta}+(1+\lambda_1)^2\right)\cos{\left(\f{2\pi x}{L}\right)}\right]^2+16\lambda^2_1\sin^2{\left(\f{2\pi x}{L}\right)}},\\
    &\cos{\varphi_{x,\tau_1,1}}=\cos{\overline{\varphi}_{x,\tau_1,1}}=\f{(1-\lambda_1)^2\sinh{4\theta}+\left(-(1-\lambda_1)^2\cosh{4\theta}+(1+\lambda_1)^2\right)\cos{\left(\f{2\pi x}{L}\right)}}{r_{x,\tau_1,1}},\\
    &\sin{\varphi_{x,\tau_1,1}}=-\sin{\overline{\varphi}_{x,\tau_1,1}}=\f{4\lambda_1\sin{\left(\f{2\pi x}{L}\right)}}{r_{x,\tau_1,1}}, \lambda_1=e^{\f{2\pi \tau_1}{L \cosh{2\theta}}},\\
\end{split}
\ee
Thus, the location in Euclidean time direction of twist and anti-twist operator is $\tau=\tau_{x,\tau_1,\alpha}$.
\
\subsubsection{Analytic continuation}
We finally perform the analytic continuation to the real time:
\be
\tau_1=it_1.
\ee

\section{Early-time evolution of entanglement entropy during the M\"obius and SSD evolution \label{App:earlytimeEE}}
\begin{figure}[htbp]
    \begin{tabular}{cc}
      \begin{minipage}[t]{0.33\hsize}
        \centering
        \includegraphics[keepaspectratio, scale=0.4]{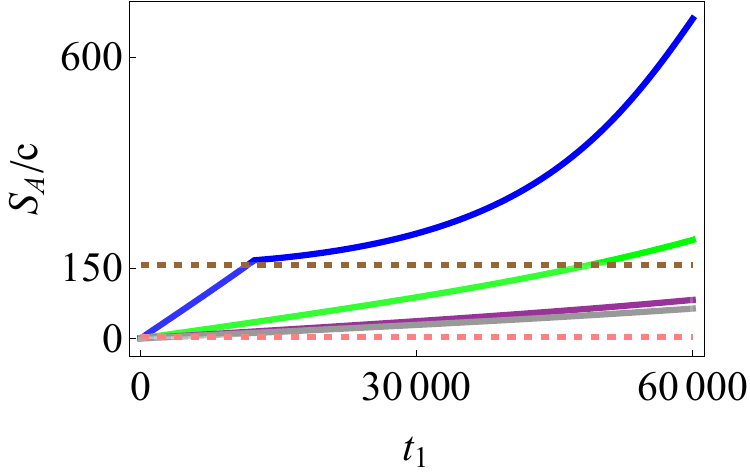}
        
    [a] $P_c=X^1_f$    
    
      \end{minipage}&
      
      \begin{minipage}[t]{0.33\hsize}
        \centering
        \includegraphics[keepaspectratio, scale=0.4]{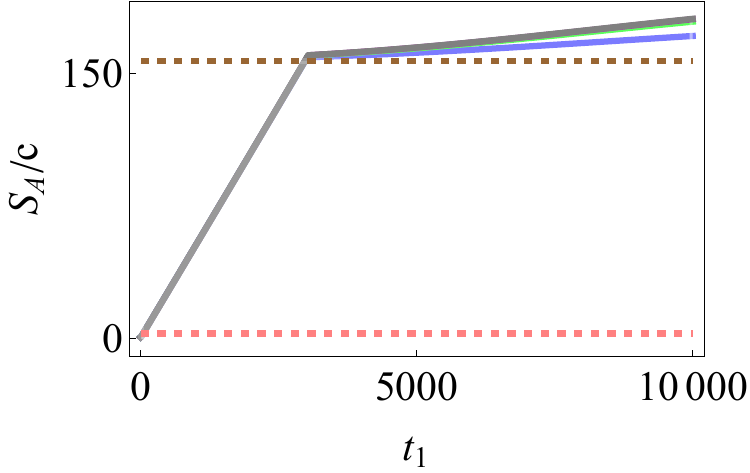}
        
          [b] $P_c=\f{L}{4}$    
          
      \end{minipage} 
      
      \begin{minipage}[t]{0.33\hsize}
        \centering
        \includegraphics[keepaspectratio, scale=0.4]{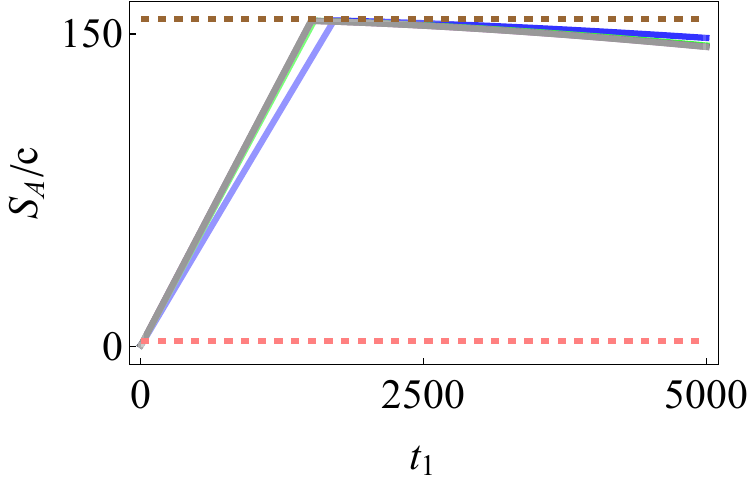}

          [c] $P_c=X^2_f$    
          
      \end{minipage} \\

 \begin{minipage}[t]{0.33\hsize}
        \centering
        \includegraphics[keepaspectratio, scale=0.4]{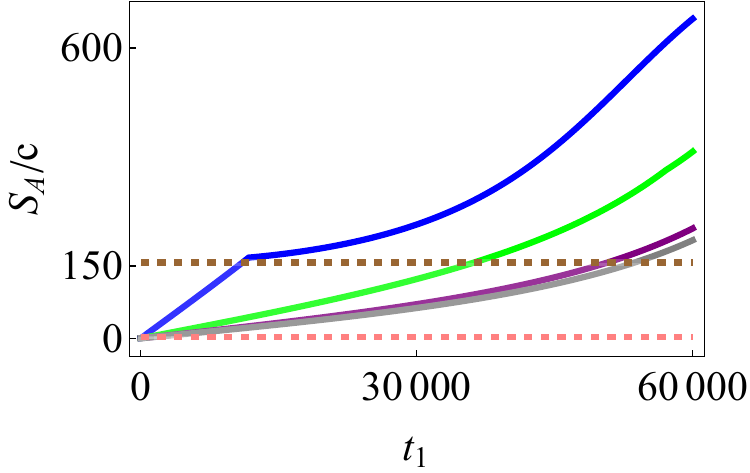}

          [d] Case 4   
        
      \end{minipage}&
      
      \begin{minipage}[t]{0.33\hsize}
        \centering
        \includegraphics[keepaspectratio, scale=0.4]{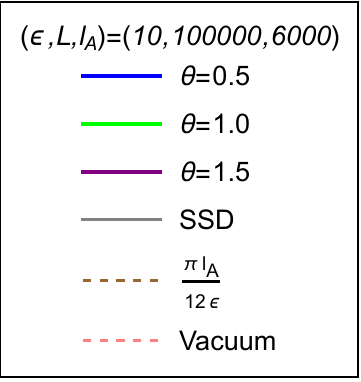}
      \end{minipage} 
    \end{tabular}
      \caption{The $t_1$-dependence of $S_{A}$ during the evolution induced by the M\"obius and SSD Hamiltonians. The panels, [a], [b], [c], and [d] correspond to cases (1), (2), (3), and (4).
      Here, $P_c$ denotes the center of $A$. For simplicity, in [a], [b], and [c], $P_c$ is taken to be $P_c=X^1_f, L/4$, and $X^2_f$, respectively.
      The blue, green, and purple curves illustrate the early time behavior of $S_A$ during the M\"obius evolution, while the grey line illustrates the early time behavior of $S_A$ during the M\"obius evolution.
      The brown and pink dashed lines illustrate $S_A$ of the thermal state with $4\epsilon$, inverse temperature, and the vacuum state.}
        \label{Fig:entanglement-entropy-SSD-M0bious-App}
  \end{figure}

We depict the early-time behavior of $S_A$ as a function of $t_1$ in Fig. \ref{Fig:entanglement-entropy-SSD-M0bious-App}. 
\section{The definition of $\mathcal{L}_{A;\text{dis},i=1,2}$ and $\mathcal{L}_{B;\text{dis},i=1,2}$ \label{App:non-universalpiece-SAB}}
We report on the definition of $\mathcal{L}_{A;\text{dis},i=1,2}$ and $\mathcal{L}_{B;\text{dis},i=1,2}$, here. 
They are defined as 
\be
\begin{split}
&\mathcal{L}_{A;\text{dis},i}=\text{Min}\left[\mathcal{L}^{(1)}_{A;\text{dis},i},\mathcal{L}^{(2)}_{A;\text{dis},i}\right], \mathcal{L}_{B;\text{dis},i}=\text{Min}\left[\mathcal{L}^{(1)}_{B;\text{dis},i},\mathcal{L}^{(2)}_{B;\text{dis},i}\right],\\
\end{split}
\ee
where if $x=X^1_f$ is nether in $A$ nor $B$, $\mathcal{L}^{(i=1,2)}_{\mathcal{V}=A,B;\text{dis},j=1,2}$ are respectively given by
\be \label{eq:netherAB}
\begin{split}
&\mathcal{L}^{(1)}_{A;\text{dis},1}=\log{\left|\sin{\left(\f{\pi}{4\epsilon}(w^{\text{New},\alpha}_{X_1}-w^{\text{New},\alpha}_{X_2}\pm iL)\right)}\right|^2}+\log{\left|\sin{\left(\f{\pi}{4\epsilon}(\overline{w}^{\text{New},\alpha}_{X_1}-\overline{w}^{\text{New},\alpha}_{X_2}\mp iL)\right)}\right|^2},\\
&\mathcal{L}^{(2)}_{A;\text{dis},1}=\log{\left|\sin{\left(\f{\pi}{4\epsilon}(w^{\text{New},\alpha}_{X_1}-w^{\text{New},\alpha}_{X_2})\right)}\right|^2}\bigg{]}+\log{\left|\sin{\left(\f{\pi}{4\epsilon}(-\overline{w}^{\text{New},\alpha}_{X_1}+\overline{w}^{\text{New},\alpha}_{X_2})\right)}\right|^2},\\
&\mathcal{L}^{(1)}_{B;\text{dis},1}=\log{\left|\sin{\left(\f{\pi}{4\epsilon}(w^{\text{New},\alpha}_{Y_1}-w^{\text{New},\alpha}_{Y_2}\pm iL)\right)}\right|^2}+\log{\left|\sin{\left(\f{\pi}{4\epsilon}(\overline{w}^{\text{New},\alpha}_{Y_1}-\overline{w}^{\text{New},\alpha}_{Y_2}\mp iL)\right)}\right|^2},\\
&\mathcal{L}^{(2)}_{B;\text{dis},1}=\log{\left|\sin{\left(\f{\pi}{4\epsilon}(w^{\text{New},\alpha}_{Y_1}-w^{\text{New},\alpha}_{Y_2})\right)}\right|^2}\bigg{]}+\log{\left|\sin{\left(\f{\pi}{4\epsilon}(-\overline{w}^{\text{New},\alpha}_{Y_1}+\overline{w}^{\text{New},\alpha}_{Y_2})\right)}\right|^2},\\
&\mathcal{L}^{(1)}_{A;\text{dis},2}=\log{\left|\sin{\left(\f{\pi}{4\epsilon}(w^{\text{New},\alpha}_{Y_2}-w^{\text{New},\alpha}_{X_1}\pm iL)\right)}\right|^2}+\log{\left|\sin{\left(\f{\pi}{4\epsilon}(-\overline{w}^{\text{New},\alpha}_{Y_2}+\overline{w}^{\text{New},\alpha}_{X_1}\mp iL)\right)}\right|^2},\\
&\mathcal{L}^{(2)}_{A;\text{dis},2}=\log{\left|\sin{\left(\f{\pi}{4\epsilon}(w^{\text{New},\alpha}_{Y_2}-w^{\text{New},\alpha}_{X_1})\right)}\right|^2}\bigg{]}+\log{\left|\sin{\left(\f{\pi}{4\epsilon}(-\overline{w}^{\text{New},\alpha}_{Y_2}+\overline{w}^{\text{New},\alpha}_{X_1})\right)}\right|^2},\\
&\mathcal{L}^{(1)}_{B;\text{dis},2}=\log{\left|\sin{\left(\f{\pi}{4\epsilon}(w^{\text{New},\alpha}_{X_2}-w^{\text{New},\alpha}_{Y_1}\pm iL)\right)}\right|^2}+\log{\left|\sin{\left(\f{\pi}{4\epsilon}(\overline{w}^{\text{New},\alpha}_{X_2}-\overline{w}^{\text{New},\alpha}_{Y_1}\mp iL)\right)}\right|^2},\\
&\mathcal{L}^{(2)}_{B;\text{dis},2}=\log{\left|\sin{\left(\f{\pi}{4\epsilon}(w^{\text{New},\alpha}_{X_2}-w^{\text{New},\alpha}_{Y_1})\right)}\right|^2}\bigg{]}+\log{\left|\sin{\left(\f{\pi}{4\epsilon}(-\overline{w}^{\text{New},\alpha}_{X_2}+\overline{w}^{\text{New},\alpha}_{Y_1})\right)}\right|^2}.\\
\end{split}
\ee
If $x=X^1_f$ is in $A$, $\mathcal{L}^{(i=1,2)}_{\mathcal{V}=A,B;\text{dis},j=1}$ is the same as (\ref{eq:netherAB}), and $\mathcal{L}^{(i=1,2)}_{\mathcal{V}=A,B;\text{dis},j=2}$ are given by
\be \label{eq:inAB}
\begin{split}
&\mathcal{L}^{(1)}_{A;\text{dis},2}=\log{\left|\sin{\left(\f{\pi}{4\epsilon}(w^{\text{New},\alpha}_{Y_2}-w^{\text{New},\alpha}_{X_2}\pm iL)\right)}\right|^2}+\log{\left|\sin{\left(\f{\pi}{4\epsilon}(\overline{w}^{\text{New},\alpha}_{Y_2}-\overline{w}^{\text{New},\alpha}_{X_2}\mp iL)\right)}\right|^2},\\
&\mathcal{L}^{(2)}_{A;\text{dis},2}=\log{\left|\sin{\left(\f{\pi}{4\epsilon}(w^{\text{New},\alpha}_{Y_2}-w^{\text{New},\alpha}_{X_2})\right)}\right|^2}\bigg{]}+\log{\left|\sin{\left(\f{\pi}{4\epsilon}(-\overline{w}^{\text{New},\alpha}_{Y_2}+\overline{w}^{\text{New},\alpha}_{X_2})\right)}\right|^2},\\
&\mathcal{L}^{(1)}_{B;\text{dis},2}=\log{\left|\sin{\left(\f{\pi}{4\epsilon}(w^{\text{New},\alpha}_{X_1}-w^{\text{New},\alpha}_{Y_1}\pm iL)\right)}\right|^2}+\log{\left|\sin{\left(\f{\pi}{4\epsilon}(\overline{w}^{\text{New},\alpha}_{X_1}-\overline{w}^{\text{New},\alpha}_{Y_1}\mp iL)\right)}\right|^2},\\
&\mathcal{L}^{(2)}_{B;\text{dis},2}=\log{\left|\sin{\left(\f{\pi}{4\epsilon}(w^{\text{New},\alpha}_{X_1}-w^{\text{New},\alpha}_{Y_1})\right)}\right|^2}\bigg{]}+\log{\left|\sin{\left(\f{\pi}{4\epsilon}(-\overline{w}^{\text{New},\alpha}_{X_1}+\overline{w}^{\text{New},\alpha}_{Y_1})\right)}\right|^2}.\\
\end{split}
\ee

\section{Vertex Operator Four-point Function in Free Dirac Fermion Boundary State}\label{VertexFourPointFunctionCalculation}
In this section, we generalize the computation of the vertex operator two-point function in \cite{Takayanagi2010} to the four-point function which we use to compute the mutual information.
\begin{equation}\label{VertexFourPointFunction}
\langle B |e^{-2\epsilon H} V_{k_L,k_R}(y_1,\Bar{y}_1))V_{l_L,l_R}(y_2,\Bar{y}_2)) V_{r_L,r_R}(y_3,\Bar{y}_3))V_{q_L,q_R}(y_4,\Bar{y}_4))|B\rangle
\end{equation}
We normal order the vertex operator in the same way as in \cite{Takayanagi2010} with the position and momenta of the zero mode in the same exponent,  
\begin{equation}
    V_{k_L,k_R}(y,\Bar{y}) = e^{ik_L(x_L+i s_L p_L y)+ik_R(x_R+i s_R p_R \bar{y})} \prod_{m>0}\left(e^{k_L \frac{\alpha_{-m}}{m}e^{my}+k_R\frac{\Tilde{\alpha}_{-m}}{m}e^{m\bar{y}}}\right)\prod_{m>0}\left(e^{-k_L \frac{\alpha_{m}}{m}e^{-my}-k_R\frac{\Tilde{\alpha}_{m}}{m}e^{-m\bar{y}}}\right).
\end{equation}
The zero mode is ordered differently from \cite{YellowBook} where the position and momentum coordinates are split between different exponentials. For Neumann boundary conditions, $k_L = -k_R = \frac{a}{N}$, while $k_L = k_R = \frac{a}{N}$ for Dirichlet boundary conditions. The symbols, $s_L$ and $s_R$, are arbitrary signs that we have introduced in front of the zero mode momentum such that under a spatial translation $\sigma\rightarrow\sigma+2\pi$, if $s_L=s_R=s$, the boson winds around the target manifold as $X(\sigma+2\pi)=X(\sigma)+2s\pi w R$ so that $s=1$ has the same periodicity as in \cite{polchinski2001string}. This is equivalent to flipping the sign of the zero mode in the Laurent expansion of the current $i\partial X$ which is an equally legitimate Laurent expansion. Following \cite{Takayanagi2010}, we also do not include any cocycle factors as explained in \cite{polchinski2001string}. Since these cocycle factors do not depend on the coordinates $y,\Bar{y}$, commuting them past the position operators can only give phases that are independent of the spacetime coordinate. Furthermore, as explained in \cite{polchinski2001string}, the cocycle factors only affect the relative signs of certain amplitudes but we are only considering a single correlation function.

Set $s_L =s_R=1$. For Neumann boundary conditions, set $\mu=-1$, $k_L = -k_R = k=-l_L=l_R=r_L=-r_R=-q_L=q_R$. For Dirichlet boundary conditions, set $k_L=k_R=k=-l_L=-l_R=r_L=r_R=-q_L=-q_R$. A calculation similar to the one in \cite{Takayanagi2010} gives 
\begin{align}\label{VertexFourPointFunction}
    &\langle B |e^{-2\epsilon H} V_{k_L,k_R}(y_1,\Bar{y}_1))V_{l_L,l_R}(y_2,\Bar{y}_2)) V_{r_L,r_R}(y_3,\Bar{y}_3))V_{q_L,q_R}(y_4,\Bar{y}_4))|B\rangle \nonumber \\
    =& \frac{1}{\eta\left(\frac{2i\epsilon}{\pi}\right)}\Bigg[\frac{\eta(\frac{2i\epsilon}{\pi})^{12}\theta_1\left(\frac{y_4-y_2}{2\pi i}\big|\frac{2i\epsilon}{\pi}\right)\theta_1\left(\frac{y_3-y_1}{2\pi i}\big|\frac{2i\epsilon}{\pi}\right)}{\theta_1\left(\frac{y_4-y_3}{2\pi i}\big|\frac{2i\epsilon}{\pi}\right)\theta_1\left(\frac{y_4-y_1}{2\pi i}\big|\frac{2i\epsilon}{\pi}\right)\theta_1\left(\frac{y_3-y_2}{2\pi i}\big|\frac{2i\epsilon}{\pi}\right)\theta_1\left(\frac{y_2-y_1}{2\pi i}\big|\frac{2i\epsilon}{\pi}\right)} \nonumber \\
    \times&
    \frac{\theta_1\left(\frac{\bar{y}_4-\bar{y}_2}{2\pi i}\big|\frac{2i\epsilon}{\pi}\right)\theta_1\left(\frac{\bar{y}_3-\bar{y}_1}{2\pi i}\big|\frac{2i\epsilon}{\pi}\right)}{\theta_1\left(\frac{\bar{y}_4-\bar{y}_3}{2\pi i}\big|\frac{2i\epsilon}{\pi}\right)\theta_1\left(\frac{\bar{y}_4-\bar{y}_1}{2\pi i}\big|\frac{2i\epsilon}{\pi}\right)\theta_1\left(\frac{\bar{y}_3-\bar{y}_2}{2\pi i}\big|\frac{2i\epsilon}{\pi}\right)\theta_1\left(\frac{\bar{y}_2-\bar{y}_1}{2\pi i}\big|\frac{2i\epsilon}{\pi}\right)} \nonumber \\
    \times& \frac{\theta_1\left(\frac{y_1+\bar{y}_2}{2\pi i}\big|\frac{2i\epsilon}{\pi}\right)\theta_1\left(\frac{y_1+\bar{y}_4}{2\pi i}\big|\frac{2i\epsilon}{\pi}\right)\theta_1\left(\frac{y_2+\bar{y}_1}{2\pi i}\big|\frac{2i\epsilon}{\pi}\right)\theta_1\left(\frac{y_2+\bar{y}_3}{2\pi i}\big|\frac{2i\epsilon}{\pi}\right)
    }
    {\theta_1\left(\frac{y_1+\bar{y}_1}{2\pi i}\big|\frac{2i\epsilon}{\pi}\right)\theta_1\left(\frac{y_1+\bar{y}_3}{2\pi i}\big|\frac{2i\epsilon}{\pi}\right)\theta_1\left(\frac{y_2+\bar{y}_2}{2\pi i}\big|\frac{2i\epsilon}{\pi}\right)\theta_1\left(\frac{y_2+\bar{y}_4}{2\pi i}\big|\frac{2i\epsilon}{\pi}\right)} \nonumber \\
    \times& \frac{\theta_1\left(\frac{y_3+\bar{y}_2}{2\pi i}\big|\frac{2i\epsilon}{\pi}\right)
    \theta_1\left(\frac{y_3+\bar{y}_4}{2\pi i}\big|\frac{2i\epsilon}{\pi}\right)\theta_1\left(\frac{y_4+\bar{y}_1}{2\pi i}\big|\frac{2i\epsilon}{\pi}\right)\theta_1\left(\frac{y_4+\bar{y}_3}{2\pi i}\big|\frac{2i\epsilon}{\pi}\right)}{\theta_1\left(\frac{y_3+\bar{y}_1}{2\pi i}\big|\frac{2i\epsilon}{\pi}\right)\theta_1\left(\frac{y_3+\bar{y}_3}{2\pi i}\big|\frac{2i\epsilon}{\pi}\right)\theta_1\left(\frac{y_4+\bar{y}_2}{2\pi i}\big|\frac{2i\epsilon}{\pi}\right)\theta_1\left(\frac{y_4+\bar{y}_4}{2\pi i}\big|\frac{2i\epsilon}{\pi}\right)}
    \Bigg]^{k^2} \nonumber \\
    \times&\begin{cases}
     |\mathcal{N}|^2 \left[ \theta_2\left( \frac{k(y_1 -y_2+y_3-y_4+\bar{y}_1-\bar{y}_2+\bar{y}_3-\bar{y}_4)}{2\pi i} \big| \frac{2i\epsilon}{\pi}\right)+\theta_3\left( \frac{k(y_1 -y_2+y_3-y_4+\bar{y}_1-\bar{y}_2+\bar{y}_3-\bar{y}_4)}{2\pi i} \big| \frac{2i\epsilon}{\pi}\right) \right]& \text{Neumann}   \\ 
     |\mathcal{N}'|^2 \theta_3\left( \frac{k(y_1 -y_2+y_3-y_4+\bar{y}_1-\bar{y}_2+\bar{y}_3-\bar{y}_4)}{2\pi i} \big| \frac{2i\epsilon}{\pi}\right) & \text{Dirichlet} 
    \end{cases}.
\end{align}

\bibliographystyle{ieeetr}
\bibliography{reference}

\newcommand{\prd}{Phys. Rev. D}\newcommand{\nat}{Nature}\newcommand{\pra}{Phys.
  Rev. A}\newcommand{\prb}{Phys. Rev. B}\newcommand{\pre}{Phys. Rev.
  E}\newcommand{\prl}{Phys. Rev. lett.}
\begin{thebibliography}{100}

\bibitem{Maldacena:1997re}
J.~M. Maldacena, ``{The Large N limit of superconformal field theories and
  supergravity},'' {\em Adv. Theor. Math. Phys.}, vol.~2, pp.~231--252, 1998.

\bibitem{2006PhRvL..96r1602R}
S.~{Ryu} and T.~{Takayanagi}, ``{Holographic Derivation of Entanglement Entropy
  from the anti de Sitter Space/Conformal Field Theory Correspondence},'' {\em
  Physical Review Letters}, vol.~96, p.~181602, May 2006.

\bibitem{2006JHEP...08..045R}
S.~{Ryu} and T.~{Takayanagi}, ``{Aspects of holographic entanglement
  entropy},'' {\em Journal of High Energy Physics}, vol.~8, p.~045, Aug. 2006.

\bibitem{Almheiri:2014lwa}
A.~Almheiri, X.~Dong, and D.~Harlow, ``{Bulk Locality and Quantum Error
  Correction in AdS/CFT},'' {\em JHEP}, vol.~04, p.~163, 2015.

\bibitem{Pastawski:2015qua}
F.~Pastawski, B.~Yoshida, D.~Harlow, and J.~Preskill, ``{Holographic quantum
  error-correcting codes: Toy models for the bulk/boundary correspondence},''
  {\em JHEP}, vol.~06, p.~149, 2015.

\bibitem{Choi:2019nhg}
S.~Choi, Y.~Bao, X.-L. Qi, and E.~Altman, ``{Quantum Error Correction in
  Scrambling Dynamics and Measurement-Induced Phase Transition},'' {\em Phys.
  Rev. Lett.}, vol.~125, no.~3, p.~030505, 2020.

\bibitem{2021PhRvB.103j4306L}
Y.~{Li} and M.~P.~A. {Fisher}, ``{Statistical mechanics of quantum error
  correcting codes},'' {\em \prb}, vol.~103, p.~104306, Mar. 2021.

\bibitem{2019PhRvB..99j4308M}
P.~{Mitra}, M.~{Ippoliti}, R.~N. {Bhatt}, S.~L. {Sondhi}, and K.~{Agarwal},
  ``{Cooling arbitrary near-critical systems using hyperbolic quenches},'' {\em
  \prb}, vol.~99, p.~104308, Mar. 2019.

\bibitem{2018PhRvL.120u0604A}
K.~{Agarwal}, R.~N. {Bhatt}, and S.~L. {Sondhi}, ``{Fast Preparation of
  Critical Ground States Using Superluminal Fronts},'' {\em \prl}, vol.~120,
  p.~210604, May 2018.

\bibitem{1995AmJPh..63..767G}
D.~J. {Griffiths} and E.~G. {Harris}, ``{Introduction to Quantum Mechanics},''
  {\em American Journal of Physics}, vol.~63, pp.~767--768, Aug. 1995.

\bibitem{PhysRevLett.101.170503}
A.~P. Young, S.~Knysh, and V.~N. Smelyanskiy, ``Size dependence of the minimum
  excitation gap in the quantum adiabatic algorithm,'' {\em Phys. Rev. Lett.},
  vol.~101, p.~170503, Oct 2008.

\bibitem{PhysRevE.84.061152}
I.~Hen and A.~P. Young, ``Exponential complexity of the quantum adiabatic
  algorithm for certain satisfiability problems,'' {\em Phys. Rev. E}, vol.~84,
  p.~061152, Dec 2011.

\bibitem{PhysRevA.86.052334}
E.~Farhi, D.~Gosset, I.~Hen, A.~W. Sandvik, P.~Shor, A.~P. Young, and
  F.~Zamponi, ``Performance of the quantum adiabatic algorithm on random
  instances of two optimization problems on regular hypergraphs,'' {\em Phys.
  Rev. A}, vol.~86, p.~052334, Nov 2012.

\bibitem{PhysRevLett.111.100502}
A.~del Campo, ``Shortcuts to adiabaticity by counterdiabatic driving,'' {\em
  Phys. Rev. Lett.}, vol.~111, p.~100502, Sep 2013.

\bibitem{PhysRevA.88.040101}
C.~Jarzynski, ``Generating shortcuts to adiabaticity in quantum and classical
  dynamics,'' {\em Phys. Rev. A}, vol.~88, p.~040101, Oct 2013.

\bibitem{1998Sci...280..421G}
S.~J. {Glaser}, T.~{Schulte-Herbreggen}, M.~{Sieveking}, O.~{Schedletzky},
  N.~C. {Nielsen}, O.~W. {S=F8rensen}, and C.~{Griesinger}, ``{Unitary Control
  in Quantum Ensembles: Maximizing Signal Intensity in Coherent
  Spectroscopy},'' {\em Science}, vol.~280, p.~421, Apr. 1998.

\bibitem{2017PNAS..114E3909S}
D.~{Sels} and A.~{Polkovnikov}, ``{Minimizing irreversible losses in quantum
  systems by local counterdiabatic driving},'' {\em Proceedings of the National
  Academy of Science}, vol.~114, pp.~E3909--E3916, May 2017.

\bibitem{2016NatSR...634187V}
S.~{van Frank}, M.~{Bonneau}, J.~{Schmiedmayer}, S.~{Hild}, C.~{Gross},
  M.~{Cheneau}, I.~{Bloch}, T.~{Pichler}, A.~{Negretti}, T.~{Calarco}, and
  S.~{Montangero}, ``{Optimal control of complex atomic quantum systems},''
  {\em Scientific Reports}, vol.~6, p.~34187, Oct. 2016.

\bibitem{PhysRevLett.117.170501}
J.~Geng, Y.~Wu, X.~Wang, K.~Xu, F.~Shi, Y.~Xie, X.~Rong, and J.~Du,
  ``Experimental time-optimal universal control of spin qubits in solids,''
  {\em Phys. Rev. Lett.}, vol.~117, p.~170501, Oct 2016.

\bibitem{PhysRevA.95.012317}
R.~R. Agundez, C.~D. Hill, L.~C.~L. Hollenberg, S.~Rogge, and M.~Blaauboer,
  ``Superadiabatic quantum state transfer in spin chains,'' {\em Phys. Rev. A},
  vol.~95, p.~012317, Jan 2017.

\bibitem{PhysRevLett.116.230503}
A.~Baksic, H.~Ribeiro, and A.~A. Clerk, ``Speeding up adiabatic quantum state
  transfer by using dressed states,'' {\em Phys. Rev. Lett.}, vol.~116,
  p.~230503, Jun 2016.

\bibitem{PhysRevE.95.012148}
G.~M. Rotskoff, G.~E. Crooks, and E.~Vanden-Eijnden, ``Geometric approach to
  optimal nonequilibrium control: Minimizing dissipation in nanomagnetic spin
  systems,'' {\em Phys. Rev. E}, vol.~95, p.~012148, Jan 2017.

\bibitem{PhysRevLett.111.260501}
G.~C. Hegerfeldt, ``Driving at the quantum speed limit: Optimal control of a
  two-level system,'' {\em Phys. Rev. Lett.}, vol.~111, p.~260501, Dec 2013.

\bibitem{2019ScPP....6...29H}
W.~W. {Ho} and T.~H. {Hsieh}, ``{Efficient variational simulation of
  non-trivial quantum states},'' {\em SciPost Physics}, vol.~6, p.~029, Mar.
  2019.

\bibitem{2018PhRvA..97f2343B}
S.~{Bao}, S.~{Kleer}, R.~{Wang}, and A.~{Rahmani}, ``{Optimal control of
  superconducting gmon qubits using Pontryagin's minimum principle: Preparing a
  maximally entangled state with singular bang-bang protocols},'' {\em \pra},
  vol.~97, p.~062343, June 2018.

\bibitem{PhysRevX.7.021027}
Z.-C. Yang, A.~Rahmani, A.~Shabani, H.~Neven, and C.~Chamon, ``Optimizing
  variational quantum algorithms using pontryagin's minimum principle,'' {\em
  Phys. Rev. X}, vol.~7, p.~021027, May 2017.

\bibitem{2018PhRvX...8c1086B}
M.~{Bukov}, A.~G.~R. {Day}, D.~{Sels}, P.~{Weinberg}, A.~{Polkovnikov}, and
  P.~{Mehta}, ``{Reinforcement Learning in Different Phases of Quantum
  Control},'' {\em Physical Review X}, vol.~8, p.~031086, July 2018.

\bibitem{Zaletel_2021}
M.~P. Zaletel, A.~Kaufman, D.~M. Stamper-Kurn, and N.~Y. Yao, ``Preparation of
  low entropy correlated many-body states via conformal cooling quenches,''
  {\em Physical Review Letters}, vol.~126, mar 2021.

\bibitem{2009arXiv0911.5506H}
T.-L. {Ho} and Q.~{Zhou}, ``{Universal Cooling Scheme for Quantum
  Simulation},'' {\em arXiv e-prints}, p.~arXiv:0911.5506, Nov. 2009.

\bibitem{PhysRevB.97.184309}
X.~Wen and J.-Q. Wu, ``Quantum dynamics in sine-square deformed conformal field
  theory: Quench from uniform to nonuniform conformal field theory,'' {\em
  Phys. Rev. B}, vol.~97, p.~184309, May 2018.

\bibitem{2019JPhA...52X5401M}
I.~{MacCormack}, A.~{Liu}, M.~{Nozaki}, and S.~{Ryu}, ``{Holographic duals of
  inhomogeneous systems: the rainbow chain and the sine-square deformation
  model},'' {\em Journal of Physics A Mathematical General}, vol.~52,
  p.~505401, Dec. 2019.

\bibitem{Goto:2021sqx}
K.~Goto, M.~Nozaki, K.~Tamaoka, M.~T. Tan, and S.~Ryu, ``{Non-Equilibrating a
  Black Hole with Inhomogeneous Quantum Quench},'' 12 2021.

\bibitem{PhysRevLett.118.260602}
W.~Berdanier, M.~Kolodrubetz, R.~Vasseur, and J.~E. Moore, ``Floquet dynamics
  of boundary-driven systems at criticality,'' {\em Phys. Rev. Lett.},
  vol.~118, p.~260602, Jun 2017.

\bibitem{2018arXiv180500031W}
X.~{Wen} and J.-Q. {Wu}, ``{Floquet conformal field theory},'' {\em arXiv
  e-prints}, p.~arXiv:1805.00031, Apr. 2018.

\bibitem{2020PhRvX..10c1036F}
R.~{Fan}, Y.~{Gu}, A.~{Vishwanath}, and X.~{Wen}, ``{Emergent Spatial Structure
  and Entanglement Localization in Floquet Conformal Field Theory},'' {\em
  Physical Review X}, vol.~10, p.~031036, July 2020.

\bibitem{Han_2020}
B.~Han and X.~Wen, ``Classification of sl2 deformed floquet conformal field
  theories,'' {\em Physical Review B}, vol.~102, Nov 2020.

\bibitem{2021PhRvR...3b3044W}
X.~{Wen}, R.~{Fan}, A.~{Vishwanath}, and Y.~{Gu}, ``{Periodically,
  quasiperiodically, and randomly driven conformal field theories},'' {\em
  Physical Review Research}, vol.~3, p.~023044, Apr. 2021.

\bibitem{2020arXiv201109491F}
R.~{Fan}, Y.~{Gu}, A.~{Vishwanath}, and X.~{Wen}, ``{Floquet conformal field
  theories with generally deformed Hamiltonians},'' {\em arXiv e-prints},
  p.~arXiv:2011.09491, Nov. 2020.

\bibitem{2021arXiv210910923W}
X.~{Wen}, Y.~{Gu}, A.~{Vishwanath}, and R.~{Fan}, ``{Periodically,
  Quasi-periodically, and Randomly Driven Conformal Field Theories (II):
  Furstenberg's Theorem and Exceptions to Heating Phases},'' {\em arXiv
  e-prints}, p.~arXiv:2109.10923, Sept. 2021.

\bibitem{PhysRevB.103.224303}
B.~Lapierre and P.~Moosavi, ``Geometric approach to inhomogeneous floquet
  systems,'' {\em Phys. Rev. B}, vol.~103, p.~224303, Jun 2021.

\bibitem{PhysRevResearch.2.023085}
B.~Lapierre, K.~Choo, C.~Tauber, A.~Tiwari, T.~Neupert, and R.~Chitra,
  ``Emergent black hole dynamics in critical floquet systems,'' {\em Phys. Rev.
  Res.}, vol.~2, p.~023085, Apr 2020.

\bibitem{Moosavi2021}
P.~Moosavi, ``Inhomogeneous conformal field theory out of equilibrium,'' {\em
  Annales Henri Poincar{\'e}}, Dec 2021.

\bibitem{PhysRevLett.122.020201}
E.~Langmann and P.~Moosavi, ``Diffusive heat waves in random conformal field
  theory,'' {\em Phys. Rev. Lett.}, vol.~122, p.~020201, Jan 2019.

\bibitem{10.21468/SciPostPhys.3.3.019}
J.~Dubail, J.-M. Stéphan, and P.~Calabrese, ``{Emergence of curved light-cones
  in a class of inhomogeneous Luttinger liquids},'' {\em SciPost Phys.},
  vol.~3, p.~019, 2017.

\bibitem{Gaw_dzki_2018}
K.~Gaw{\k{e}}dzki, E.~Langmann, and P.~Moosavi, ``Finite-time universality in
  nonequilibrium {CFT},'' {\em Journal of Statistical Physics}, vol.~172,
  pp.~353--378, mar 2018.

\bibitem{10.21468/SciPostPhys.2.1.002}
J.~Dubail, J.-M. Stéphan, J.~Viti, and P.~Calabrese, ``{Conformal field theory
  for inhomogeneous one-dimensional quantum systems: the example of
  non-interacting Fermi gases},'' {\em SciPost Phys.}, vol.~2, p.~002, 2017.

\bibitem{PhysRevB.83.060414}
T.~Hikihara and T.~Nishino, ``Connecting distant ends of one-dimensional
  critical systems by a sine-square deformation,'' {\em Phys. Rev. B}, vol.~83,
  p.~060414, Feb 2011.

\bibitem{2009PThPh.122..953G}
A.~{Gendiar}, R.~{Krcmar}, and T.~{Nishino}, ``{Spherical Deformation for
  One-Dimensional Quantum Systems},'' {\em Progress of Theoretical Physics},
  vol.~122, pp.~953--967, Oct. 2009.

\bibitem{2011PhRvA..83e2118G}
A.~{Gendiar}, M.~{Dani{\v{s}}ka}, Y.~{Lee}, and T.~{Nishino}, ``{Suppression of
  finite-size effects in one-dimensional correlated systems},'' {\em \pra},
  vol.~83, p.~052118, May 2011.

\bibitem{PhysRevB.84.165132}
I.~Maruyama, H.~Katsura, and T.~Hikihara, ``Sine-square deformation of free
  fermion systems in one and higher dimensions,'' {\em Phys. Rev. B}, vol.~84,
  p.~165132, Oct 2011.

\bibitem{2012JPhA...45k5003K}
H.~{Katsura}, ``{Sine-square deformation of solvable spin chains and conformal
  field theories},'' {\em Journal of Physics A Mathematical General}, vol.~45,
  p.~115003, Mar. 2012.

\bibitem{2015JPhA...48E5402I}
N.~{Ishibashi} and T.~{Tada}, ``{Infinite circumference limit of conformal
  field theory},'' {\em Journal of Physics A Mathematical General}, vol.~48,
  p.~315402, Aug. 2015.

\bibitem{2016IJMPA..3150170I}
N.~{Ishibashi} and T.~{Tada}, ``{Dipolar quantization and the infinite
  circumference limit of two-dimensional conformal field theories},'' {\em
  International Journal of Modern Physics A}, vol.~31, p.~1650170, Nov. 2016.

\bibitem{2016arXiv160309543O}
K.~{Okunishi}, ``{Sine-square deformation and Mobius quantization of
  two-dimensional conformal field theory},'' {\em arXiv e-prints},
  p.~arXiv:1603.09543, Mar. 2016.

\bibitem{2017arXiv170906238T}
S.~{Tamura} and H.~{Katsura}, ``{Zero-energy states in conformal field theory
  with sine-square deformation},'' {\em arXiv e-prints}, p.~arXiv:1709.06238,
  Sept. 2017.

\bibitem{2018PTEP.2018f1B01T}
T.~{Tada}, ``{Conformal quantum mechanics and sine-square deformation},'' {\em
  Progress of Theoretical and Experimental Physics}, vol.~2018, p.~061B01, June
  2018.

\bibitem{fan2021floquet}
R.~Fan, Y.~Gu, A.~Vishwanath, and X.~Wen, ``Floquet conformal field theories
  with generally deformed hamiltonians,'' 2021.

\bibitem{wen2018floquet}
X.~Wen and J.-Q. Wu, ``Floquet conformal field theory,'' 2018.

\bibitem{Goto:2023wai}
K.~Goto, M.~Nozaki, S.~Ryu, K.~Tamaoka, and M.~T. Tan, ``{Scrambling and
  Recovery of Quantum Information in Inhomogeneous Quenches in Two-dimensional
  Conformal Field Theories},'' 2 2023.

\bibitem{deBoer:2023lrd}
J.~de~Boer, V.~Godet, J.~Kastikainen, and E.~Keski-Vakkuri, ``{Quantum
  information geometry of driven CFTs},'' 5 2023.

\bibitem{Caputa:2022zsr}
P.~Caputa and D.~Ge, ``{Entanglement and geometry from subalgebras of the
  Virasoro algebra},'' {\em JHEP}, vol.~06, p.~159, 2023.

\bibitem{10.21468/SciPostPhys.14.5.108}
S.~Datta, B.~Lapierre, P.~Moosavi, and A.~Tiwari, ``{Marginal quenches and
  drives in Tomonaga-Luttinger liquids},'' {\em SciPost Phys.}, vol.~14,
  p.~108, 2023.

\bibitem{Kudler-Flam:2023ahk}
J.~Kudler-Flam, M.~Nozaki, T.~Numasawa, S.~Ryu, and M.~T. Tan, ``{Bridging two
  quantum quench problems -- local joining quantum quench and M\"obius quench
  -- and their holographic dual descriptions},'' 9 2023.

\bibitem{Liu:2023tiq}
X.~Liu, A.~McDonald, T.~Numasawa, B.~Lian, and S.~Ryu, ``{Quantum Quenches of
  Conformal Field Theory with Open Boundary},'' 9 2023.

\bibitem{Goto:2023yxb}
K.~Goto, T.~Guo, T.~Nosaka, M.~Nozaki, S.~Ryu, and K.~Tamaoka, ``{Spatial
  deformation of many-body quantum chaotic systems and quantum information
  scrambling},'' 5 2023.

\bibitem{Wen:2022pyj}
X.~Wen, R.~Fan, and A.~Vishwanath, ``{Floquet's Refrigerator: Conformal Cooling
  in Driven Quantum Critical Systems},'' 10 2022.

\bibitem{Calabrese:2005in}
P.~Calabrese and J.~L. Cardy, ``{Evolution of entanglement entropy in
  one-dimensional systems},'' {\em J. Stat. Mech.}, vol.~0504, p.~P04010, 2005.

\bibitem{Miyaji:2014mca}
M.~Miyaji, S.~Ryu, T.~Takayanagi, and X.~Wen, ``Boundary states as holographic
  duals of trivial spacetimes,'' {\em Journal of High Energy Physics},
  vol.~2015, p.~152, May 2015.

\bibitem{Margolus:1997ih}
N.~Margolus and L.~B. Levitin, ``{The Maximum speed of dynamical evolution},''
  {\em Physica D}, vol.~120, pp.~188--195, 1998.

\bibitem{Wen_2018}
X.~Wen and J.-Q. Wu, ``Quantum dynamics in sine-square deformed conformal field
  theory: Quench from uniform to nonuniform conformal field theory,'' {\em
  Physical Review B}, vol.~97, May 2018.

\bibitem{wen2021periodically}
X.~Wen, R.~Fan, A.~Vishwanath, and Y.~Gu, ``Periodically, quasi-periodically,
  and randomly driven conformal field theories: Part i,'' 2021.

\bibitem{Takayanagi2010}
T.~Takayanagi and T.~Ugajin, ``Measuring black hole formations by entanglement
  entropy via coarse-graining,'' {\em Journal of High Energy Physics},
  vol.~2010, p.~54, Nov 2010.

\bibitem{Takayanagi2022}
T.~Takayanagi and T.~Tsuda, ``Free fermion cyclic/symmetric orbifold cfts and
  entanglement entropy,'' {\em Journal of High Energy Physics}, vol.~2022,
  p.~4, Dec 2022.

\bibitem{Witten:1998zw}
E.~Witten, ``{Anti-de Sitter space, thermal phase transition, and confinement
  in gauge theories},'' {\em Adv. Theor. Math. Phys.}, vol.~2, pp.~505--532,
  1998.

\bibitem{Ryu:2006bv}
S.~Ryu and T.~Takayanagi, ``{Holographic derivation of entanglement entropy
  from AdS/CFT},'' {\em Phys. Rev. Lett.}, vol.~96, p.~181602, 2006.

\bibitem{Headrick:2013zda}
M.~Headrick, ``{General properties of holographic entanglement entropy},'' {\em
  JHEP}, vol.~03, p.~085, 2014.

\bibitem{2016JSMTE..06.4003C}
P.~{Calabrese} and J.~{Cardy}, ``{Quantum quenches in 1+1 dimensional conformal
  field theories},'' {\em Journal of Statistical Mechanics: Theory and
  Experiment}, vol.~6, p.~064003, June 2016.

\bibitem{doi:10.1073/pnas.1703516114}
V.~Alba and P.~Calabrese, ``Entanglement and thermodynamics after a quantum
  quench in integrable systems,'' {\em Proceedings of the National Academy of
  Sciences}, vol.~114, no.~30, pp.~7947--7951, 2017.

\bibitem{2005JSMTE..04..010C}
P.~{Calabrese} and J.~{Cardy}, ``{Evolution of entanglement entropy in
  one-dimensional systems},'' {\em Journal of Statistical Mechanics: Theory and
  Experiment}, vol.~4, p.~04010, Apr. 2005.

\bibitem{Hartman:2013qma}
T.~Hartman and J.~Maldacena, ``{Time Evolution of Entanglement Entropy from
  Black Hole Interiors},'' {\em JHEP}, vol.~05, p.~014, 2013.

\bibitem{Ryu:2006ef}
S.~Ryu and T.~Takayanagi, ``{Aspects of Holographic Entanglement Entropy},''
  {\em JHEP}, vol.~08, p.~045, 2006.

\bibitem{Haegeman:2011uy}
J.~Haegeman, T.~J. Osborne, H.~Verschelde, and F.~Verstraete, ``{Entanglement
  Renormalization for Quantum Fields in Real Space},'' {\em Phys. Rev. Lett.},
  vol.~110, no.~10, p.~100402, 2013.

\bibitem{Nozaki:2012zj}
M.~Nozaki, S.~Ryu, and T.~Takayanagi, ``{Holographic Geometry of Entanglement
  Renormalization in Quantum Field Theories},'' {\em JHEP}, vol.~10, p.~193,
  2012.

\bibitem{Mollabashi:2013lya}
A.~Mollabashi, M.~Nozaki, S.~Ryu, and T.~Takayanagi, ``{Holographic Geometry of
  cMERA for Quantum Quenches and Finite Temperature},'' {\em JHEP}, vol.~03,
  p.~098, 2014.

\bibitem{Takayanagi:2011zk}
T.~Takayanagi, ``{Holographic Dual of BCFT},'' {\em Phys. Rev. Lett.},
  vol.~107, p.~101602, 2011.

\bibitem{Fujita:2011fp}
M.~Fujita, T.~Takayanagi, and E.~Tonni, ``{Aspects of AdS/BCFT},'' {\em JHEP},
  vol.~11, p.~043, 2011.

\bibitem{Ohmori:2014eia}
K.~Ohmori and Y.~Tachikawa, ``{Physics at the entangling surface},'' {\em J.
  Stat. Mech.}, vol.~1504, p.~P04010, 2015.

\bibitem{Kusuki:2022ozk}
Y.~Kusuki and Z.~Wei, ``{AdS/BCFT from conformal bootstrap: construction of
  gravity with branes and particles},'' {\em JHEP}, vol.~01, p.~108, 2023.

\bibitem{2007PhRvL..99v0405V}
G.~{Vidal}, ``{Entanglement Renormalization},'' {\em \prl}, vol.~99, p.~220405,
  Nov. 2007.

\bibitem{2009arXiv0912.1651V}
G.~{Vidal}, ``{Entanglement Renormalization: an introduction},'' {\em arXiv
  e-prints}, p.~arXiv:0912.1651, Dec. 2009.

\bibitem{2011arXiv1109.5334E}
G.~{Evenbly} and G.~{Vidal}, ``{Quantum Criticality with the Multi-scale
  Entanglement Renormalization Ansatz},'' {\em arXiv e-prints},
  p.~arXiv:1109.5334, Sept. 2011.

\bibitem{2015JHEP...05..152M}
M.~{Miyaji}, S.~{Ryu}, T.~{Takayanagi}, and X.~{Wen}, ``{Boundary states as
  holographic duals of trivial spacetimes},'' {\em Journal of High Energy
  Physics}, vol.~2015, p.~152, May 2015.

\bibitem{Chandra:2022pgl}
A.~R. Chandra, J.~de~Boer, M.~Flory, M.~P. Heller, S.~H\"ortner, and A.~Rolph,
  ``{Cost of holographic path integrals},'' {\em SciPost Phys.}, vol.~14,
  no.~4, p.~061, 2023.

\bibitem{Chandra:2021kdv}
A.~R. Chandra, J.~de~Boer, M.~Flory, M.~P. Heller, S.~H\"ortner, and A.~Rolph,
  ``{Spacetime as a quantum circuit},'' {\em JHEP}, vol.~21, p.~207, 2021.

\bibitem{Gainutdinov_2013}
A.~M. Gainutdinov, J.~L. Jacobsen, N.~Read, H.~Saleur, and R.~Vasseur,
  ``Logarithmic conformal field theory: a lattice approach,'' {\em Journal of
  Physics A: Mathematical and Theoretical}, vol.~46, p.~494012, nov 2013.

\bibitem{KOO1994459}
W.~Koo and H.~Saleur, ``Representations of the virasoro algebra from lattice
  models,'' {\em Nuclear Physics B}, vol.~426, no.~3, pp.~459--504, 1994.

\bibitem{GAINUTDINOV2013245}
A.~Gainutdinov, N.~Read, and H.~Saleur, ``Continuum limit and symmetries of the
  periodic g$\ell(1|1)$ spin chain,'' {\em Nuclear Physics B}, vol.~871, no.~2,
  pp.~245--288, 2013.

\bibitem{2020arXiv200310106Y}
R.~{Yoshii}, S.~{Yamashika}, and S.~{Tsuchiya}, ``{Entanglement propagation in
  thermalization of an isolated quantum system},'' {\em arXiv e-prints},
  p.~arXiv:2003.10106, Mar. 2020.

\bibitem{YellowBook}
P.~Di~Francesco, P.~Mathieu, and D.~Senechal, {\em {Conformal Field Theory}}.
\newblock Graduate Texts in Contemporary Physics, New York: Springer-Verlag,
  1997.

\bibitem{polchinski2001string}
J.~Polchinski, {\em String Theory}, vol.~1 of {\em Cambridge Monographs on
  Mathematical Physics}.
\newblock Cambridge University Press, 1998.

\end{thebibliography}

\end{document}